\documentclass{vgtc}                          

\ifpdf
  \pdfoutput=1\relax                   
  \usepackage{graphicx}                
  \DeclareGraphicsExtensions{.pdf,.png,.jpg,.jpeg} 
\else
  \usepackage{graphicx}                
  \DeclareGraphicsExtensions{.eps}     
\fi%

\graphicspath{{figures/}{pictures/}{images/}{./}} 

\usepackage{microtype}                 
\PassOptionsToPackage{warn}{textcomp}  
\usepackage{textcomp}                  
\usepackage{mathptmx}                  
\usepackage{times}                     
\usepackage[noadjust]{cite}                      
\usepackage{tabu}                      
\usepackage{booktabs}                  

\usepackage[mathcal]{eucal}
\usepackage{amsmath} 
\usepackage{bm}
\usepackage{balance}
\usepackage[font=footnotesize,labelfont=bf]{caption}
\usepackage{multirow}
\usepackage{multicol}
\usepackage{colortbl}       
\usepackage{enumitem}
\usepackage{xr}
\usepackage{xcolor}
\usepackage{xspace}
\usepackage{placeins}
\usepackage{float}
\usepackage{tabularray}
\usepackage{tablefootnote}
\usepackage{hyperref}
\usepackage{xr-hyper}
\usepackage{soul}
\soulregister\cite7 
\usepackage[noadjust]{cite}
\soulregister\ref7
\soulregister\item7
\soulregister\itemize7
\soulregister\description7
\soulregister\enumerate7
\soulregister\footnote7
\soulregister\todo7
\soulregister\underline7
\soulregister\bad7
\soulregister\rev7
\soulregister\fixed7

\newcommand{\todo}[1]{{\color{red} [#1]}}

\newcommand{\toreview}[2][yellow]{%
  {\sethlcolor{#1}\hl{\leavevmode#2}}%
}
\definecolor{revision_color}{HTML}{FFFFFF}
\newcommand{\rev}[1]{\toreview[revision_color]{#1}}
\definecolor{fixed_color}{HTML}{FFFFFF}
\newcommand{\fixed}[1]{\toreview[fixed_color]{#1}}
\definecolor{bad_color}{HTML}{FF5555}
\newcommand{\bad}[1]{\toreview[bad_color]{#1}}

\vgtccategory{Research}

\vgtcinsertpkg

\title{SketchXplain: Intuitive Visual Explanations of Image Classifiers with Sketches}

\author{
Wencan Zhang\thanks{These authors contributed equally to this work.} \and
Mario Michelessa\footnotemark[\value{footnote}] \and
Xuejun Zhao \and
Brian Y. Lim\thanks{Corresponding author. e-mail: brianlim@nus.edu.sg}
}
\affiliation{\scriptsize National University of Singapore}

\teaser{
  \centering
  \includegraphics[width=0.82\textwidth]{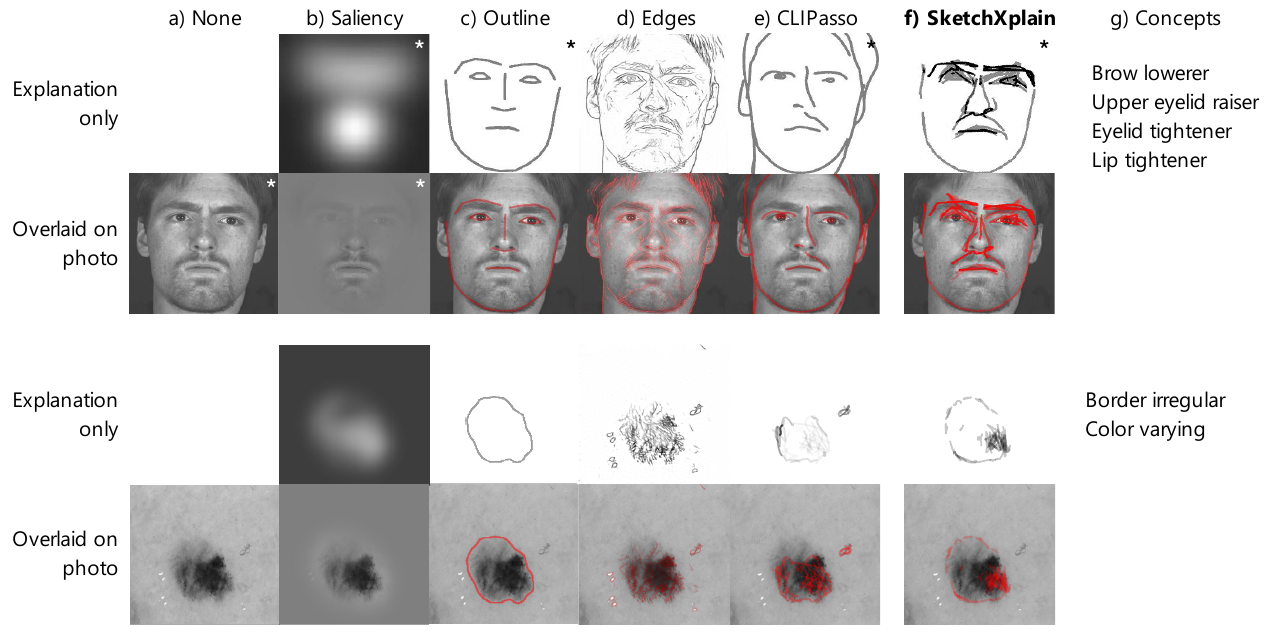}
  \vspace{-0.3cm}
  \captionsetup{width=1.09\textwidth}
  \caption[width=\textwidth]{ 
    Visual \fixed{saliency} (b), sketch (\fixed{c--f}) and verbal explanations 
    of image classifications for two domains, face expression (angry) and skin lesion (melanoma):
    a) None,
    b) Saliency map~\cite{selvaraju2017grad} 
    c) Outline \fixed{from facial landmarks~\cite{kazemi2014one} or lesion segmentation masks~\cite{tschandl2020human}},
    d) Edges detected~\cite{chan2020deep} from photo,
    e) CLIPasso~\cite{vinker2022clipasso} sketch that neglects explanatory cues,
    f) our SketchXplain that 
    leverages sketches for intuitive explanations,
    g) Concepts labels~\cite{koh2020concept}.
    We compared a subset * of visualizations in quantitative user studies.
  }
  \vspace{-0.2cm}
  \label{fig:use-case-demo}
}

\abstract{
Saliency map \fixed{visualizations} explain image-based AI predictions by pointing to regions, but these are often unintuitive and semantically unclear, leaving an interpretability gap. 
We argue that AI explanations should be intuitive--—coherent to user knowledge\fixed{, }yet simple and selective to accelerate interpretation. 
Inspired by artistic drawings, we propose SketchXplain to generate sketch-based \fixed{visual} explanations for intuitive image-based explainable AI (XAI). 
Combining techniques in saliency maps, concept-bottleneck models, and sketch optimization, SketchXplain integrates saliency to select coherent observation artifacts, concepts for knowledge coherence, cues to represent them, and abstraction for simplicity. 
Evaluating on face expression recognition, modeling and user studies showed that SketchXplain supported quicker interpretation with more aligned \fixed{visualizations} than saliency maps or simple drawings. 
Further evaluation on skin lesion diagnosis found that SketchXplain more coherently \fixed{visualized} disease symptoms, better supporting lay diagnosis. 
Thus, this work illustrates the value of sketches for intuitive, simple, coherent, and quick image-based XAI \fixed{visualizations}.
} 


\begin{document}

\maketitle
\raggedbottom
\setlength{\parskip}{0pt plus 0pt minus 0pt}
\firstsection{Introduction}



\section{Introduction}


Artificial Intelligence (AI) has achieved impressive performance on image prediction tasks~\cite{ren2016faster, esteva2017dermatologist}.
However, the complexity of AI models has raised concerns about their lack of transparency~\cite{lipton2018mythos, miller2019explanation}. This has driven the growing need for Explainable AI (XAI)~\cite{lim2009and, ribeiro2016should, lundberg2017unified, abdul2018trends, olah2017feature}, which aims to make model predictions more understandable to users.

Saliency maps~\cite{selvaraju2017grad, zhou2016learning, ramaswamy2020ablation, bach2015pixel, fong2017interpretable, montavon2017explaining} are among the most popular XAI techniques to explain visual tasks.
They attribute a model's prediction to specific pixels~\cite{bach2015pixel}
or intermediate neurons in the model~\cite{selvaraju2017grad, zhou2016learning}. While saliency maps faithfully represent the model's attention by highlighting relevant regions, they often fail to convey meaningful semantics, rendering explanations unintuitive~\cite{selbst2018intuitive}.
For instance, a user might know where to focus but fail to understand why that region is relevant (Fig.~\ref{fig:use-case-demo}b). 
This limited intuitiveness increases cognitive effort~\cite{boggust2025abstraction}, and makes it difficult to interpret~\cite{kaur2022sensible, zhang2022towards} or anticipate~\cite{alqaraawi2020evaluating} model behavior.

\begin{figure}[!t]
    \centering
    \includegraphics[width=6.5cm]{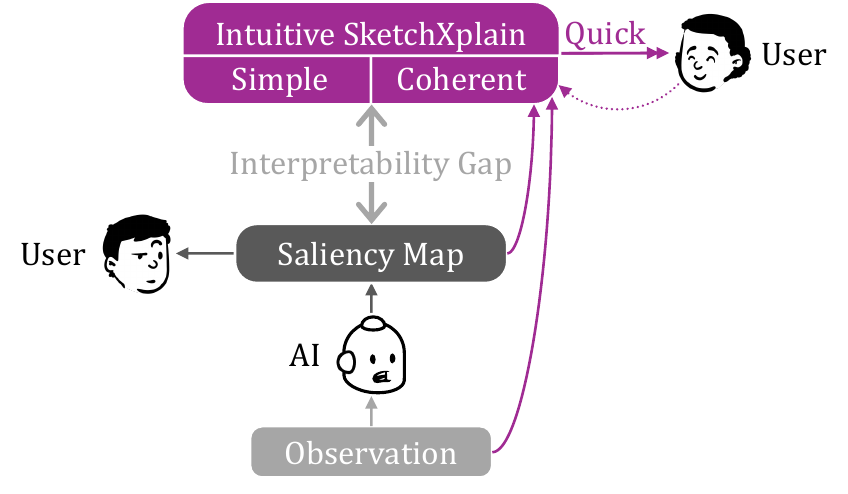}
    \vspace{-0.3cm}
    \caption{
        Interpretability gap in saliency maps is addressed by intuitive sketch-based explanations that are simple and coherent, leading to quicker interpretation.
    }
    \label{fig:conceptual-gap}
    \vspace{-0.5cm}
\end{figure}

One intuitive way to convey how images are perceived is through \fixed{\textit{simplified}} abstract \fixed{\textbf{sketches}} that emphasize salient strokes,
omitting less important contour lines~\cite{fish1990amplifying}\fixed{, while remaining \textit{coherent} to key concepts}.
Since sketches are perceived as approximately realistic images~\cite{hertzmann2020line}, 
people can \fixed{\textit{quickly}} and accurately categorize scenes and recognize objects from them~\cite{walther2011simple}.
\fixed{Beyond artistic abstraction, this simplicity, coherency, and quickness, enables sketches to be used as a shared boundary object to communicate between disciplines~\cite{retelny2016embedding}.}
This opens a new opportunity for communicating \fixed{semantic explanations \textit{visually}} beyond verbalizing concept-based explanations~\cite{kim2018interpretability, koh2020concept}.
%
Thus, we identify three desiderata of \textit{Intuitive Explanations}---\textit{Simplicity}~\cite{miller2019explanation, abdul2020cogam, wang2025less}, \textit{Coherence}~\cite{thagard1989explanatory, miller2019explanation, nauta2023anecdotal} and \textit{Quick Interpretation}~\cite{abdul2020cogam, swaroop2024accuracy, wang2025less}\fixed{---to close the interpretability gap of saliency maps (Fig.~\ref{fig:conceptual-gap})}.
We operationalize these desiderata to 
\fixed{define design objectives and evaluation measures for}
intuitive sketch explanations that align with human knowledge, reflect the visual observation, and incorporate saliency-informed \fixed{semantic cues}.

Guided by these desiderata, we propose \textbf{SketchXplain}, a deep neural network to generate simple yet coherent sketch explanations (Fig.~\ref{fig:use-case-demo}f). 
SketchXplain consists of 
1) base image encoder,
2a) concept bottleneck model to predict intermediate concepts and combine them to predict the class label,
2b) cue localization to locate each concept in the image,
3a) stroke initialization to extract edges from the image and filter them based on localization, and
3b) stroke optimization to refine the sketch to faithfully represent the image and prediction label.
Although the sketch explanation is model-agnostic, like LIME~\cite{ribeiro2016should}, it is faithful to the model's input and prediction while providing plausible, human-accessible explanations---like AI rationalization from human labels~\cite{ehsan2018rationalization} or large language models~\cite{gat2024faithful}.

We \fixed{first} evaluated SketchXplain on a facial expression task using various intuitiveness measures.
We compared SketchXplain explanations against saliency maps and other line-drawing methods across multiple studies \fixed{(Section~\ref{sec:face_eval})}:
a) \fixed{preliminary} modeling studies assessing \fixed{proxies of} explanation visual complexity and coherence; 
b) qualitative user study identifying interpretation benefits; 
c) quantitative user study evaluating quick interpretation under time constraints.
For generalization, we extended SketchXplain to explain skin lesion predictions \fixed{(Section~\ref{sec:evaluation_skin_lesion}) and general images (in Appendix)}.
\fixed{All user studies were approved by our university institutional review board (IRB).}
Overall, we demonstrate that SketchXplain produces sketches that are coherent and sufficiently simple, enabling more intuitive visual explanations.

\rev{Hence, we frame SketchXplain visualizations as articulating AI predictions to \textit{intuitively} guide final user decisions.}
Our \textbf{contributions} are:
\begin{enumerate}[label=\arabic*), leftmargin=*, itemsep=0.em, topsep=0.2em, parsep=0pt, partopsep=0pt]
    \item First to use sketches as an \fixed{intuitive} explanatory modality for visual explanation to close \fixed{the} interpretability gap of \fixed{saliency maps}. 
    \fixed{We steer sketch generation toward explanatory coherence.}
    \item \fixed{SketchXplain, an explanation method to generate intuitive sketch visualizations, satisfying the desiderata of simplicity, coherence, and quick interpretation for image classifiers.}
    \item Demonstration of sketch explanations on \fixed{multiple domains (}faces and skin lesions\fixed{), with proxy and user evaluations}.
\end{enumerate}

\section{Background and Related Work}
\fixed{We discuss the landscape of image-based XAI and its cognitive challenges, providing the basis for using visual sketches as intuitive explanations.}

\subsection{Explainable AI for Computer Vision}
\label{sec:related_work_xai_for_cv}

\rev{Many visualizations have been developed to understand how CNNs infer concepts and prediction labels from images
~\cite{bau2017network, hohman2020summit, liu2017towards, kahng2018activis, wang2021cnnexplainer, wang2020genealogy}.
While these help students or developers better learn about or debug machine learning, they are less accessible to lay users.
Instead, attribution and concept-based explanations are more popular to support end-user understanding of AI decisions.}

Attribution explanation methods \fixed{report} feature importance by attributing model outputs to input features~\cite{ribeiro2016should, lundberg2017unified, krause2017workflow}, allowing \fixed{users} to focus on \fixed{salient} features for reasoning.
In image-based AI prediction tasks, saliency map techniques highlight influential pixels using gradient~\cite{selvaraju2017grad}, perturbation~\cite{fong2017interpretable}, decomposition~\cite{montavon2017explaining}, ablation~\cite{ramaswamy2020ablation}, or relevance propagation~\cite{bach2015pixel}.
Although pointing to specific regions can direct a user's attention, 
\fixed{saliency maps lack semantic contextualization, limiting intuitive interpretation.}

In contrast, 
concept-based XAI examines the influence of human-interpretable concepts. 
\fixed{Techniques include verbally explaining via concept bottlenecks~\cite{koh2020concept},
interactively inspecting post hoc concepts~\cite{kim2018interpretability, zhao2021conceptextract, cai2019human}, and 
visualizing learned neurons or filters~\cite{olah2017feature, zeiler2014visualizing}.}
While these explanations are more semantically-aligned with 
\fixed{domain knowledge,
they are not visually linked to the representation or cues in the image, leaving a coherence gap}.

In this work, we are the first to investigate the usability and \fixed{usefulness} of sketches as an explanation modality, moving beyond simple attribution of influential features or verbal concept descriptions.
It is important to note that our approach does not share the same goal as techniques to explain sketch generation~\cite{qu2023sketchxai, bandyopadhyay2024sketch}. By treating concepts and cues as foundational elements, we use saliency methods to adjust their combination, achieving coherent yet simple sketches that are intuitive for human interpretation.

\subsection{Visualization for End-User Human Interpretability}
\label{sec:related_work_vis_for_interpretability}

\rev{Various XAI visualizations, including rule matrices~\cite{ming2019rulematrix}, beam search trees~\cite{strobert2019seq2seq}, simplified network graphs~\cite{hohman2020summit}, network traces~\cite{wang2021cnnexplainer}, or interactive distribution plots~\cite{kahng2019ganlab},
have been developed to assist data scientists in debugging data issues and model behavior.
Although some help end-users to understand
models through saliency maps~\cite{boggust2022shared} and class tree hierarchies~\cite{boggust2025abstraction},
the visual formats remain complex for lay users.}


\fixed{In practice, XAI techniques have faced usability challenges,}
including misleading or incomplete explanations~\cite{bansal2021does, ehsan2024xai},
misinterpretations~\cite{kaur2020interpreting}, and over-reliance~\cite{buccinca2021trust, wang2021explanations}. 
To address these issues, researchers have advocated for human-centered XAI~\cite{miller2019explanation, wang2019designing, liao2021human}.
\fixed{Approaches include 
moderating cognitive load and improving memorability~\cite{abdul2020cogam, bo2024incremental},
aligning AI architectures to human reasoning processes~\cite{zhang2022towards, matsuyama2023iris, lim2025diagrammatization},
anticipating human misconceptions~\cite{buccinca2025contrastive}, 
forcing deliberation~\cite{buccinca2021trust},
progressive disclosure on demand~\cite{lim2011design, lim2013evaluating, springer2019progressive}, and
selective or tailored explanations~\cite{wang2019designing, lai2023selective, lim2009assessing, liao2020questioning}.

In contrast, we propose a complementary approach 
that leverages sketches as a selective visual abstraction for intuitive visual explanations.
We hypothesize that such explanations would engage System 1 thinking~\cite{kahneman2011thinking} to help users rapidly assimilate an understanding of the AI decision for visual tasks.}

\subsection{Illustrative Visual Abstraction for Semantic Communication}
Unlike photographs with full visual detail, illustrations~\cite{male2017illustration} are flexible media used to explain or enrich ideas, concepts, or narratives, intentionally abstracting or stylizing information to guide attention~\cite{dwyer1970exploratory}, clarify concepts~\cite{cook2008students}, or convey meaning beyond literal imagery~\cite{male2017illustration}\fixed{, making them highly effective for visualizations~\cite{agrawala2011design}.
To computationally replicate this communicative abstraction, non-photorealistic rendering (NPR)~\cite{gooch2001non} methods synthesize illustrations from photographs or 3D models.}
Unlike traditional rendering which focuses on realistic lighting and shading, NPR emphasizes artistic styles or structural information, such as contour extraction~\cite{hertzmann1999introduction}, stroke-based rendering (SBR)~\cite{kazi2012vignette}
and stylization~\cite{kyprianidis2012state}. 
\fixed{However, these neglect cognitive principles for communicating illustrations~\cite{agrawala2011design}.

Aligning with Agrawala et al.'s paradigm of abstraction~\cite{agrawala2011design}, we leverage illustrative visual sketching not to assert an artistic style, but to emphasize semantic cues.
By using a sparse set of} strokes to emphasize a subject's core semantic essence~\cite{hertzmann2020line, chamberlain2016genesis}, 
sketches bridge the gap between spatially depictive geometry and structurally descriptive semantic concepts~\cite{fish1990amplifying}.
\fixed{However, traditional sketching guidelines and classical NPR rendering pipelines operate purely as static post-processing on raw visual geometry, 
lacking the capacity to dynamically extract, represent, or prioritize the underlying activations to explain predictions of image-classifier models.}

\subsection{AI Sketch Generation from Photographs}
Sketch generation differs from edge-map extraction~\cite{canny1986computational},
 which is purely geometric, by producing abstracted drawings that \fixed{prioritize} semantics. 
\fixed{Recent AI-based methods
employ} image-to-image transformation between photos and sketches~\cite{yi2019apdrawinggan}
or unpaired domain transfer~\cite{isola2017image, wang2018high}, which can be trained with cycle consistency losses~\cite{chan2022learning}.
Although these methods generate sketches in various styles, they rely heavily on curated sketch datasets. 
\fixed{However, target sketches corresponding to source images may not exist.

Instead, without requiring supervised learning via target sketches, graphical} vector-based \fixed{(vs. bitmap-based)} approaches \fixed{have been proposed to} generate sketches~\cite{ha2017neural, ribeiro2020sketchformer, chen2017sketch}
via differentiable rendering~\cite{li2020differentiable}.
For example, recent work~\cite{vinker2022clipasso, vinker2023clipascene} define sketches as sets of B{\'e}zier curves and optimizes stroke parameters and leverage CLIP~\cite{radford2021learning} with a joint image-text latent space to align visual and textual representations. 
\fixed{While these methods do achieve simplified sketches, faithful to the photos and textual descriptions, they neglect explanatory cues and semantics important for user understanding of image-AI predictions, which we integrate seamlessly in SketchXplain.}

\section{\rev{Desiderata for Intuitive Visual Explanations}}
\label{sec:conceptual}

To bridge the interpretability gap of XAI and reduce misinterpretations, we argue that AI explanations should be more intuitive.
\fixed{Cognitive psychologists describe intuition as 
thinking \textit{``automatically and quickly, with little or no effort''}~\cite{kahneman2011thinking},
\textit{``focusing on the relevant and deliberately ignoring the rest''}~\cite{todd2000precis}, 
such that \textit{``once experienced intuitive decision makers see the pattern, any decision they have to make is usually obvious''}~\cite{klein2004power}.}
Therefore, information that is \textit{coherent} to user expertise, yet \textit{simple} and focused, helps facilitate \textit{quick} intuitive reasoning.
%
Consolidating XAI literature, we target the following desiderata for intuitive visual explanations:
\begin{enumerate}[label=\arabic*), leftmargin=*, itemsep=0.em, topsep=0.1em, parsep=0pt, partopsep=0pt]
    \item \textit{Simple}~\cite{miller2019explanation, abdul2020cogam, wang2025less} to convey salient information clearly and concisely.
    \item \textit{Coherent}~\cite{miller2019explanation, nauta2023anecdotal, thagard1989explanatory} to align with beliefs, evidence and hypotheses.
    \item \textit{Quick}~\cite{abdul2020cogam, swaroop2024accuracy, wang2025less} to allow easy understanding and avoid confusion.
\end{enumerate}
From these three desiderata, we derive explanation design objectives, proxy modeling evaluation metrics, and user-study measures, which
we apply to sketch-based explanations of image-based AI predictions.

\subsection{Explanation Design Objectives}
In Section~\ref{sec:technicalApproach}, we describe our technical approach toward intuitive sketch explanations.
To support \textit{simplicity}, we optimize sketches for \textbf{abstraction} to ignore less relevant fine details, with \textbf{smooth strokes} rather than jagged ones. 
To support \textit{coherence}, we optimize sketches to be representative of the AI \textbf{prediction task}, explanatory \textbf{concept} labels from human knowledge, \textbf{cues} that visually represent concepts, and the visual \textbf{observation} of the instance. 

\subsection{Proxy Evaluation Metrics}
\label{sec:conceptual_proxy_metrics}
In Section~\ref{sec:eval_face_modeling}, we evaluated each design objective with corresponding \fixed{proxy} metrics.
For simplicity, we evaluated its opposite \textit{visual complexity} in terms of information in the spatial domain (e.g., local entropy) and frequency domain (e.g., discrete cosine transform in JPEG XL).
For \textit{coherence}, we measured the alignment of the sketch's 
\fixed{spatial representation to the original photo and extracted visual cues, 
and embedding representation to the embeddings for the AI prediction and explanatory concept labels.}

\subsection{User Study Measures}
\label{sec:conceptual_user_measures}
\fixed{Since explanations are for human interpretation, we conducted user studies to evaluate explanation intuitiveness.}
Sections~\ref{sec:eval_face_qualitative} and \ref{sec:eval_lesion_qualitative} report qualitative user studies to examine the inherent \textit{coherence} and \textit{usability} of sketch explanations.
Section~\ref{sec:eval_face_quantitative} \fixed{reports} a quantitative study under time constraints to assess if users can \textit{quickly interpret} 
\fixed{the sketch explanation to recall relevant concepts and infer
prediction labels.}

\section{SketchXplain: Technical Approach}
\label{sec:technicalApproach}

\begin{figure*}[!t]
    \centering
    \includegraphics[width=13.0cm]{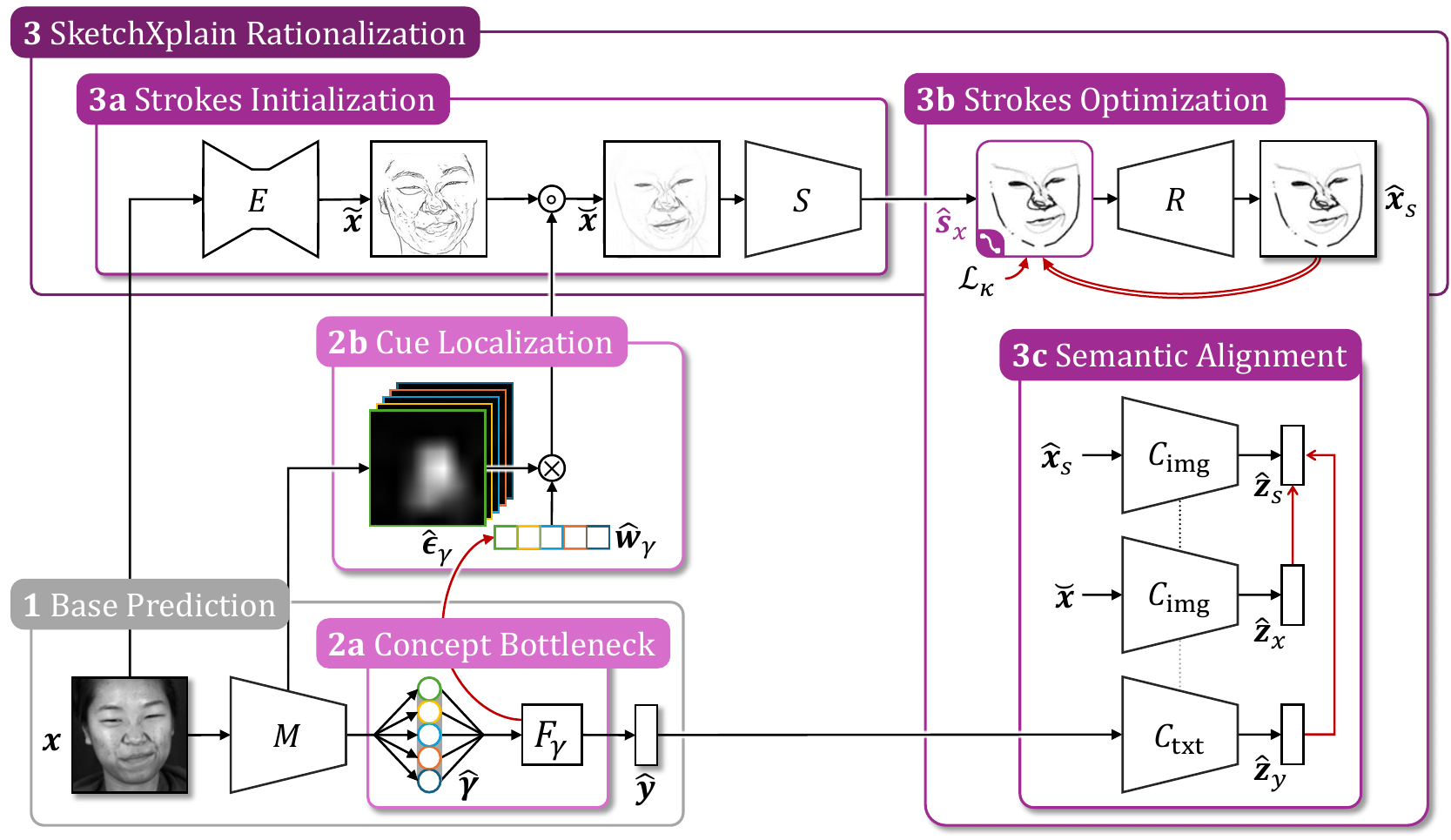}
    \vspace{-0.3cm}
    \caption{
        SketchXplain architecture comprising:
        1) base prediction, 
        2) base explanations in terms of concepts (2a) and cues (2b),
        3) sketch explanation to generate initial strokes (3a), and optimize them (3b) to align them with the predicted class label (3c).
        Instance shown for a face expression use case (Section~\ref{sec:face_eval}).
    }
    \label{fig:architecture}
    \vspace{-0.5cm}
\end{figure*}

We propose SketchXplain, an explainable deep neural network with three main modules (Fig.~\ref{fig:architecture}): 1) base prediction, 2) concept explanation, and 3) sketch rationalization.
\fixed{It satisfies two desiderata 
to generate \textit{simple} strokes by abstracting visual information into smoothed curves,
which are \textit{coherent} to the image prediction task and explanatory concepts by aligning visual cues to semantic embedding representations and the observed photo.}


\subsection{Base Prediction}
We begin with the base task (Fig.~\ref{fig:architecture}, Step 1) of predicting class label $\hat{y}$ (e.g., expression) from input image $\bm{x}$ (e.g., face) \fixed{using an image encoder $M$}.

\subsection{Concept Explanations}
We provide a base explanation of concepts \fixed{via a concept bottleneck network} (Fig.~\ref{fig:architecture}, Step 2a) and \fixed{localizations} of associated cues in the image \fixed{via concept-specific saliency maps} (Fig.~\ref{fig:architecture}, Step 2b),
\fixed{which are subsequently used to constrain the coherence of the sketch explanation}.

\subsubsection{Concept Bottleneck}
\label{sec:technical_approach_cbm}
Explanations should be relatable~\cite{zhang2022towards} to concepts~\cite{kim2018interpretability, koh2020concept}. 
SketchXplain first predicts concepts $\hat{\bm{\gamma}}$ \fixed{via} multi-label binary classification \rev{as multiple concepts can co-occur}\footnote{\rev{Multi-label binary classification handles multiple non-exclusive binary labels, while multi-class classification only selects one.} 
For example, Facial action unit AU1 Inner Brow Raiser, AU2 Outer Brow Raiser, AU5 Upper Eyelid Raiser, and AU26 Jaw Drop co-occur for Surprised expression; see Table~\ref{table:ausbyemo}.}. 
\fixed{Concepts are then used as intermediate features} in a concept bottleneck model $F_\gamma$~\cite{koh2020concept} to predict the label $\hat{y}$ \fixed{via} multi-class classification. 

\subsubsection{Cue Localization} \label{sec:cue_localization}
For image-based predictions, saliency maps are widely used to explain which pixels \fixed{are} important. 
We implemented Grad-CAM for transformers~\cite{jacobgilpytorchcam} to generate one CAM per \rev{multi-label} concept $\hat{\bm{\epsilon}}_\gamma$, which 
\fixed{highlights important pixels for predicting each}
corresponding concept $\hat{\bm{\gamma}}$. 
\fixed{We extracted importance weights $\hat{\bm{w}}_{\gamma}$ (i.e., gradients of prediction label w.r.t. concepts) through the fully connected layers of the concept bottleneck $F_\gamma$.}
Unlike concept predictions $\hat{\bm{\gamma}}$ that communicate whether a concept is \textit{present}, $\hat{\bm{w}}_{\gamma}$ \textit{explains} how important each concept is to determine the class label.

To ensure more sensible saliency maps, we regularize the saliency prediction by making Grad-CAM a secondary prediction task trained via self-supervision~\cite{zhang2022debiased}.
Specifically, we \fixed{penalize} salient pixels that are far away from where each concept is located. 

\subsection{Sketch Rationalization}
With the semantic information of concept-based explanations, and the localization information from saliency maps, we can develop more intuitive explanations using sketches.
Specifically, our approach extracts and abstracts informative \textit{Concept lines} relevant to concepts to communicate key details as explanations.
This extraction and sketching process involves extra pathways \fixed{to rationalize} beyond the scope of the original prediction model $M$, so it is not a mechanistic interpretability approach~\cite{conmy2023towards}.
Nevertheless, \fixed{explanation} rationalization is a plausible approach~\cite{ehsan2018rationalization} for integrating astute observations from a sketch expert (as for artists) with attentive feedback from the target predictor (as for patrons or clients).

We generate sketch explanations with two steps: 
a) initialize informative strokes extracted from the photo (Fig.~\ref{fig:architecture}, Step 3a), 
b) draw the sketch starting from those strokes and optimize it to best represent the original photo and its prediction label (Fig.~\ref{fig:architecture}, Step 3b).
Prior work, CLIPasso~\cite{vinker2022clipasso}, draws a small set of curved lines that capture the semantics of a photo (via CLIP~\cite{radford2021learning}) by optimizing the similarity between the final sketch and the input photo.
\fixed{This serves the goal of \textit{simplicity}, but not coherence to explanation.}
\fixed{We \textbf{extend} CLIPasso to additionally
i) extract more detailed lines than just object outlines, 
ii) infer explanatory concepts relevant to the prediction label, and
iii) align strokes to visual cues representing corresponding concepts.
Thus, this serves the goal of explanation \textit{coherence}.}

\subsubsection{Strokes Initialization}
\label{sec:stroke_init}
To generate the sketch, CLIPasso starts with an initial set of strokes whose locations are sampled from salient regions. 
However, CLIPasso's stroke initialization often fails to produce semantically meaningful sketches (see Table~\ref{table:demo-face}, CLIPasso). 
\fixed{To address this, we use \textit{detailed lines} to capture finer details, such as skin folds and dynamic wrinkles in faces, or roughness in textured surfaces.}
These \fixed{detailed lines} are extracted into a bitmap image $\tilde{\bm{x}}$ using the pretrained image-to-image generator $E$ by Chan et al.~\cite{chan2022learning}. 
\rev{See Appendix Tables~\ref{table:demo-face-intermediate} and~\ref{table:demo-lesion-intermediate}, last two columns, for a comparison between initialization with detailed lines $\hat{\bm{z}}_x$ and with original photo $\bm{x}$.}

\fixed{As in CLIPasso, we prioritize the most salient lines of $\tilde{\bm{x}}$ by
leveraging} saliency maps $\hat{\bm{\epsilon}}_\gamma$ and importance weights $\hat{\bm{w}}_\gamma$ from \fixed{Section~\ref{sec:cue_localization}}.
Using Grad-CAM, we merge all CAMs as a weighted-sum $\hat{\bm{\epsilon}} = ReLU \left( \hat{\bm{w}}_\gamma^\top \hat{\bm{\epsilon}}_\gamma \right)$, and use the aggregated saliency map to mask the detailed lines based on their importance, i.e., $\Breve{\bm{x}} = \hat{\bm{\epsilon}} \odot \tilde{\bm{x}}$, where $\odot$ is the Hadamard element-wise multiplication.
\fixed{This aims to improve the explanation \textit{coherence} to the explanatory conceptual cues.
These weighted detailed lines are used in the next stage (stroke optimization) to sample the stroke lines.}

\subsubsection{Strokes Optimization}

As in CLIPasso, we seek to generate \textit{parametric strokes} $\hat{\bm{s}}_x$ that can be updated using gradient-based optimization.
We initialize these strokes by sampling $S$ endpoints for quadratic B{\'e}zier curves from $\Breve{\bm{x}}$, treating it as a probability density function where more salient pixels have a higher probability of being sampled.
The parametric strokes $\hat{\bm{s}}_x$ can be rasterized into a sketch $\hat{\bm{x}}_s$ using a differentiable rasterizer $R$~\cite{li2020differentiable}.
Initially, the rasterized sketch $\hat{\bm{x}}_s$ will resemble a sparse fragmented rendition\footnote{See Appendix Tables~\ref{table:demo-face-intermediate} and~\ref{table:demo-lesion-intermediate}, column Init. strokes.} of $\Breve{\bm{x}}$, 
but can be improved through iteratively updating $\hat{\bm{s}}_x$ during inference.
\fixed{Unlike model training where weights are updated, we freeze the weights in $R$, and use iterative optimization via gradient descent at inference time to update the input $\hat{\bm{s}}_x$, like with activation maximization~\cite{erhan2009visualizing}.}

%

To guide the sketch optimization, we \fixed{perform \textit{Semantic Alignment} (Fig.~\ref{fig:architecture}, Step 3c) of the sketch explanation $\hat{\bm{x}}_s$ to the input photo $\breve{\bm{x}}$ and predicted label $\hat{y}$.}
We leverage CLIP~\cite{radford2021learning}, which encodes images and text into a shared feature space \fixed{to enable direct comparison between visual and textual representations.
Specifically, we obtain embedding representations for the sketch $\hat{\bm{z}}_s$, input photo $\hat{\bm{z}}_x$, and prediction label $\hat{\bm{z}}_y$.
The original CLIPasso only aligns the sketches toward the visual semantics $\hat{\bm{z}}_x$, but we add alignment toward the prediction label $\hat{\bm{z}}_y$.
We then penalize large cosine distances between embedding vectors.
This aims to improve the explanation \textit{coherence} to the image and prediction label.
To improve the explanation \textit{simplicity}, we also include a smoothness constraint $\kappa(\hat{\bm{s}}_x)$ for less curvy strokes.}
Thus, the full training loss is
\begin{equation}
L = 
-\cos(\hat{\bm{z}}_s, \hat{\bm{z}}_x) 
- \lambda_t \cos(\hat{\bm{z}}_s, \hat{\bm{z}}_y) 
+ \lambda_{\kappa} \kappa(\hat{\bm{s}}_x),
\label{eq:loss}
\end{equation}
where $\kappa(\hat{\bm{s}}_x) =
\frac{|\hat{\bm{s}}_{x,1}''\hat{\bm{s}}_{x,2}'-\hat{\bm{s}}_{x,1}'\hat{\bm{s}}_{x,2}''|}{(\hat{\mathbf{s}}_{x,1}'^2+\hat{\mathbf{s}}_{x,2}'^2)^{3/2}}$
is the curvature of a stroke $\hat{\bm{s}}_x = (\hat{\mathbf{s}}_{x,1},\hat{\mathbf{s}}_{x,2})$ in 2D, with primes indicating first- and second-order derivatives. $\lambda_t$ and $\lambda_{\kappa}$ are the corresponding hyperparameters. 
\fixed{Tables~\ref{table:demo-face} and~\ref{table:demo-lesion} show example SketchXplain visualizations of face expressions and skin lesions, respectively.}

\subsection{Implementation Details}
\label{sec:implementation}
We implemented SketchXplain in PyTorch. 
For concept inference, we fine-tuned the final layer of the concept bottleneck model with Adam~\cite{kingma2014adam} for 100 epochs (batch size 32), and used cross-entropy loss for classifications. 
For sketch optimization, we also used Adam, with a learning rate of 1.0 \rev{(applying the full gradient magnitude)} for stroke positions and $10^{-2}$ for stroke opacity. 
\fixed{Like \cite{vinker2022clipasso, frans2022clipdraw},} to improve stability, gradients were averaged over four sketches \fixed{as data augmentions via} random affine transformations \fixed{on each sketch instance}. 
Hyperparameters in Eq.~\ref{eq:loss} were set to $\lambda_t = 0.1$ and $\lambda_{\kappa} = 0.01$ based on grid search. To encourage finer details, $\lambda_t = 0$ was set during the final 30\% of training. 
For each image, we optimized sketch parameters over 1000 iterations (\fixed{taking about 10 sec} per image), as in \cite{vinker2022clipasso, frans2022clipdraw}. 
\fixed{To improve simplicity, in the rasterizer~\cite{li2020differentiable}, we enabled stroke opacity to de-emphasize less important lines, and discretized them to two levels\footnote{\fixed{Pilot evaluations found $\geq 3$ levels less legible, with messy information overload.}}.}
All experiments were conducted on a server with 8 NVIDIA RTX 3090 GPUs.

\begin{table*}[!t]
    \centering
    \caption{
        Typical Action Units (AUs) are the basis for concept-based facial expression explanations. 
        12 AUs are arranged in two rows, where the first row shows upper face AUs, while the second row shows lower face AUs. 
        Pictograms adapted from Blasberg et al.~\cite{blasberg2023you}.
    }
    \vspace{-0.3cm}
    \includegraphics[width=16.0cm]{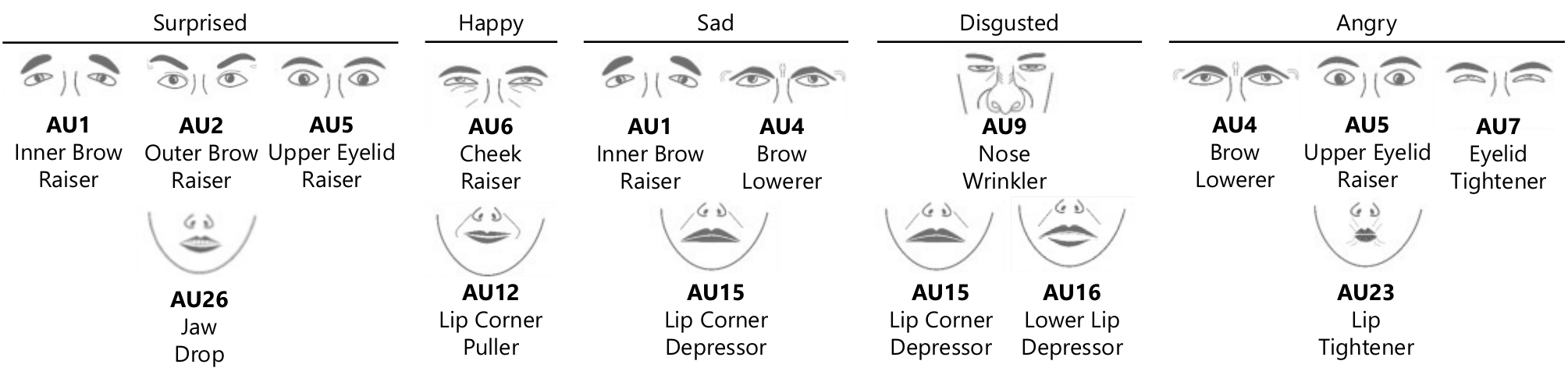}
    \label{table:ausbyemo}
    \vspace{-0.6cm}
\end{table*}

\section{Evaluation on Face Expression}
\label{sec:face_eval}
\fixed{We investigated the usability and usefulness of sketch explanations in user studies of two application domains: facial expression recognition (this section) and skin lesion cancer diagnosis (described later in Section~\ref{sec:evaluation_skin_lesion})\footnote{We also investigated sketch explanations of general images; see Appendix~\ref{sec:general_image}.}.
We chose face expression recognition since it is accessible to lay users 
and important for applications like human-AI communication~\cite{bartlett2003real}, mental health therapy~\cite{thieme2023designing}, educational tools~\cite{rahiman2024revolutionizing}. 
Also, face sketches are a universally familiar medium~\cite{fish1990amplifying}.
To explain facial expressions, we used facial Action Units (AU) as explanatory concepts.
AUs are specific muscle movements, such as raised inner eyebrows and lip tightener, to express emotions\footnote{We focused on recognizing expressed emotions, not internal emotional states.}~\cite{ekman1978facial},
which people observe as cues to infer the emotion of the subject~\cite{sayette2001psychometric, cohn2007observer}.
Research questions: 
Are sketch explanations more ...}
\begin{enumerate}[label={RQ\arabic*)}, leftmargin=*, itemsep=0.em, topsep=0.2em, parsep=0pt]
  \item Intuitive (simpler, more coherent) than baseline \fixed{visual explanation}s (line drawings, descriptive sketches, saliency maps)?
  \item Interpretable and coherent to human intuition (qualitatively)?  
  \item Intuitively \fixed{(quickly)} interpretable to understand AI predictions?
\end{enumerate} 
\fixed{We answer these questions with a preliminary modeling study with proxy metrics (RQ1), qualitative study of user interpretation of various visualizations (RQ2), and quantitative user study for quick interpretation (RQ3).}

\rev{Furthermore, due to the abstraction in sketch explanations, sketches can also provide the benefit of privacy protection by not exposing identifiable information in the original photo.
We investigated this in another quantitative user study and, for brevity\footnote{\rev{Quick results on privacy: we found that sketch explanations could protect identity, and gender and ethnicity information almost as well as Saliency maps.}}, present details in Appendix~\ref{sec:privacy_evaluation}.}

\subsection{Data Preparation and Selection}
We used the Binghamton-Pittsburgh 3D Dynamic Spontaneous Facial Expression Database (BP4D-Spontaneous)~\cite{zhang2013high} dataset to train and test all models. This contains videos of face expressions from 41 actors (23 Female, 18 Male) with diverse ethnicities (\fixed{20 White}, 11 Asian, \fixed{6 Black}, 4 Hispanic).
Each video consists of a continuous series of images, totaling 147.5k images.
This dataset \fixed{is} primarily used for predicting action units, but it is also very suitable for expression prediction. However, it lacks expression labels, so we used HSEmotion~\cite{savchenko2023facial} to automatically annotate expressions.
Furthermore, due to the transient face changes, many images do not fully represent expressions, so we filtered out images that had low prediction confidence ($<$70\%).
To obtain a balanced dataset with equal numbers of instances per expression, we excluded two classes with low frequency (Contemptuous, Fearful), resulting in 6 expressions---Surprised, Happy, Neutral, Sad, Disgusted, and Angry. 
Table~\ref{table:ausbyemo} shows the typical associations between facial expressions and action units (AUs).
Consequently, we selected a subset of 4900 images of face photos with pseudo-labels for 6 expression classes, which we split into 80\% training and 20\% test.

Due to the sequential nature of image frames in the BP4D dataset, many test images are redundantly similar. Therefore, we selected two examples of each actor expressing a different emotion, arriving at 110 images. After balancing for ethnicity \fixed{(White, Asian, Black)}\footnote{\fixed{We omitted Hispanic due to data sparsity (too few identifies).}} and gender \fixed{(Female, Male)}, we obtained 72 images \fixed{(6 expressions $\times$ 3 ethnicities $\times$ 2 genders $\times$ 2 examples)}, which we used in our modeling and user studies.

\subsection{Preliminary Modeling Study} 
\label{sec:eval_face_modeling}

\begin{table*}[!t]
    \centering
    \caption{
        Examples comparing visualizations (cols) for explaining face expressions (rows). 
    }
    \vspace{-0.3cm}
    \includegraphics[width=12.5cm]{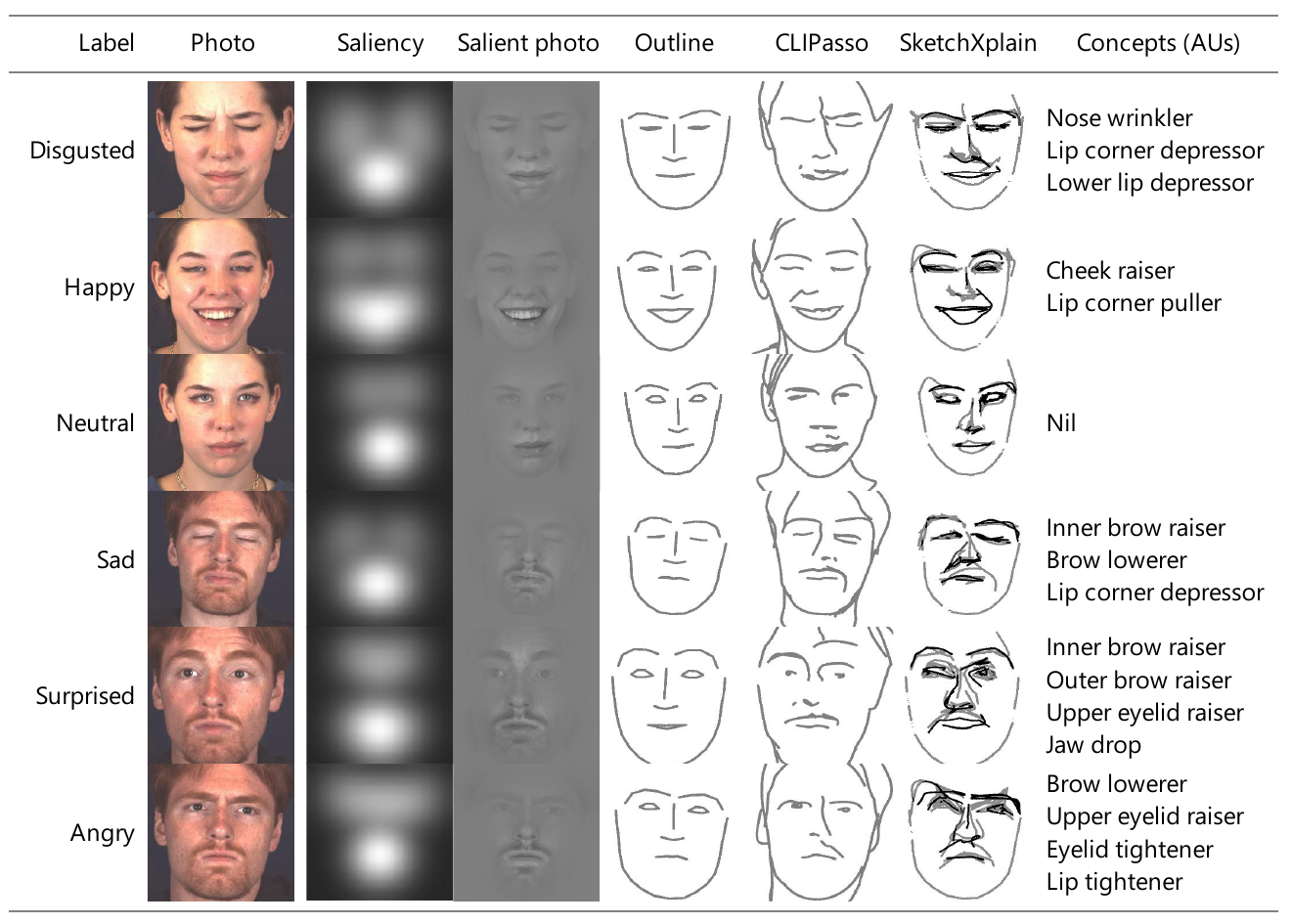}
    \label{table:demo-face}
    \vspace{-0.3cm}
\end{table*}

\fixed{Using pretrained OpenGraphAU~\cite{luo2022learning} as the backbone $M$, on the BP4D dataset, 
SketchXplain
achieved good accuracy 82.0\% and F1 Score 71.3 on face expression classification, 
per-AU F1 scores 44.3--89.2 (M = 59.9)\footnote{Comparable to prior work (M = 65.5 in~\cite{luo2022learning}).}.}
\rev{We conducted a modeling study as a preliminary check on the extent that SketchXplain generates sketch explanations toward the desiderata of simplicity and coherence with proxy metrics.}


\subsubsection{Proxy Computational Metrics}
\label{sec:proxy_metrics}
\fixed{We briefly introduce proxy metrics for simplicity and coherence.
See Appendix~\ref{sec:metrics} for more details of how specific metrics were selected.}

\textbf{\textit{Visual complexity:}}
\fixed{To estimate simplicity across heterogeneous image types, we examined various information-theoretic metrics of visual complexity:
i) \textbf{Local Shannon Entropy}~\cite{wu2013local} to 
indicate the variability of pixel intensities while accounting for spatial relationships between pixel blocks; and
ii) \textbf{JPEG XL file size}~\cite{machado2015computerized, yu2013image} to indicate information density in the \textit{spatial} (lossless) and \textit{frequency} (lossy) domains
to account for human perception of spatial frequency~\cite{sachs1971spatial}, and serving as a plausible estimator of human complexity perception~\cite{machado2015computerized}.}

\textbf{\textit{Coherence:}}
\fixed{To estimate explanation coherence with \textit{knowledge}
\rev{(\textbf{AI Alignment} to $\hat{y}$, \textbf{Concept Alignment} to $\hat{\gamma}$)} and
with \textit{observation} 
\rev{(\textbf{Cue Alignment} to $\breve{\bm{x}}$, and \textbf{Photo Alignment} to $\bm{x}$)},
we encoded each of them in a joint vision-language embedding representation
(using an image and sketch-specific evaluator model, CLIP~\cite{radford2021learning} and TASK-Former~\cite{sangkloy2022sketch}, respectively)\footnote{We also examined other evaluators (DINOv2~\cite{oquab2023dinov2} and Grounding-DINO~\cite{liu2024grounding}), but they were insensitive to all AUs, likely due to limited pretraining.}, and measured their cosine similarity~\cite{hu2023tifa}.} 

\subsubsection{Visual Explanation Comparisons}
\label{sec:compare_xai}

\begin{figure*}[!t]
    \centering
    \includegraphics[width=13.4cm]{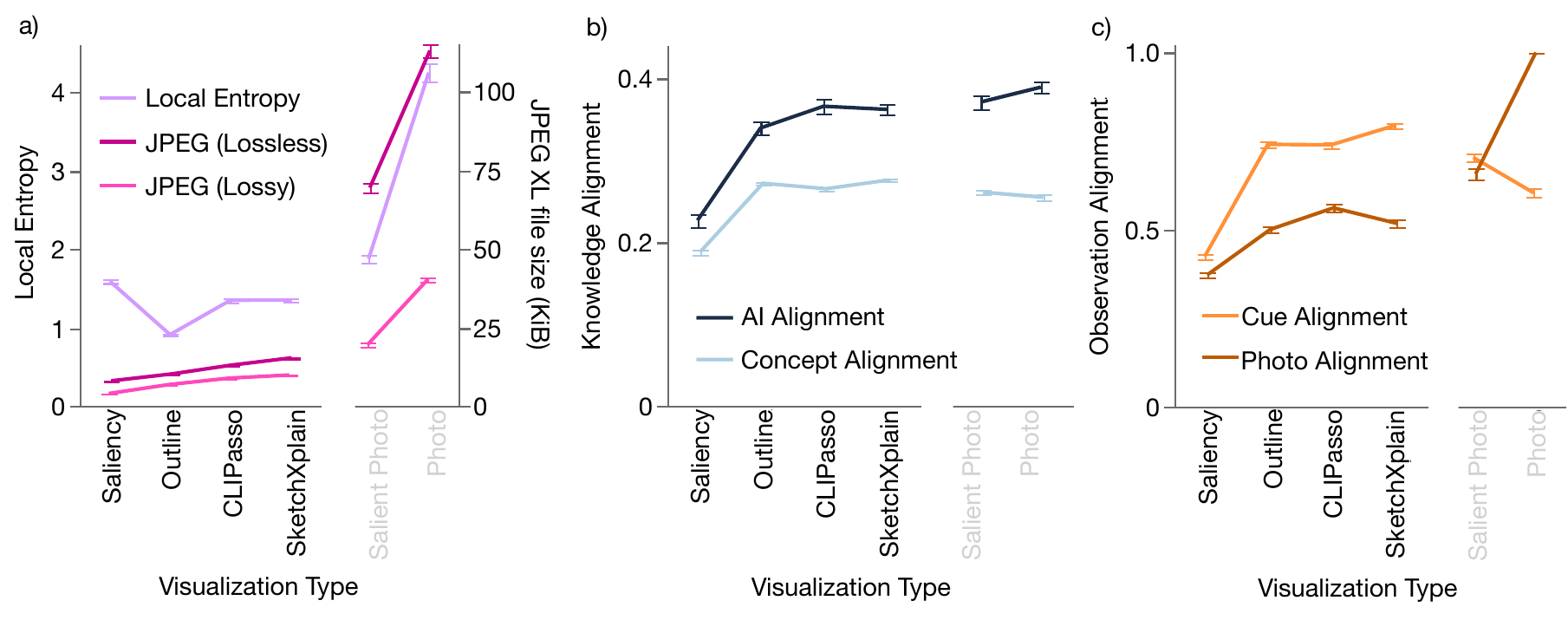}
    \vspace{-0.3cm}
    \caption{
        \fixed{Results of modeling proxy evaluation
        of visualization a) simplicity and CLIP-based coherence to b) knowledge and c) observation across line drawings and saliency explanations.
        See Appendix Fig.~\ref{fig:model-face-all-taskformer} for TASK-Former cosine similarity coherence results.}
        Saliency, Outline, and CLIPasso are baseline explanations,
        \rev{Photo is a gold standard reference and not an explanation, and Salient Photo is a hybrid explanation that partially includes Photo information.}
        Error bars indicate 95\% confidence intervals.
        }
    \label{fig:model-face-all}
    \vspace{-0.5cm}
\end{figure*}

We compared SketchXplain against 
other visualizations:
i) \underline{\smash{Saliency map}} (Grad-CAM~\cite{selvaraju2017grad}) which highlights important \fixed{explanatory} pixels but as nebulous blobs;
ii) \fixed{\underline{\smash{Outline}}} drawing which traces key facial features based on \fixed{landmarks~\cite{kazemi2014one}};
iii) \underline{\smash{CLIPasso}}~\cite{vinker2022clipasso} which abstracts lines as a descriptive sketch 
\fixed{but not an explanatory one, since it does not explicitly consider explanatory concepts.
All sketches were rendered with equal 24 strokes\footnote{See Appendix~\ref{sec:ablation} for our ablation study on stroke count.}.}
\rev{We included iv) \underline{\smash{Salient Photo}}, which overlays a saliency mask on the original image to improve usability,
but it leaks source information of the image pixels, giving it an unfair advantage to other explanations.
v) \underline{\smash{Photo}} was included only as a reference baseline for the gold standard of maximum information, but it is \textit{not} an explanation; it is the input photo independent of the AI logic.}

\subsubsection{Proxy Evaluation Results}
\fixed{Fig.~\ref{fig:model-face-all} shows the results across metrics for the compared visualizations.}

\fixed{\textbf{Visual complexity (Fig.~\ref{fig:model-face-all}a).}
Both local entropy and JPEG XL size exhibited similar trends. 
\underline{\smash{Photo}} and \underline{\smash{Salient Photo}} were most complex with pixel-level details. 
Conversely, \underline{\smash{Outline}} was the simplest, since it had the fewest strokes and visual features.
Sketch visualizations (\underline{\smash{CLIPasso}}, \underline{\smash{SketchXplain}}) occupied a middle ground. 
\underline{\smash{Saliency}} had the lowest JPEG XL file sizes, yet higher local entropy than line drawings; this reflects some inconsistency in the metrics.
Nevertheless, \underline{\smash{SketchXplain}} with intuitive facial cues is significantly simpler than the popular, usable Salient Photo.}

\fixed{\textbf{Coherence with knowledge (Fig.~\ref{fig:model-face-all}b).} 
All line drawings were highly aligned to the AI predicted labels $\hat{y}$ and Concepts $\hat{\gamma}$,
\underline{\smash{CLIPasso}} and \underline{\smash{SketchXplain}} were highest, close to the \underline{\smash{Photo}} gold standard; while \underline{\smash{Saliency}} was least aligned, likely due to its amorphous shapes.
\underline{\smash{SketchXplain}} had the best \textbf{Concept Alignment}, indicating its explanatory power.}

\fixed{\textbf{Coherence with observation (Fig.~\ref{fig:model-face-all}c).}
\underline{\smash{SketchXplain}} had the best \textbf{Cue Alignment} 
to $\breve{\bm{x}}$,
consistent with its stroke initialization and optimization constraints. \underline{\smash{Outline}} retained moderate alignment due to preserved facial structure, while \underline{\smash{Saliency}} was poorly aligned. 
For \textbf{Photo Alignment} 
to $\bm{x}$,
\underline{\smash{Salient Photo}} was highest as expected due to preserved photo pixels.
\underline{\smash{CLIPasso}} had higher alignment than SketchXplain, indicating a trade-off for photo fidelity against explanatory cues.}

\rev{In summary, these 
results
suggest that sketch explanations---particularly SketchXplain---preserve task-relevant concepts and cues while reducing visual complexity compared to pixel-based visualizations.
However, we acknowledge that the proxy metrics may not fully capture perceived simplicity and semantics, especially across diverse image modalities. 
Therefore, we further evaluated human interpretation next.}

\subsection{Qualitative User Study}
\label{sec:eval_face_qualitative}
Having investigated that SketchXplain can provide coherent yet simple visual explanations for face expressions in the modeling study, we next aim to \fixed{validate} these effects with real users.
We conducted a qualitative user study with the think-aloud protocol to understand: 
1) how people inherently identify and interpret expressions from face photos, and 
2) how various visualizations could help people to interpret face expressions.

\subsubsection{Method and Procedure}
We recruited 19 participants from a university mailing list. They were 9 males, 10 females, with ages 19--25 (Median = 22.0). We conducted the study via a recorded Zoom session\fixed{, with consent}. The experiment took 50--65 minutes and each participant was compensated \$11.50 USD in local currency. 

Participants viewed separate face visualizations across 9 trials without time constraint.
For each trial, they were asked to identify the facial expression 
from a randomly selected visualization---Photo, Outline, CLIPasso, SketchXplain, Saliency, or Salient photo. 
In separate trials, visualizations were shown alone or overlaid on the photo, with different types presented for the same original image to support reflective comparison.
While \fixed{identifying the} expression in the visualization, \fixed{participants were} instructed to think aloud 
\fixed{about cue helpfulness and difficulties.}
\fixed{Furthermore, they} could freely return to different trials for comparison.

\subsubsection{Findings}

We conducted a thematic analysis on recorded interviews, focusing on:
i) how people recognize face expressions from photos and 
ii) how they use or misuse various visualizations to interpret expressions.

In general, across different visualization types, participants \fixed{interpreted} expressions to \fixed{constituent} \textbf{facial features}, such as inferring happy from \textit{``a smiling mouth with teeth showing and smiley eyes''} [Participant P8], or identifying a sad face by \textit{``a droopy mouth, watering eyes with slanted eyebrows''} [P8]. They also described expressions in terms of \textbf{dynamic wrinkles} and \textbf{muscle contraction}. 
For example, P7 perceived surprise from \textit{``the wrinkles on forehead and the widely opened eyes''}, while P11 identified happiness based on \textit{``upward tension around his cheeks''}. 
Some participants quickly interpreted expressions based on the \textbf{shapes} of facial features, as P8 noted \textit{``look at the shape of his mouth, especially over here. (the target's mouth)''}.
Others performed a more in-depth analysis by referring to \textbf{action units cues} and \textbf{weighing} their contribution toward plausible expressions. Contemplating a sad face, P3 noted \textit{``his eyebrows are raised and his eyes are downturn, so I wasn't sure whether this part suggested sad or disgust. However, the pout here is quite prominent.''}
Participants often \textbf{pointed at} or \textbf{annotated} on key areas of the face when they were \textbf{unable to verbally articulate} the cue. While pointing at the cheeks, P7 remarked that \textit{``the skin on her face is generally expanding outwards, though I don't really know how to justify it''}. P13 combined annotation with verbal explanation: \textit{``the differentiating factor is the mouth like this''}, while drawing a curvy line around mouth, \textit{``when you're disgusted after eating something, your mouth looks like this''}. 
\textbf{Take-away:} Interpretation went beyond verbal Concepts and Saliency pointing, and required complex shapes and visual annotations that were hard to articulate. This justifies the need for sketch explanations.

Participants identified various advantages and limitations when interpreting expressions from different visualizations. 
\underline{\smash{Salient photo}}, which masked out less important regions on the face, allowed participants to \textbf{verify alignment} between their mental model and salient regions. P13 felt \textit {``I should be correct since I see the curvy mouth highlighted by the AI''}. 
However, participants generally found \underline{\smash{Saliency maps}} unhelpful since they were not overlaid on the photo, as they \textit {``don't see any semantics about the face or the expression''} [P4]. 
\underline{\smash{Outline}}, which captured the shape of facial features, were considered \textit{``\textbf{concise} and easy to interpret the surprised expression ... clearly captures the opened mouth and upturned eyebrows''} [P7].
However, participants were also concerned about its \textbf{limited details} from the original face. P13 noted \textit{``[couldn't] see the wrinkles around cheeks and nose''}. 
and was also \textbf{misled} to think \textit{``the photo seems more surprised, but the visualization looks more happy to me.''}
Compared to the Outline, \underline{\smash{CLIPasso}} had more \textbf{expressive} lines that were \textbf{evocative} of expressions.
P2 described it as \textit{``intuitive and giving a feeling of negative expressions''}. 
However, the drawings were sometimes \textbf{unfaithful} to the original face, with P4 commenting \textit{``because the lines are incomplete and don't align with facial features''} and P13 noting \textit{``seems to match the face well, but too many lines make it hard to tell the surprised expression''}.


\underline{\smash{SketchXplain}} 
\fixed{was \textbf{intuitive} to users; for example, P3 found that ``based on the \underline{\smash{SketchXplain}} visualization, it's more obvious she's feeling happy''.}
Unlike CLIPasso, SketchXplain had more \fixed{\textbf{coherent}} strokes that \textit{``highlighted what I was referring to when I identified happy from the photo''} [P1]. This helped P7 \textit{``realize [a face] is not happy but angry if I have paid attention to the combo of arched eyes and frowning eyebrows.''}
Like Saliency, participants appreciated the \fixed{\textbf{simplicity}} to emphasize on \fixed{\textbf{salient lines}} by SketchXplain, noting that \textit{``from these darkest strokes, I can tell it's clearly a happy face because the upward motion [of her lip corner], the scrunched-up eyes and arched eyebrows indicate she's smiling''} [P1].
By sourcing from fine lines, \underline{\smash{SketchXplain}} could also effectively convey \textbf{subtle cues}, such as \textit{``the darker wrinkle lines around nose are straightforward to indicate disgust''} [P8].
This even helped to augment their perception, for example, P1 \textit{``didn't notice these lines until I saw the visualization.''}
Meanwhile, participants also noted some limitations with \underline{\smash{SketchXplain}}, such as missing information like \textit{``I can't tell the eyeballs and the gaze direction''} [P14]. 
To convey relative importance between AUs, SketchXplain varied stroke tone darkness~\cite{lu2012combining}, but this confused some users who were \textit{``distracted by other shallow lines''} [P12].
\textbf{Take-away:} SketchXplain successfully convey\fixed{ed} concepts through recognizable facial features (like Outline), cues through expressive lines (like CLIPasso), and relevance using darker strokes (like Saliency)\fixed{; thus supporting several desiderata for explanatory intuitiveness}.

\subsection{Quantitative User Study}
\label{sec:eval_face_quantitative}

\fixed{Having found that participants in the qualitative user study could intuitively interpret sketch explanations, 
we next quantitatively evaluate if sketch explanations can be intuitively interpreted, 
where a user can \textit{quickly} assimilate \textit{relevant} cues and concepts, and \textit{correctly} infer the AI's prediction label.}

Inspired by Kendall et al.~\cite{kendall2016iconic} which evaluated rapid human recognition of icon-based face expressions, 
\fixed{to determine how quickly the relevant information is correctly interpreted, 
we displayed visualizations in very brief durations ($<$ 500ms).}
We aimed to answer the research question:
How well can participants interpret Face Expression predictions when viewing different Visualizations, for varying Display Durations?

\subsubsection{Experiment Design and Apparatus}
\label{sec:eval_face_quantitative_design}
\begin{figure}[!t]
    \centering
    \includegraphics[width=7.0cm]{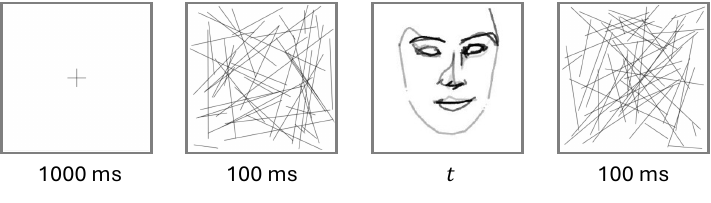}
    \vspace{-0.4cm}
    \caption{
        Experiment apparatus of image sequence shown to participants per trial in the quantitative user study:
        1) Centered crosshair shown for 1000.0ms to focus the participant's attention,
        2) Random lines shown for 100.0ms as distraction,
        3) Visualization of randomly chosen Visualization Type and expression shown for randomly chosen Display Duration $t$ (33.3--500.0ms) as main effect test,
        4) Random lines shown for 100.0ms as distraction, 
        before showing questions to label expression and AUs.
    }
    \label{fig:experiment-trial-sequence}
    \vspace{-0.4cm}
\end{figure}

We conducted a within-subjects experiment with three independent variables (IVs): 
Visualization Type (Photo only\footnote{\fixed{We included the original photo as a gold standard visualization of maximum information, though it is \textit{not} an AI explanation, since it is just the model input.}}, Saliency, Salient photo\footnote{\fixed{We included this as it is a popular XAI method, though this leaks information from the input photo, leading to unfair advantage.}}, Outline, CLIPasso, SketchXplain\footnote{We mostly excluded visualizations overlaid on the photo, to compare information gained purely from the visualization \fixed{without leakage} from the input photo.}; 
see \fixed{Table~\ref{table:demo-face} for examples}), 
Target expression $y$ (Surprised, Happy, Neutral, Sad, Disgusted, Angry), and
Display Duration $t$ (33.3ms\footnote{\fixed{The display durations were chosen to be as rapid as possible for modern computers. Since computer monitors may have a slow refresh rate of 30Hz~\cite{4Kmarket}, we limited the shortest duration to be 1 frame, i.e., $33.3\text{ms} = 1\text{s}/30$.}}, 66.7ms, 100.0ms, 133.3ms, 166.7ms, 
\fixed{200.0--500.0ms}\footnote{\fixed{We included pilot results for 200.0--500.0ms to show results are consistent even with less time pressure.
We had conducted a preliminary study of 80 participants for a wider range of durations 33.3--500.0ms, and  
found no significant differences from $\geq$~166.7ms onward, so limited the main study to $\leq$~166.7ms.}}). 
%
%
\fixed{Fig.~\ref{fig:experiment-trial-sequence} shows the experiment apparatus with timed visualization.}
Our \fixed{apparatus} user interface measured the screen refresh rate of each user's monitor to ensure the correct duration for each online participant.
%
\fixed{We evaluated with} the same 72 
\fixed{instances as} in the modeling study.

For dependent variables (DVs), 
\fixed{after each viewed visualization,
we measured} \rev{\textbf{AI Alignment} 
of the perceived label}\footnote{\rev{Participants were asked to recognize face expression \textit{as conveyed by the visualization} to measure its \textit{communication value}, rather than infer or learn what the AI would have predicted.
This is subtly different from the forward simulatability task in XAI user studies that focus on AI understanding than explanation communication.}}
\rev{(multiple-choice question) to the AI label $\hat{y}$,
and \textbf{Concept Recall} (multiple-response question of up to three face AUs explanation cues), calculated as TP/(TP+FN), for each AU concept $\hat{\gamma}$}.

\subsubsection{Experiment Procedure}
Each participant completed the following procedure:
\begin{enumerate}[label=\arabic*), itemsep=0.em, topsep=0.2em, parsep=0pt, partopsep=0pt]
    \item Introduction to the study.
    \item Consent to participate.
    \item Tutorial on interpreting all face visualizations.
    \item Screening questions to ensure ability to correctly:
    \begin{enumerate}[label=\alph*), itemsep=0.em, topsep=0.1em, parsep=0pt]
        \item match expressions to various straightforward photos, 
        \item recognize AUs from another photo, and 
        \item interpret visualizations.
    \end{enumerate}
    \item Main study with 72 trials, each involving viewing a rapid series of images with the test instance (Fig.~\ref{fig:experiment-trial-sequence}), and \fixed{answering} questions. 
    A reference of AUs to expressions is provided to aid users.
    \item Answer demographic questions.
    \item Acknowledge bonus calculations and exit.
\end{enumerate}
See Appendix Figs.~\ref{appendix-figure:Tutorial_shared}--\ref{appendix-figure:Trial_speed_test_q} for questionnaire details.

\begin{figure*}[!t]
    \centering
    \includegraphics[width=16.2cm]{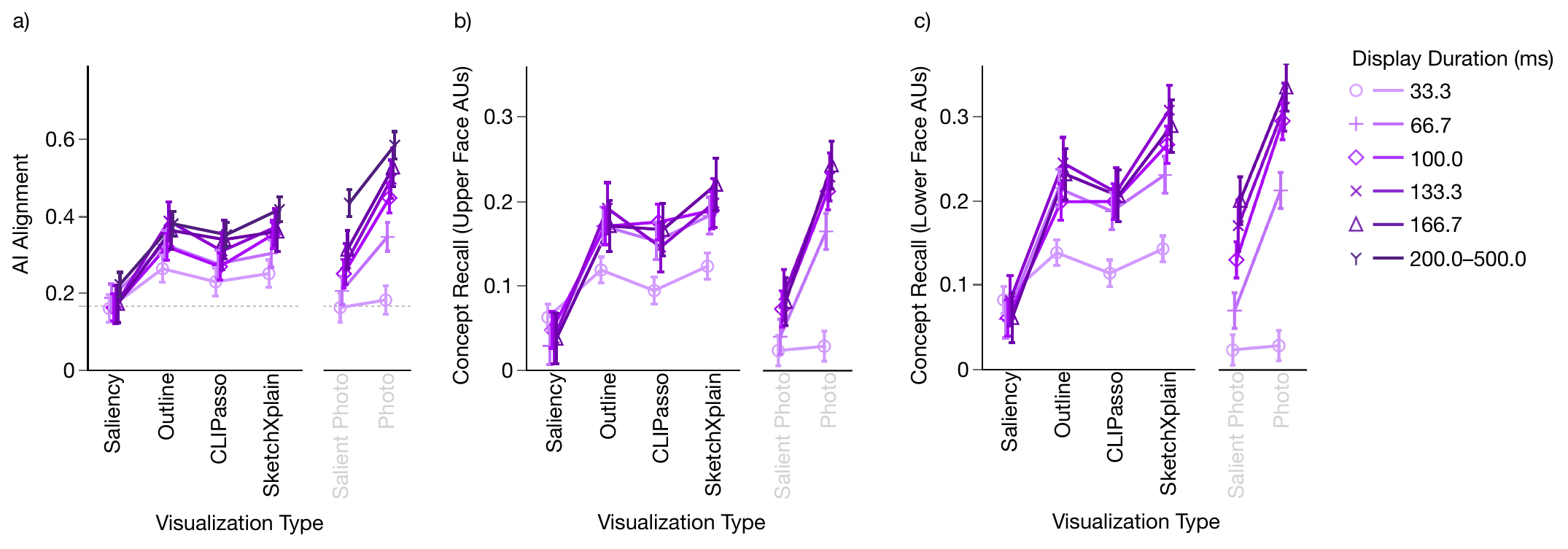}
    \vspace{-0.4cm}
    \caption{
        Results \fixed{of the quantitative user study on face expression quick interpretation} across Visualization Types \fixed{and Display Duration} for measures:
        a) AI \fixed{Alignment},
        \fixed{b) Concept Recall (Upper Face AUs),
        c) Concept Recall (Lower Face AUs).}
        \rev{Gray Visualization Types: 
        Photo is a gold standard of maximum information, not a competitor explanation method since it is the input image, and 
        Salient Photo is more usable than Saliency but unfairly leaks photo input data.}
        Gray dotted line in (a) represents correctness of a random guess from 6 MCQ choices (16.7\%) \fixed{for AI Alignment}.
    }
    \label{fig:human_face_all}
    \vspace{-0.4cm}
\end{figure*}

\subsubsection{Participants}
\label{sec:participants}
We recruited 466 participants from Prolific with high qualification rates ($\geq$1000 completed HITs, $>$97\% approval rate). 176 participants passed our screening test. They were \rev{89 males, 84 females and 3 preferred not to say}, with ages 21--71 (Median = 37.0). They completed the survey in median 32.8 minutes and were compensated \pounds4.00. We incentivized effort with up to \pounds1.00 for more correct expression labeling.

\subsubsection{Statistical Analysis and Quantitative Results}


We fit a linear mixed effects model for each dependent variable as the response, Visualization Type, Expression and Display Duration, with other confounding variables as fixed effects, some interaction effects among the factors, and Participant as a random effect. See Appendix Table~\ref{table:stat_intuitiveness} for details.

Participants' \fixed{perceived labels showed higher \textit{AI Alignment}} as the Display Duration increased across all visualizations, except for Saliency (Fig.~\ref{fig:human_face_all}a). 
When the Display Duration was extremely short ($t = 33.3$ms\fixed{; light purple line}), both Outline (p $<$ .0011) and SketchXplain (p = .0070) achieved significantly higher \fixed{alignment} compared to Photo (Fig.~\ref{fig:human_face_all}a), indicating their \fixed{\textbf{quickness} in conveying information about the AI predictions}. 
Photo was harder for participants to interpret at this short duration, perhaps due to excessive details,
\fixed{though at longer durations, this gold standard served as an upper limit indication of how much information participants could acquire at each time duration.}
In general, CLIPasso sketches were significantly poorer in supporting \fixed{AI-aligned user perception} compared to SketchXplain (\fixed{p $<$ .0032}).
\fixed{Saliency was the worst in helping users to align their perception, perhaps due to the lack of featural context of its amorphous blobs.
Conversely, with the added context from the retained regions of the input photo, Salient photos were more interpretable (higher AI alignment) than Saliency alone.}

\fixed{\textit{Concept Recall} also} improved with increased Display Duration, 
and had similar trends across Visualization Types as for AI Alignment
\fixed{(Fig.~\ref{fig:human_face_all}b, c)}.
Interestingly, participants could recall Lower Face AUs (about mouth, nose) better than Upper Face AUs (eyes, eyebrows); \fixed{contrast t-test: p $< .0001$}. 
\fixed{With sufficient Display Duration ($t > 33.3$ms),} SketchXplain had the highest recall among non-Photo Visualization Types (\fixed{p $< .0001$}), indicating the usefulness of its semantically-aligned, cue-aware sketches.
Outline was weaker \fixed{than SketchXplain} at supporting Lower Face AUs \fixed{(p $< .0001$)}, possibly due to the similar mouth shape (open or closed) across expressions.
CLIPasso had even weaker concept recall \fixed{(p $< .0001$)} due to not explicitly encoding AU information.
Saliency explanations were \fixed{most} unhelpful \fixed{(p $< .0001$)}.
\fixed{Salient Photo did improve recall for longer durations ($>100.0$ms, p $< .0001$), but still worse than SketchXplain (p $< .0001$).}


\section{Evaluation on Skin Lesion Classification}
\label{sec:evaluation_skin_lesion}

\begin{table*}
\footnotesize
\centering
\vspace{-0.1cm}
\setlength{\tabcolsep}{3pt}
\caption{
ABCD criteria for melanoma diagnosis.
Given a skin lesion, let $A$ denote its area, $P$ its perimeter and $\bm{L}$ its outline.
Higher values indicate more suspicious features.}
\vspace{-0.2cm}
\begin{tabular}{
    >{\raggedright\arraybackslash\hspace{0pt}}m{1.5cm}  
    m{10.0cm}  
    m{7.5cm}  
}
\toprule
Criteria & Concept Description and Label Quantification Method & Visual Cue Feature Extraction Method \\

\midrule
Asymmetry
  & A lesion is \fixed{labeled} symmetrical if it resembles its reflection across its major and minor axes. 
  \fixed{We define non-overlapping pixels as the XOR ($\oplus$) between the mask ($M$) and its reflection ($M_r$) along these axes~\cite{garnavi2012computer, she2007combination, kasmi2016classification} and quantify asymmetry as the proportion of non-overlapping pixels, i.e.,
  $\lVert M \oplus M_r\rVert_1 / A$.}
  & \fixed{We highlight non-overlapping pixels ($M \oplus M_r$) between the lesion mask and its reflections along major and minor axes.}\\
\midrule

Border Irregularity
  & Perimeter length $P$ is larger for lesions with irregular borders. We use the perimeter–area ratio $\frac{P}{A}$, and the non-circularity index $\frac{P^{2}}{4\pi A}$ to measure deviation from a circular shape~\cite{garnavi2012computer, she2007combination, majumder2019feature}. 
  & We highlight boundary segments corresponding to \fixed{high} curvature along $\bm{L}$ with a sliding window and marking boundary segments whose curvature exceeds a threshold. \\
\midrule

Color Variation
  & More colorful lesions are more suspicious. 
  We capture overall RGB color variation using the standard deviation of each channel~\cite{she2007combination}.
  & We highlight lesion pixels whose color deviates from the mean lesion color by more than one standard deviation in any RGB channel.\\
\midrule

Diameter
  & Larger lesions are more likely melanoma. We estimate the equivalent diameter for irregular shapes, i.e., $2\sqrt{A/\pi}$, which is the diameter of a circle with the same area~\cite{majumder2019feature, ali2020towards}.
  & We highlight the lesion outline $\bm{L}$ as a cue to overall lesion size. \\
  
\bottomrule
\end{tabular}
\vspace{-0.2cm}
\label{tab:abcd}
\end{table*}

\begin{table*}[!t]
    \centering
    \caption{
        Examples comparing visualizations (cols) for explaining suspicious melanoma (rows). 
    }
    \vspace{-0.3cm}
    \includegraphics[width=12.5cm]{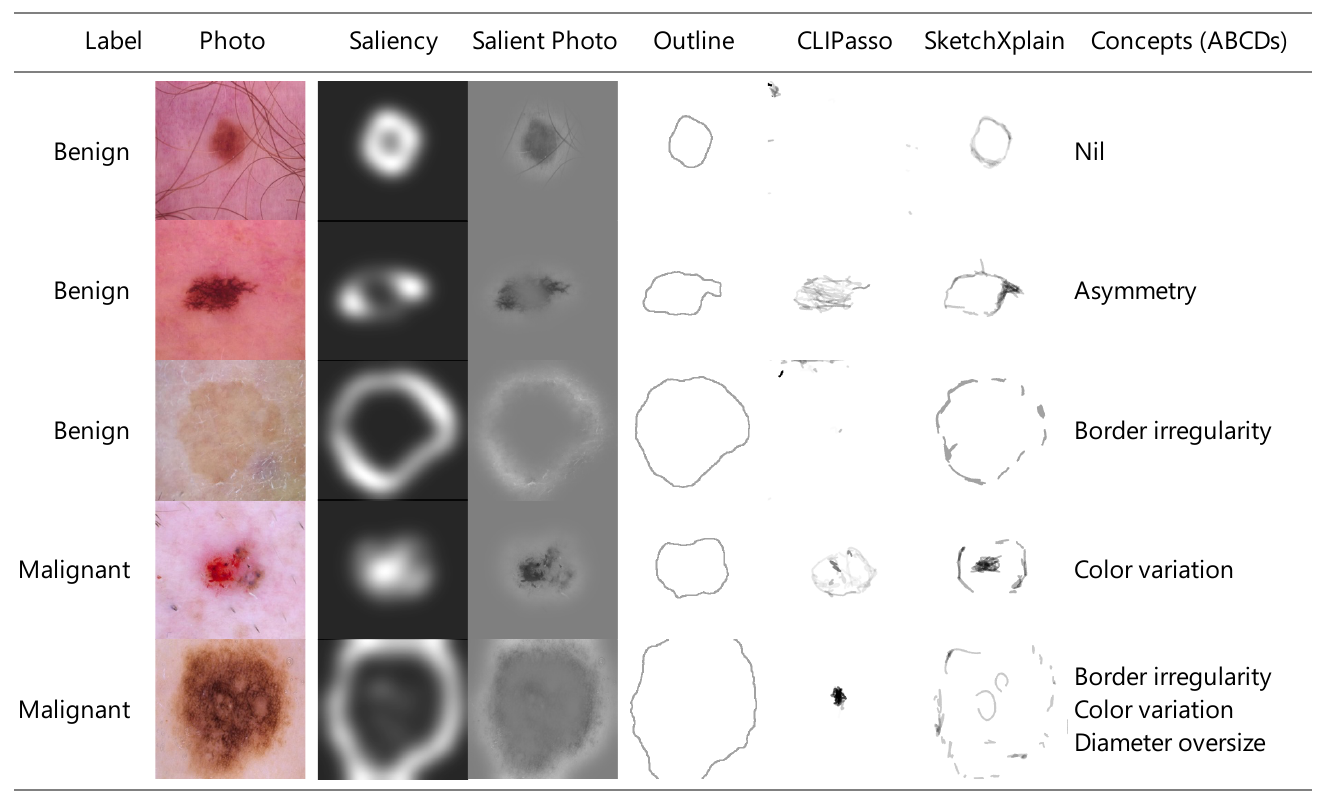}
    \label{table:demo-lesion}
    \vspace{-0.5cm}
\end{table*}

To test the generalizability of sketch explanations, 
we \fixed{also} applied SketchXplain to skin lesion classification, a domain with well-established, lay-accessible concepts\fixed{---}the ABCDE criteria~\cite{abbasi2004early} for melanoma diagnosis. 
\fixed{We investigate how} explainable sketches could help non-specialists identify suspicious melanoma at an early stage~\cite{yu2021early}, enhance clinical trust and facilitate communication between clinicians and patients~\cite{chanda2024dermatologist}.

We describe the dermoscopic image preparation, the SketchXplain extension to extract \rev{ABCD}\footnote{\rev{Full criteria is ABCDE including E for evolution, which requires repeated measures of multiple photos to observe, so we omitted it for our single-image application.}} concepts, and a qualitative \fixed{user study to learn} insights into the potential use and benefits of sketch explanations for communicating melanoma risk. 
We omitted a modeling study with CLIP scores, because CLIP was not trained on medical images, leading to unreliable out-of-distribution measurements.
We also omitted a quantitative user study \fixed{with tight display durations}
as ABCD concepts are not inherently familiar like facial action units and thus require substantial user training.

\subsection{Data Preparation}
\label{sec:skin_lesion_data_prep}
We used 11,720 images of skin lesions from the HAM10000 dataset \fixed{with diagnosis labels}~\cite{tschandl2018ham10000} and 
\fixed{segmentation masks of lesion locations~\cite{tschandl2020human}.}
We simplified the lesion prediction to binary classification (i.e., benign or melanoma) by grouping all benign classes\footnote{Benign: Benign Keratoses, Dermatofibroma, Melanocytic Nevi, Vascular lesions.} and excluding malignant classes\footnote{Malignant: Basal Cell Carcinoma, Melanoma, and Pigmented Actinic Keratoses.} other than melanoma.
Future work could extend explanations to other malignant lesions with distinct clinical features.

\subsection{Concept Labeling and Cue Extraction}
\label{sec:skin_lesion_concept_extraction}

We reused the SketchXplain architecture, with additional data labeling and feature extraction. 
Since the skin-lesion dataset lacked ABCD labels, we estimated these concepts using simple image-processing heuristics adapted from prior work~\cite{she2007combination, majumder2019feature, kasmi2016classification, ali2020towards}:
\fixed{Asymmetry (uneven halves), Border irregularity (ragged edges), Color variation (multiple shades), Diameter (larger than 6 mm).
Table~\ref{tab:abcd} describes the ABCD concepts and visual cue feature extraction methods}. 
The resulting ABCD scores served as pseudo-labels for training the concept extractor in our concept bottleneck model ($M$ and $F_\gamma$).
\fixed{For} inference, SketchXplain first predicts the concepts $\hat{\gamma}$ (Fig.~\ref{fig:architecture}, \textit{Step 2a}) and then the class $\hat{y}$.
For cue localization (Fig.~\ref{fig:architecture}, \textit{Step 2b}), we regularized the Grad-CAM with ABCD heuristic heatmaps.
\rev{We used the same pipeline to generate sketches as for face expressions\footnote{\rev{See Appendix Table~\ref{table:demo-lesion-intermediate}, column Init. strokes for examples of initialization.}}}. 
For semantic alignment, we disabled text-based guidance (setting $\lambda_t = 0$ in Eq.~\ref{eq:loss}) because CLIP is not trained on medical images.
Table~\ref{table:demo-lesion} compares visualization types (columns) across different cases (rows) with varying diagnoses and ABCD features.

\rev{Like~\cite{hou2024concept}, we fined-tuned ResNet18~\cite{he2016resnet} as the backbone $M$, on the HAM1000 dataset.
SketchXplain achieved good accuracy 85.8\% and F1 Score 71.2 on melanoma classification, 
and ABCD per-criteria F1 scores 71.6--82.1 (M = 78.5)}\footnote{\rev{Better than face expression classification in Section~\ref{sec:eval_face_modeling} and prior work~\cite{luo2022learning}.}}\rev{.

While the aforementioned results demonstrated good in-distribution prediction performance,
the pretrained image-to-image generator $E$~\cite{chan2022learning} may be out-of-distribution.
Nevertheless, the generated Detailed lines (see Appendix Table~\ref{table:demo-lesion-intermediate}) were qualitatively faithful to the photos.
Next, we further examined their plausibility in a formative qualitative user study.}



\subsection{Qualitative user study}
\label{sec:eval_lesion_qualitative}
We conducted a qualitative user study using the think-aloud protocol to investigate how different visualizations \fixed{(Saliency, Salient photo, Outline drawing, CLIPasso sketch, or SketchXplain)} could help people identify concepts and melanoma cases.

\subsubsection{Visual Explanation Comparisons}
\fixed{We compared SketXplain against the same baseline visualizations described in Section~\ref{sec:compare_xai}, with one difference:
\underline{\smash{Outline}} was obtained from tracing the edge of the segmentation mask~\cite{tschandl2020human}.}

\subsubsection{Method and Procedure}
\label{sec:eval_lesion_qualitative_method}
We recruited 10 non-clinical participants through university mailing lists and personal contacts. The sample included 3 females \rev{and 7 males}, ages 23--34 \fixed{(Median = 29.0)}. 
From images of skin lesions, they were asked to identify ABCD concepts and determine the melanoma type based on various visualizations. 
To prepare participants, we provided a brief tutorial describing how ABCD concepts relate to melanoma, and familiarized them with 7--10 examples across all visualizations.
In the main study, participants viewed 15--20 image instances, each with a random visualization.

Since skin lesions are less familiar to lay people than faces, we overlaid visualizations on the photos for added context, rather than using sketches in isolation on a white background.
For generality, this also allowed us to evaluate sketch explanations as annotations.
Participants described what they saw, decided on the diagnosis, and explained their reasoning.
All sessions were conducted over Zoom, and audio and screen interactions recorded with participant consent.
The study took 32 minutes on average, and participants were compensated \$6.20 USD in local currency for their time.

\subsubsection{Findings}
We \fixed{performed} a thematic analysis on participant utterances and interactions, focusing on how participants used or misused visualizations to identify concepts and diagnose melanoma.
Unlike participants who perceived facial action units inherently in face expressions, here, participants were more deliberative and analytical.
Participant P2 focused on
\textit{``balancing the number of abnormal concepts and the severity of each abnormality''}. 
P1 \textit{``prioritized certain concepts such as color variation and diameter''}.
\fixed{However, when viewing \underline{\smash{Photo}} alone, without explicit ABCD indicators or annotations, participants found it hard to distinguish between normal from abnormal features.}
P4 found it \textit{``hard to decide whether this is small or larger diameter without a threshold''}.
P6 \textit{``[could] not tell whether this [lesion] [was] [as]symmetrical if I can't even see a clear boundary''}.
Next, we articulate differences across visualization types.

\underline{\smash{Saliency map}} or \underline{\smash{Salient photo}} were perceived as less helpful than \fixed{all} line drawings, because they imprecisely depicted the boundary and size. P4 
felt the smooth boundary \textit{``too blurred, I can't see the true border''}.
P4 also found \underline{\smash{Salient photo}} misleading, since \textit{``the \fixed{[grayscale]} uneven color caused by the mask misleads me into thinking the lesion has color variation''}.

In contrast, line drawing annotations on the photo gave participants a strong first impression and encouraged them to \textbf{relate observations with the clinical ABCD concepts} to make more informed judgments.
P1 appreciated \underline{\smash{Outline}} because [it] provided \textit{``simple and straightforward annotations on the boundary''}.
However, P2 disagreed, noting \fixed{its limitation} that \fixed{\textit{``it always has the same intensity. It might be good for asymmetry and border, but in fact we can see the border alone so it is not very useful.''}
Conversely, simplifying lines presented other concerns;} \underline{\smash{CLIPasso}}'s \textbf{imprecise spatial annotations} often misled interpretation: P7 commented that \textit{``the visualization cuts it in half, so it might be asymmetry''} and P8 noted, 
\textit{``from the photo [the] left part of the lesion is bad, but it isn't highlighted''}.

Nonetheless,
\underline{\smash{CLIPasso}} stimulated more reflection.
P3 noted that its \textit{``strokes inside the region reminded me to examine the potential color abnormality''}. 
Furthermore, \underline{\smash{SketchXplain}} was perceived as more \textbf{aligned} to the original photo and more \textbf{coherent} with the underlying concepts. 
P2 remarked, \textit{``the emphasized scribbles around the corner help me judge asymmetry"} and
P6 appreciated that \textit{``the disconnected lines help me confirm border irregularity''}, whereas \fixed{if he had viewed} \underline{\smash{Outline}}, he \textit{``would judge the border differently when viewing a smooth annotation''}. 
However, \fixed{the} increased interpretability with both \underline{\smash{CLIPasso}} and \underline{\smash{SketchXplain}} came at the cost of \fixed{increased} cognitive load.
P1 found that \textit{``the lines are more complex than the [\underline{\smash{Outline}}] contour, which takes time to comprehend''}. 

\textbf{Take-away}: SketchXplain produced interpretable strokes that were more coherent with the \fixed{exlanatory} ABCD concepts and \fixed{lesion} observations, better supporting users to make \fixed{confident} judgments on the task.

\section{Discussion}
Having introduced sketch explanations as a new paradigm for intuitive visual explanations,
we discuss their generalization\fixed{, contextualization,} and further development.

\subsection{Extending Sketch Explanations}
\fixed{Our work was the first to propose sketch visualizations as explanations of AI predictions, and we had investigated it for a limited scope.
Here, we discuss extensions to other visual properties, explanatory concepts, and application domains.}

\fixed{Through SketchXplain we had investigated the simplicity--coherence trade-off with the visual properties of} \fixed{stroke count, smoothness and tone. 
However, other stroke properties, such as stroke length, thickness, and tapering, could be leveraged to convey salient explanatory cues.
Future work could even compare the difference in perceived acuity, intuitiveness and explanation cue recall across these visual features.}

\fixed{We had used concept bottleneck models (CBMs)~\cite{koh2020concept} with pre-deteremined concepts for facial action units (AUs) and skin lesion criteria, but this can be extended to semi-supervised Label-Free Concept Bottleneck Model (LF-CBM)~\cite{oikarinen2023label} and open-domain cue localization with with Owl-ViT~\cite{minderer2022simple} and Segment Anything Model (SAM)~\cite{kirillov2023segment} that we investigated in Appendix~\ref{sec:general_image_sketchxplain} for general images.}
Other methods to obtain concepts include TCAV~\cite{kim2018interpretability}, crowdsourced elicitation~\cite{mishra2021crowdsourcing} or unsupervised discovery~\cite{ghorbani2019towards}.

While sketches are commonly used by artists to convey \fixed{facial} expressions, we have demonstrated that they can also be applied to other domains, such as 
\fixed{annotating skin lesions.
Since biology and medicine heavily employ sketches~\cite{baldwin2010art, hay2013using}, sketch explanations can provide new opportunities to explain predictions on medical images~\cite{chan2020deep}}
with sketched annotations on pathology or radiology slides to articulate fine physical or anatomical features, rather than merely pointing at them with saliency maps~\cite{rajpurkar2018deep, sayres2019using}.
\fixed{Furthermore, future work could investigate the tunability of sketches toward simplicity or coherence for everyday lay tasks or scientific, medical, and engineering tasks. 
In the latter tasks, diagrammatic constraints could also be added to enforce domain alignment~\cite{lim2025diagrammatization}.}
\subsection{From Analytical to Intuitive AI Explanations}
\label{sec:discussion_intuitive_xai}

\rev{While many explainable AI (XAI) methods facilitate analytical reasoning through charts or network visualizations~\cite{caruana2015intelligible, ribeiro2016should, lundberg2017unified, krause2017workflow, krause2016interacting, hohman2019gamut, zhang2026comparables},
users typically conserve cognitive effort~\cite{buccinca2021trust} and satisfice their understanding~\cite{kaur2024interpretability}, leading to premature misinterpretations and decision errors~\cite{wang2019designing}.
Hence, XAI must be intuitive.

Prior works have sought this through simplicity via sparsity~\cite{ribeiro2016should, doshi2017towards} and smoothness~\cite{caruana2015intelligible, abdul2020cogam}, 
yet these properties alone are insufficient for intuitiveness. 
Others indirectly address explanatory coherence via relatability~\cite{zhang2022towards} and faithfulness~\cite{ribeiro2016should}. 
In feature-based XAI on tabular data, methods replace technical jargon with human-interpretable terms or concepts~\cite{zytek2022need}, or signifier icons~\cite{lim2011design}. 
Moreover, concept-based explanations~\cite{kim2018interpretability, koh2020concept} could be made more intuitive with analogies using prior knowledge~\cite{he2022like}.

In contrast,
methods for non-tabular data go beyond semantic remapping to adopt domain-specific cues and conventions. 
For instance, audio-based XAI can be made relatable through counterfactual cases and contrastive cues~\cite{zhang2022towards}. 
In image-based XAI, saliency maps (e.g.,~\cite{selvaraju2017grad}) remain the dominant modality due to their intuitive highlighting, yet they can be incoherent to model inference~\cite{adebayo2018sanity} or human belief priors~\cite{erion2021improving}. 
While some methods improve coherence through regularizing attribution priors~\cite{erhan2009visualizing} and diagrammatic conventions~\cite{lim2025diagrammatization}, 
we introduce the sketch as a new visual modality that leverages lay familiarity to enhance both explanation simplicity~\cite{miller2019explanation}, coherence~\cite{thagard1989explanatory}, and interpretation speed~\cite{swaroop2024accuracy, wang2025less}. 
However, as current methods rarely prioritize rapid interpretation, we employ speed-constrained tests from cognitive psychology~\cite{kendall2016iconic} to precisely evaluate the perceptual efficiency of explanations.

Therefore, by prioritizing simplicity, coherence, and speed, sketch-based XAI accommodates the human tendency to satisfice, thereby limiting cognitive load and streamlining explanation transmission to the viewer. 
This set of desiderata provides a basis for developing and evaluating image-based XAI that shifts the focus from purely analytical auditing toward intuitive interpretation.}

\subsection{Standalone Abstraction vs. Annotative Overlay Sketches}
\label{sec:standalone_annotative}
\rev{In two domains, we have examined the usability and usefulness of sketches as explanations in different presentation paradigms.
Sketch explanations of facial expressions were intuitively interpretable as standalone, since 
humans possess an innate familiarity with facial geometry, allowing them to decode line shapes and spatial relationships without external context.
Methodologically, evaluating these sketches in isolation allowed us to examine their independent explanatory value while preventing information leakage from source photos (further studied in Appendix~\ref{sec:privacy_evaluation}).

Conversely, for skin lesion sketch explanations, to accommodate limited familiarity or domain expertise, we included the photo as contextual grounding for the overlaid sketches.
This dependency is analogous to feature attribution explanations $a(x)$ that are less interpretable without corresponding feature values $x$, and to saliency maps that are obscure without the underlying source pixels being visible.
Consequently, it is vital for visual explanations to be demonstrably coherent~\cite{thagard1989explanatory} or aligned with observation~\cite{lim2025diagrammatization}. 
Overlaying sketches on photos facilitates this by allowing users to perform sanity checks against the input grounding~\cite{boggust2022shared, lim2025diagrammatization}.

Future work should investigate this extended usage of overlaid sketch explanations with domain experts. 
A salient research direction is the trade-off between sketch abstraction and spatial alignment to the photograph. While highly abstracted sketches may facilitate rapid interpretation, they may compromise the detailed examination required for scientific, clinical, or engineering rigor, suggesting a need for dynamic levels of detail in sketch-based XAI interfaces.}

\subsection{Evaluating Intuitive Understanding}


We had evaluated visualizations under tight time limits \fixed{in the quantitative user study on quick perception of face expression visualizations}. This is common in human perception psychology studies~\cite{willis2006first}, and allows precise measurement of users' intuitive impression (System 1 thinking~\cite{kahneman2011thinking}). 
We do not argue that sketch explanations will only be beneficial under rapid exposures, which future work can evaluate, but differences across visualizations may \fixed{diminish} when users rationalize slowly (System 2)~\cite{ehsan2018rationalization}.

While poor 
\fixed{label} recognition or \fixed{concept} recall under time pressure could indicate visual complexity, we did not explicitly measure perceived visual complexity. 
Nevertheless, prior work has shown the congruency between the objective measure of JPEG file size and perception~\cite{machado2015computerized}, \fixed{supporting to our approach.}
Furthermore, we did not measure participants' subjective self-reported satisfaction of each visual explanation, or their ability to debug AI prediction errors. 
The latter is especially important for mitigating AI over-reliance~\cite{chen2023understanding, schoeffer2024explanations}, which future work can explore \fixed{with technical users}. 

Emojis and icon-based faces are popular for illustrating face expressions and can be quickly perceived by users~\cite{kendall2016iconic}. However, we did not evaluate them as baselines since they are static per expression and act more like labels \fixed{rather than explanations contextual to the instance features or observation.}

Finally, although SketchXplain explanations convey conceptual information via \fixed{visual} cues, 
\fixed{they could be conveyed verbally by labeling which concept are present~\cite{koh2020concept}.}
Future work could investigate this further, though we hypothesize that visual explanations will remain more intuitive because they exploit higher human visual bandwidth instead of slower general language understanding, and \fixed{avoid AU terminology (e.g., ``lip corner depressor'') that} is known to be difficult for lay users to understand~\cite{pearson1985visual, donato1999classifying}.
\fixed{Nevertheless, text-based concepts can be complementary to sketches, and should be investigated in future work.}

\section{Conclusion}
We have introduced SketchXplain as a new XAI paradigm to \fixed{visually} explain \fixed{image-based AI predictions} intuitively and faithfully. 
It incorporates concept-based and saliency explanations to generate cue-aware, semantically-aligned sketches from relevant source lines.
Through modeling and user studies on face expression \fixed{and melanoma} image prediction \fixed{tasks}, we demonstrated that sketch explanations support intuitive interpretation,
\fixed{through simpler and more coherent visual explanations that are quicker to interpret compared to saliency maps and other outline drawings or non-explanatory sketches.}
This work contributes to the diverse options of \fixed{visual} XAI by offering intuitive and expressive sketch rationalizations to improve interpretability.


\bibliographystyle{IEEEtran}
\balance
\bibliography{ref}

\appendix
\twocolumn
\captionsetup[figure]{labelfont={bf},font={footnotesize},name={Appendix Fig.},labelsep=period}
\captionsetup[table]{labelfont={bf},font={footnotesize},name={Appendix Table},labelsep=period}
\label{sec:appendix-example}

\section{Evaluations on Face Expression Image Domain}

\fixed{We provide additional details on the SketchXplain evaluation for facial expressions, covering the modeling study used as a preliminary check, the user study on Intuitive Interpretation discussed in the main paper (Section~\ref{sec:eval_face_quantitative}).
Here, we also discuss an additional user study to investigate the Privacy Protection benefits of sketch explanations.} 

\subsection{Modeling Study}
\label{sec:metrics}
\fixed{In Section~\ref{sec:eval_face_modeling}, we evaluated whether SketchXplain generates sketch-based explanations that satisfy the desiderata of simplicity and coherence using proxy metrics. 
We further justify these proxy metrics and provide additional analyses, including a privacy evaluation, an ablation study on the number of strokes, and intermediate outputs for illustration.}

\subsubsection{Justifications for Evaluation Metrics} 
\fixed{We justify the choice of proxy metrics for Visual Complexity and Coherence, which we used to evaluate whether the visualizations satisfy the desiderata in Section~\ref{sec:proxy_metrics}.}

\vspace{1em}
\noindent \textbf{Visual Complexity:}
To be more accessible, AI explanations should be visually simple for people to comprehend. However, there is no reliable computational metric to compare visual complexity across heterogeneous image domains (e.g., photos, saliency maps, and line drawings). Therefore, we used two entropy-based measures---local Shannon entropy and JPEG XL file size---as complementary proxy indicators of visual complexity.

Previous work~\cite{gonzalez2009digital} used Shannon entropy to indicate the variability of pixel intensities, but it ignores spatial relationships between pixels.
Hence, we used \textit{local Shannon entropy}~\cite{wu2013local}, which divides the image into blocks, computes the entropy of each block and aggregates their average.
For our images of $256 \times 256$ pixels, we chose a block size of $11 \times 11$, which can adequately contain variations in curved lines and shapes.
Although local entropy captures variability among image blocks, it does not encode local spatial information within each block. 
Image compression, in contrast, can capture this information and represent it concisely~\cite{yu2013image}, making it a convenient and plausible estimator of human-perceived complexity~\cite{machado2015computerized}. 
Following this approach~\cite{machado2015computerized, yu2013image}, we used JPEG file size to complement local entropy, assuming larger file sizes correlate with higher information density, which requires greater effort to interpret.
However, standard lossy compression of JPEG is not suitable for line drawings, since the discrete cosine transform (DCT) method would require high-frequency coefficients to represent the hard edges, causing file sizes to be unfairly large.
Therefore, we also used ``modular'' \textit{lossless JPEG XL}~\cite{alakuijala2019jpeg} which encodes images in the spatial domain (not the frequency domain) using predictive coding~\cite{kobayashi1974image}. This allowed us to better predict the value of each pixel using neighboring ones and compute the residual between the predictions and actual intensities. These pixel-based residuals represent the surprisingness of the image, which is captured via entropy encoding. 
Thus, the resulting file size represents the visual complexity in terms of the amount of surprise from predictable patterns.
Furthermore, to account for the human ability to perceive spatial frequency in images~\cite{sachs1971spatial},
we used \textit{lossy JPEG XL} as an additional metric by setting the quality parameter to 99/100 to switch to the ``variable DCT'' encoding regime.
In summary, we used local Shannon entropy and lossless and lossy JPEG XL file sizes to estimate visual complexity.

\vspace{1em}
\noindent \textbf{Coherence:}
An explanation visualization should be \rev{coherent with the AI prediction label, i.e., have \textit{AI Alignment}.
Beyond this, it should also align with
the underlying explanatory concepts (\textit{Concept Alignment}) to avoid spurious reasoning.
Moreover, explanations should correspond to representative visual cues of the concepts (\textit{Cue Alignment}) and the source image (\textit{Observation Alignment}) so that users can relate what they see with relevant concepts and their mental knowledge.}
To assess coherence, we leveraged the CLIP model~\cite{radford2021learning} 
\rev{to obtain joint vision-language embedding representations
allowing alignment evaluation across modalities.
We developed coherence metrics based} on cosine similarity scores from CLIP~\cite{hu2023tifa}.
These similarity scores were computed by embedding both the visualization and its comparator (e.g., baseline drawing) and then taking their cosine similarity, where 1 indicates identical semantics and 0 indicates complete dissimilarity.
As in~\cite{wang2023reprompt}, those scores indicate \rev{whether the visualization aligns with the predicted label $\hat{y}$
and underlying concepts $\hat{\gamma}$.
Since CLIP can also map cross-domain images into a shared semantic space, we used CLIP scores to evaluate alignment with visual cues $\breve{\bm{x}}$ or the original image $\bm{x}$.}

\fixed{For generality, we also performed the same analyses with TASK-Former (Text And SKetch transformer)~\cite{sangkloy2022sketch} as an additional evaluator metric for coherence.
It is more suited to sketch-based representations, since it was co-trained on images, sketches and text. 
See results in Appendix Fig.~\ref{fig:model-face-all-taskformer} which is consistent with results for CLIP in Fig.~\ref{fig:model-face-all}b--c).}

\begin{figure}[tbp]
    \centering
    \includegraphics[width=9.0cm]{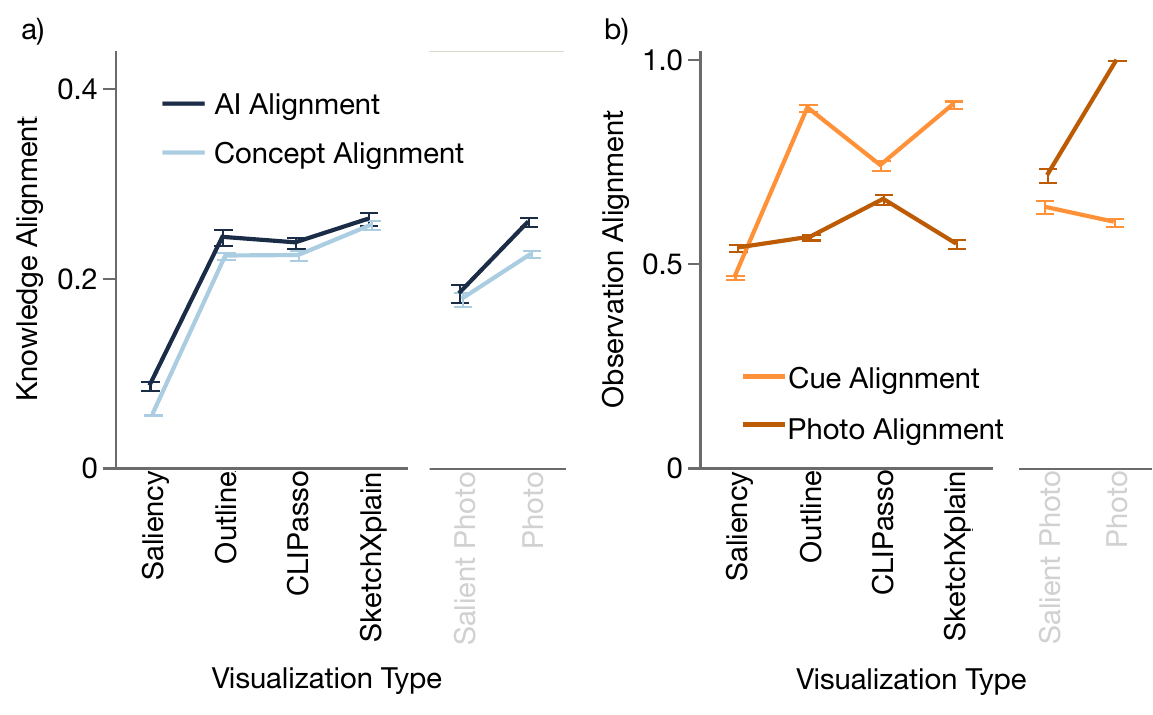}
    \vspace{-0.3cm}
    \caption{
        \fixed{Coherence measured with TASK-Former~\cite{sangkloy2022sketch} cosine similarity towards} 
        a) Knowledge,
        b) Observation alignment. 
    }
    \label{fig:model-face-all-taskformer}
    \vspace{-0.3cm}
\end{figure}

\subsubsection{Privacy Evaluation}
\label{sec:privacy_evaluation_modeling}
\begin{figure}[tbp]
    \centering
    \includegraphics[width=4.8cm]{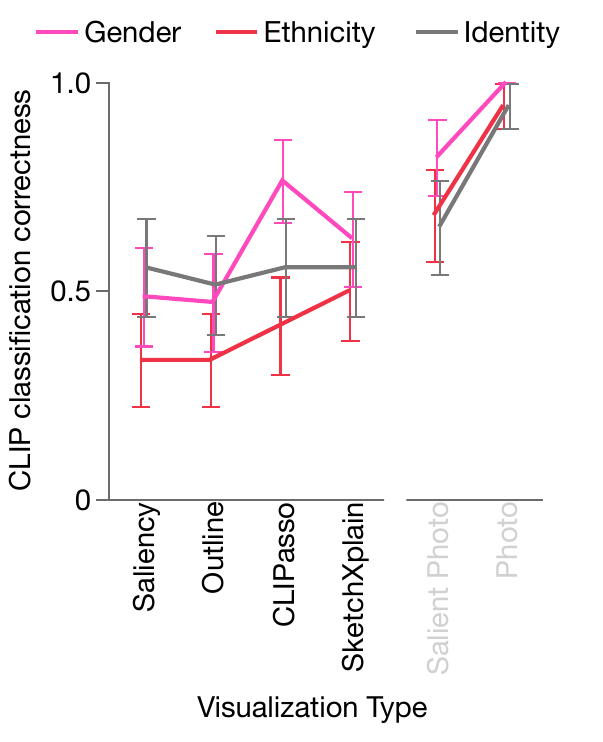}
    \vspace{-0.3cm}
    \caption{
        Privacy measured as CLIP classification correctness (i.e. re-identification risk) toward Gender, Ethnicity and Identity (as the original Photo).
    }
    \label{fig:model-results-privacy}
    \vspace{-0.3cm}
\end{figure}

A side-effect of abstract sketches is the omission of less relevant details in the face photos. This is useful for privacy, as it helps mitigate privacy attacks that re-identify people by exploiting the explanations without seeing the original photos~\cite{zhao2021exploiting}.

\onecolumn
\begin{multicols}{2}
We evaluated re-identification risk by how similar the visualization was to sensitive attributes---gender, ethnicity, identity. Gender and ethnicity labels were encoded as text and then converted to CLIP embeddings. Identity was encoded from the CLIP embedding of the original face photo.
We modeled a privacy attack as a classification, where we calculated the cosine similarity of the visualization CLIP embedding to all sensitive attribute labels, and chose the label with the highest CLIP similarity score.

Appendix Fig.~\ref{fig:model-results-privacy} shows the results of \textit{privacy} attack classification correctness. 
\underline{\smash{Photo}} contained the most sensitive information, followed by \underline{\smash{Salient Photo}} that still retained key details of faces.
All other visualizations provided better privacy protection (lower attack correctness).
Thus, \underline{\smash{SketchXplain}} provided stronger privacy protection than Salient Photo.

\subsubsection{Ablation Analysis}
\label{sec:ablation}
\begin{figure}[!t]
    \centering
    \noindent\makebox[\textwidth]{%
    \includegraphics[width=14.5cm]{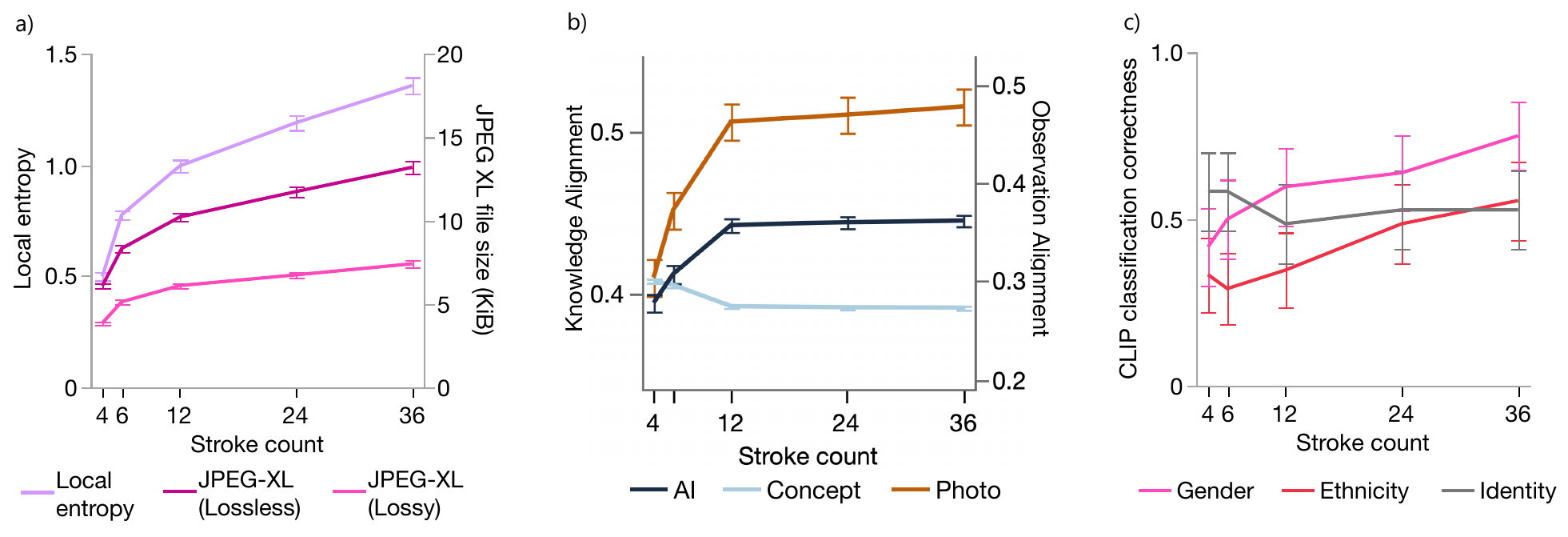}
    }
    \vspace{-0.3cm}
    \caption{
        Ablation results across various SketchXplain stroke settings:
        a) Simplicity,
        b) CLIP-based coherence,
        c) Privacy in explanations measured as CLIP classification correctness (i.e. re-identification risk).
    }
    \label{fig:model-results-face-ablation}
    \vspace{-0.3cm}
\end{figure}

We conducted an ablation analysis on \textbf{stroke count}.
The stroke initialization in SketchXplain requires explicitly setting the stroke count. We compared 4, 6, 12, 24, and 36 strokes to assess their impact on sketch coherence, simplicity, and privacy.
Appendix Fig.~\ref{fig:model-results-face-ablation} shows quantitative results and Appendix Table~\ref{table:demo-face-ablation} shows example sketches with varying strokes.

As expected, adding strokes raised the visual complexity of SketchXplain (Appendix Fig.~\ref{fig:model-results-face-ablation}a). 
Photo and AI Alignment increased with stroke count, but the trend plateaued after 12 strokes (Appendix Fig.~\ref{fig:model-results-face-ablation}b).
Strangely, Concept Alignment decreased with stroke count despite starting very low.
Perhaps, because AU concepts are not well-learned in the pre-trained CLIP model, the fewest strokes were poorly related to the concepts and increasing stroke count made the sketch more distinct and differentiated from the unclear (to CLIP) concepts.
Nonetheless, Concept Alignment could be improved by adding a loss term to regularize the sketch embedding to align with AU embeddings (in Fig.~\ref{fig:architecture}, Step 3c).
There was a slight increase in privacy risk as stroke count increased for gender and ethnicity, but there was no effect on the more challenging identity recognition (Appendix Fig.~\ref{fig:model-results-face-ablation}c). 
\end{multicols}

\begin{table}[!t]
    \centering
    \caption{
        Examples of SketchXplain sketches with increasing stroke count (cols) for 6 expressions (rows). 
    }
    \label{table:demo-face-ablation}
    \vspace{-0.3cm}
    \includegraphics[width=13.0cm]{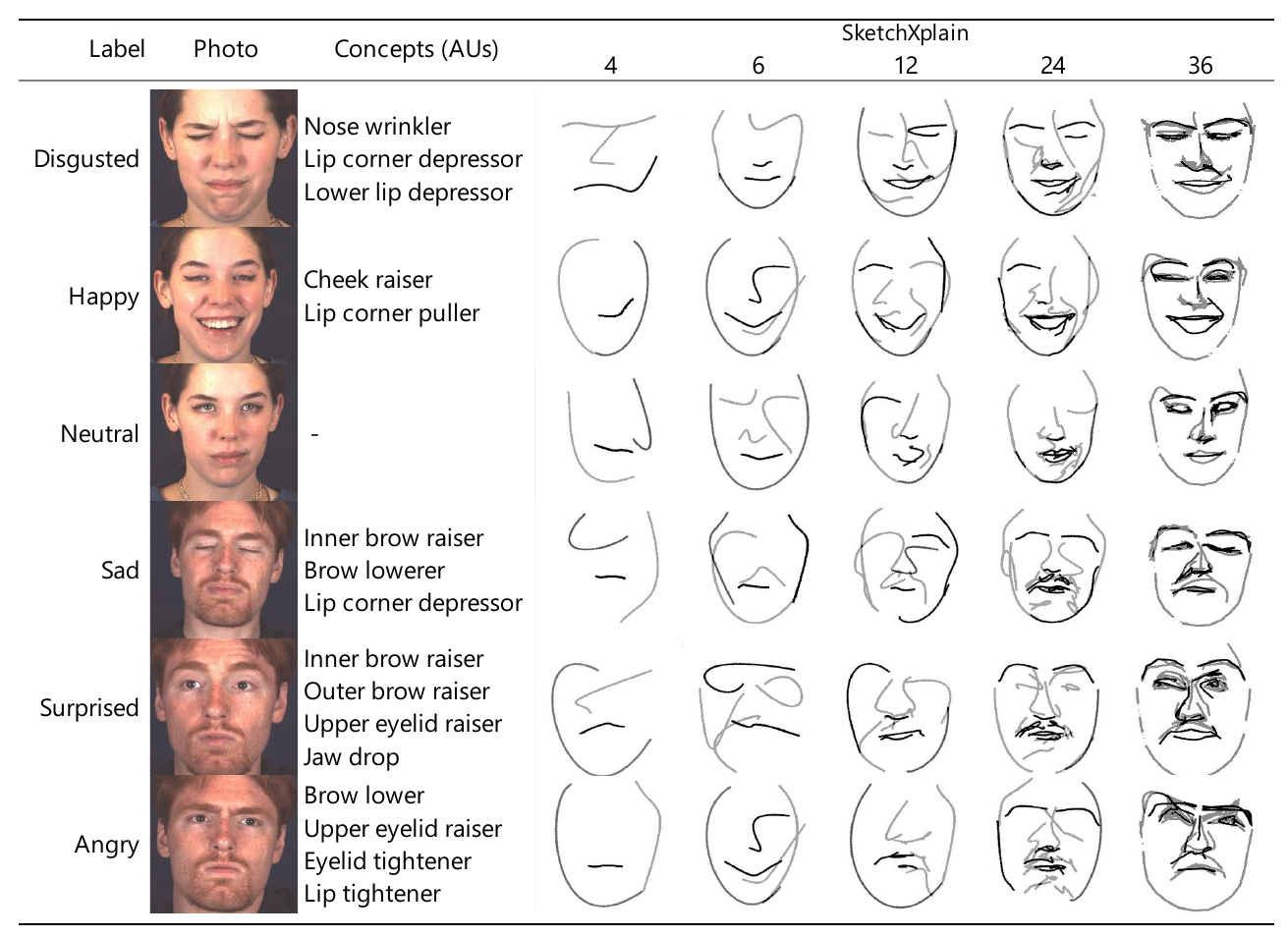}
    \vspace{-0.3cm}
\end{table}

\begin{multicols}{2}
\subsubsection{SketchXplain Intermediate Outputs}
\fixed{To illustrate how SketchXplain generates sketch explanations for facial expressions, we show examples of intermediate outputs in Appendix Table~\ref{table:demo-face-intermediate} corresponding to modular architecture steps in Fig.~\ref{fig:architecture}.}

\subsection{Quantitative User Study on Quick Interpretation}
\fixed{We provide more details of the statistical analysis for the quantitative user study reported in Section~\ref{sec:eval_face_quantitative} to evaluate quick, intuitive interpretation.}

\subsubsection{Statistical Analysis}
\fixed{We fit a linear mixed effects model for each dependent variable as the response, Visualization type, Expression and Display Duration, with other confounding variables as fixed effects, some interaction effects among the factors, and Participant as a random effect. 
Appendix Table~\ref{table:stat_intuitiveness} reports the model fit ($R^2$) and statistical effects of the linear mixed effects model.}

\end{multicols}
\begin{table}[!t]
    \centering
    \caption{
        Examples of intermediate outputs of SketchXplain for 6 expressions (rows).
    }
    \label{table:demo-face-intermediate}
    \vspace{-0.3cm}
    \includegraphics[width=14.0cm]{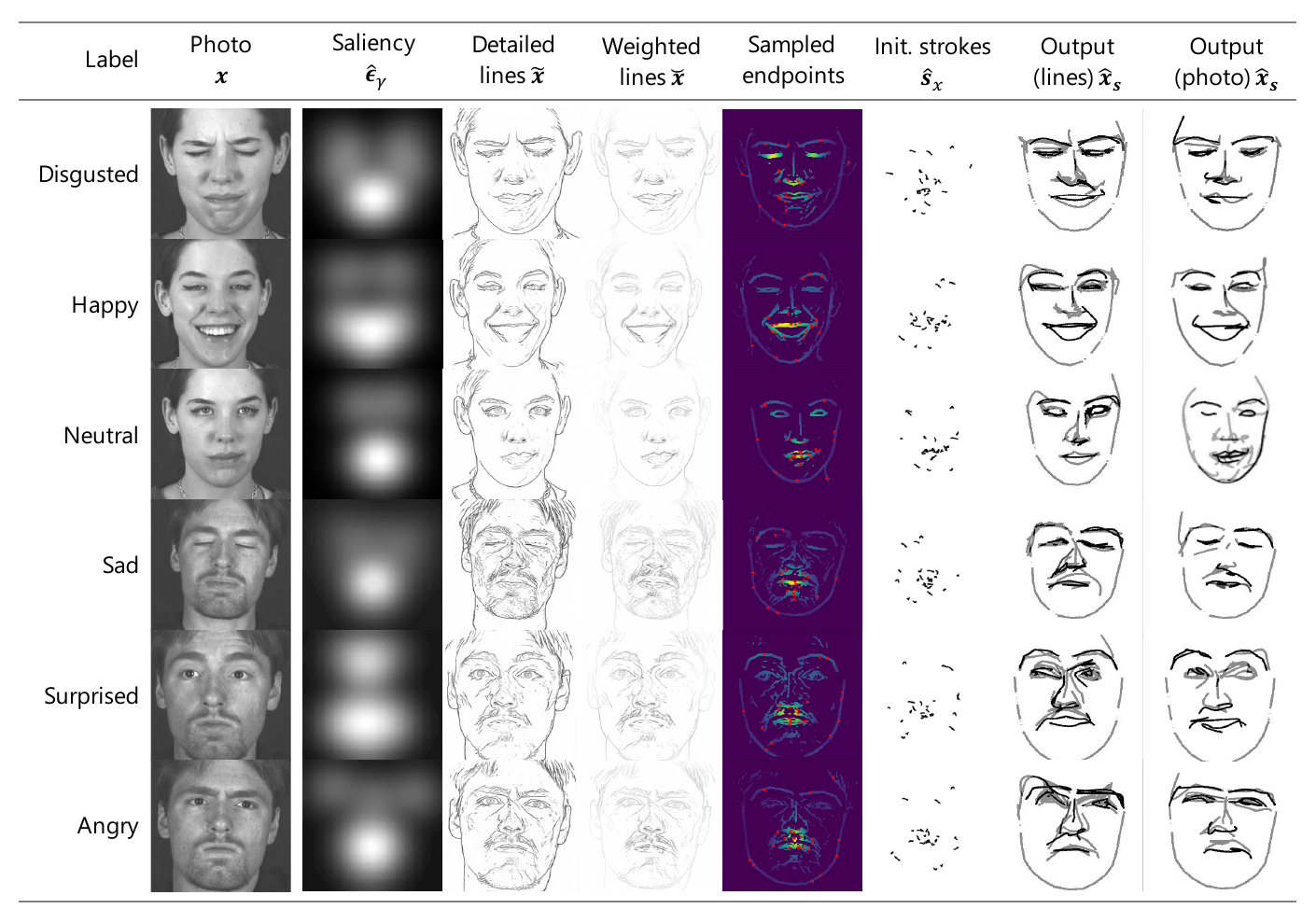}
    \vspace{-0.3cm}
\end{table}

\begin{table}[!t]
\footnotesize
\centering
\caption{
    Statistical analysis of responses due to effects one per row as linear mixed effects models for \fixed{quantitative user study on quick interpretation of face expression visualizations}. All models had various fixed main and interaction effects (shown as one effect per row) and Participant as a random effect. $F$ and $p$ values indicate ANOVA tests and $R^2$ indicates model goodness-of-fit.
}
\begin{tabular}{llrrc}
\hline
\textbf{Response}                      & \textbf{\begin{tabular}[c]{@{}l@{}}Linear \fixed{Mixed} Effects Model\\ (Participant as random effect)\end{tabular}} & \textbf{F} & \textbf{p$>$F}            & \textbf{R\textsuperscript{2}} \\ \hline
\multirow{7}{*}{AI Simulatability} & Visualization Type +                                                                                   & 6.6        & $<$.0001                     & \multirow{7}{*}{.196}        \\
                                       & Expression Truth +                                                                                        & 44.3      & $<$.0001                     &                               \\
                                       & Display Duration +                                                                                     & 42.5       & $<$.0001                     &                               \\
                                       & Visualization Type $\times$ Display Duration +                                                                & 5.7        & $<$.0001                     &                               \\
                                       & Visualization Type $\times$ Expression Truth +                                                                   & 9.0       & $<$.0001                     &                               \\
                                       & Display Duration $\times$ Expression Truth +                                                                     & 11.7       & $<$.0001                     &                               \\
                                       & Visualization Type $\times$ Display Duration $\times$ Expression Truth                                                  & 2.2        & $<$.0001                     &                               \\ \arrayrulecolor{lightgray} \hline
\multirow{11}{*}{Recall (AUs)}         & Visualization Type +                                                                                   & 22.5       & $<$.0001                     & \multirow{11}{*}{.352}        \\
                                       & Expression Truth +                                                                                        & 117.9      & $<$.0001                     &                               \\
                                       & Display Duration +                                                                                     & 85.3       & $<$.0001                     &                               \\
                                       & AU Type +                                                                                              & 15.8       & $<$.0001                     &                               \\
                                       & Visualization Type $\times$ Display Duration +                                                                & 9.3        & \multicolumn{1}{l}{$<$.0001} &                               \\
                                       & Visualization Type $\times$ Expression Truth +                                                                   & 52.4       & $<$.0001                     &                               \\
                                       & Display Duration $\times$ Expression Truth +                                                                     & 20.8       & $<$.0001                     &                               \\
                                       & Expression Truth $\times$ AU Type +                                                                              & 1237.7     & \multicolumn{1}{l}{$<$.0001} &                               \\
                                       & Display Duration $\times$ AU Type +                                                                           & 41.0       & \multicolumn{1}{l}{$<$.0001} &                               \\
                                       & Visualization Type $\times$ Display Duration $\times$ AU Type +                                                      & 22.6       & \multicolumn{1}{l}{$<$.0001} &                               \\
                                       & Visualization Type $\times$ Expression Truth $\times$ AU Type                                                           & 11.6       & $<$.0001                     &                               \\ \arrayrulecolor{black} \hline
\end{tabular}
\label{table:stat_intuitiveness}
\end{table}

\clearpage

\begin{multicols}{2}
\subsubsection{Supplementary Results on Multiclass Recognition}
\label{sec:results_confusion_matrices}
\begin{figure}[!t]
    \centering
    \includegraphics[width=16.5cm]{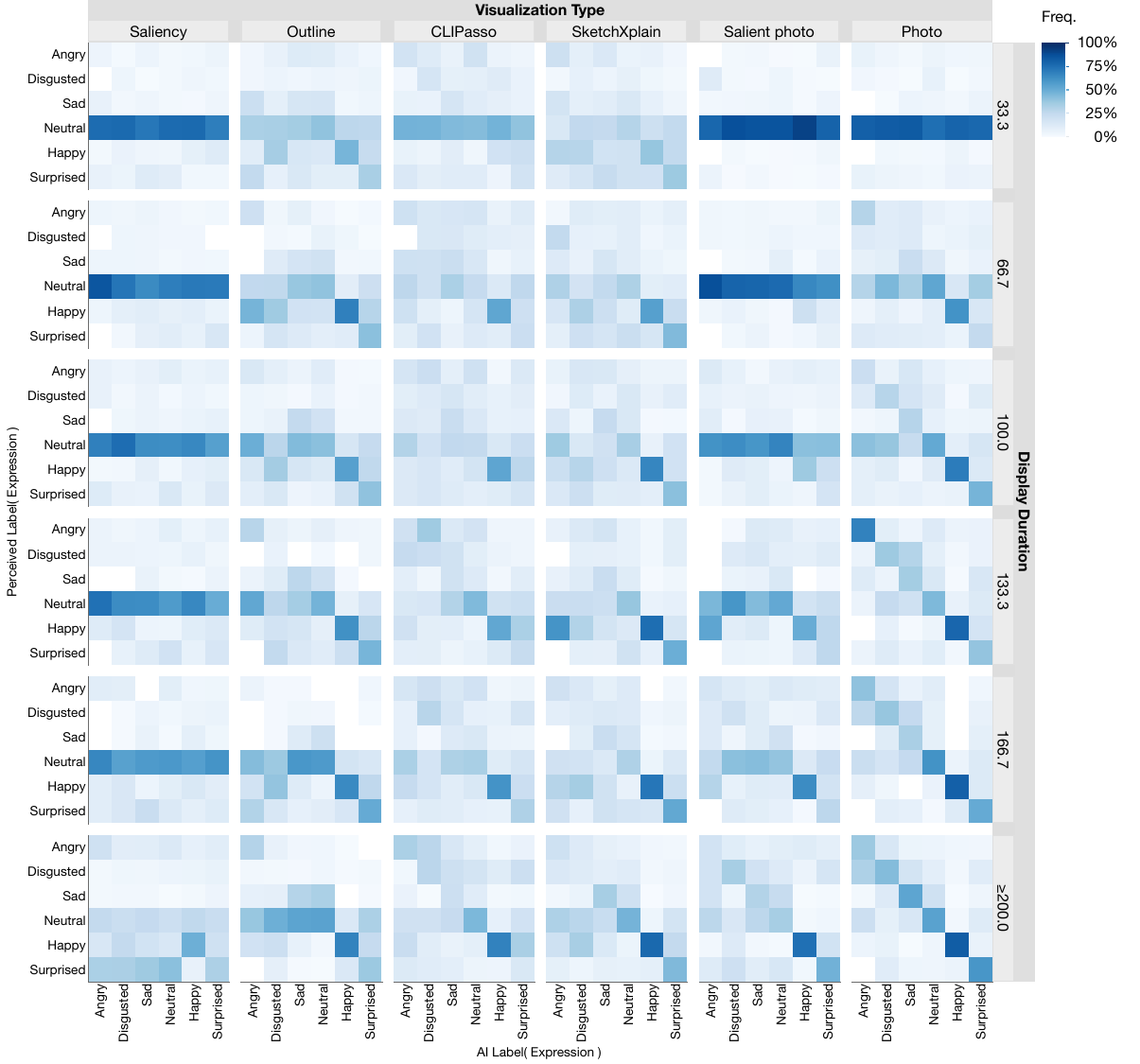}
    \caption{
        Confusion matrices of \fixed{AI alignment detailed results from the quantitative user study on quick interpretation of face expression visualizations} across Visualization Type and Display Duration. 
        The y-axis represents participants' \fixed{perceived labels}, while the x-axis shows \fixed{AI-predicted labels}. Darker blue indicates a higher number of responses in the corresponding cell.
    }
    \label{appendix-figure:intuitive_cm}
\end{figure}

\rev{To illustrate which face expressions are preserved or confused by visualizations, we present confusion matrices between participant perceived labels and AI predicted labels across Visualization Type and Display Duration (Appendix Fig.~\ref{appendix-figure:intuitive_cm}). 

Participants viewing \underline{\smash{Saliency}} exhibited a strong bias toward rating expressions as Neutral, regardless of the true Expression or Display Duration. 
This was similar for \underline{\smash{Salient Photo}}, but diminished for longer Display Durations ($t > 133.3$ms) due to supplementary information from Photo.
Interestingly, this bias toward Neural expression was also evident for participants viewing \underline{\smash{Photo}} under very tight time constraint ($t = 33.3$ms).

In contrast, line-drawing visualizations were less susceptible to this bias, likely because their salient, simple strokes were more intuitive than detailed pixels or vague saliency blobs. 
Though, \underline{\smash{Outline}} and \underline{\smash{CLIPAsso}} still demonstrated mild Neural expression bias at the shortest Display Duration ($t = 33.3$ms).
Overall, \underline{\smash{SketchXplain}} achieved strong recognition performance across all Display Durations by combining the \textit{simplicity} needed for \textit{quick} interpretation with the salient \textit{coherence} required for expression recognition.}
\\
\end{multicols}

\clearpage

\subsubsection{Survey Screenshots}
We present screenshots of the survey questionnaire in the \fixed{quantitative user study on quick interpretation of face expression visualizations} (Section~\ref{sec:eval_face_quantitative}).


\vspace{-0.3cm}
\begin{figure*}[!h]
    \centering    
    \includegraphics[width=11.0cm]{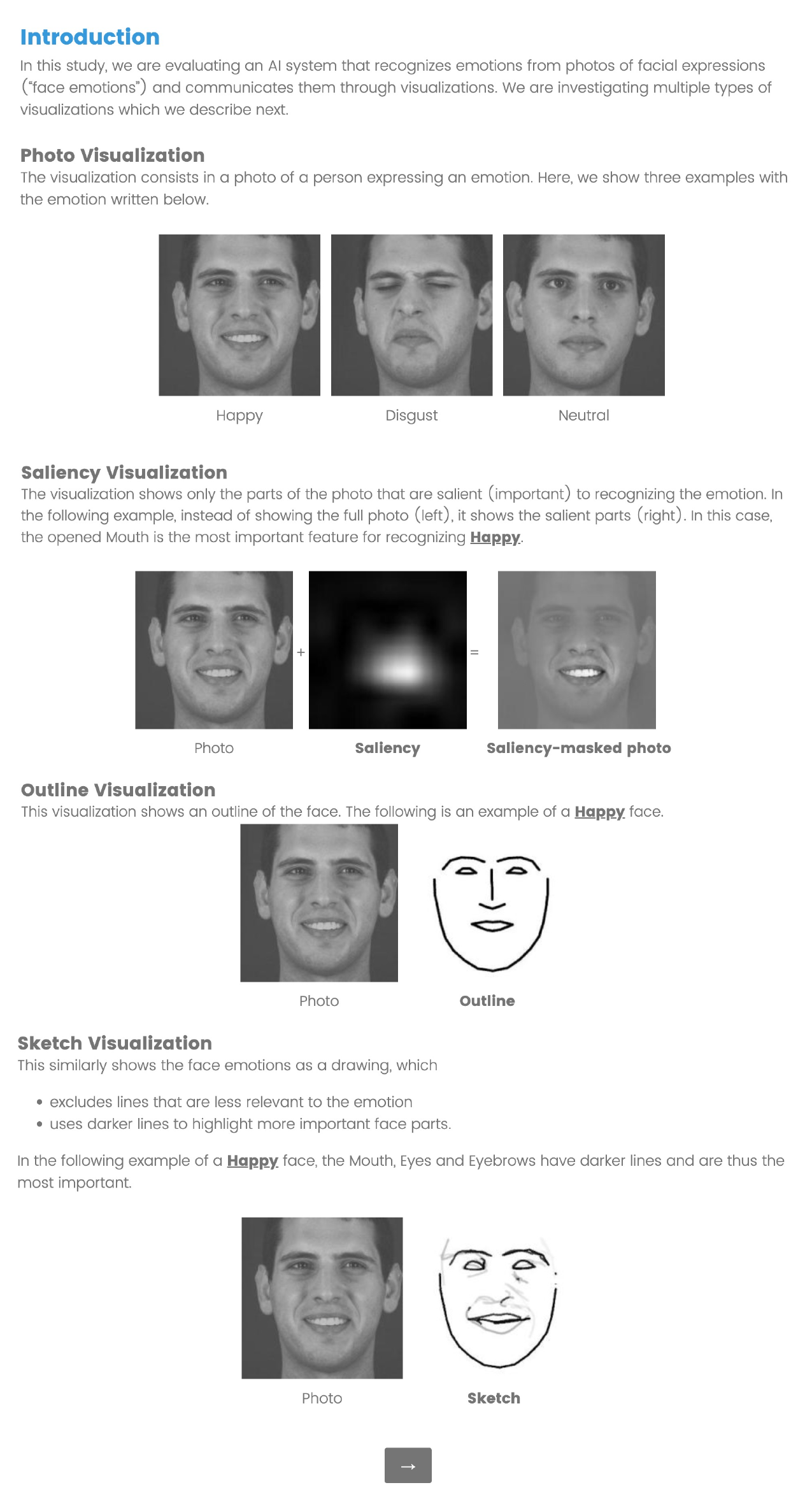}
    \caption{Tutorial to clarify users' tasks and showcase visualizations. (Shared for both quantitative studies)}
    \label{appendix-figure:Tutorial_shared}
\end{figure*}

\begin{figure*}[h!]
    \centering    
    \includegraphics[width=14cm]{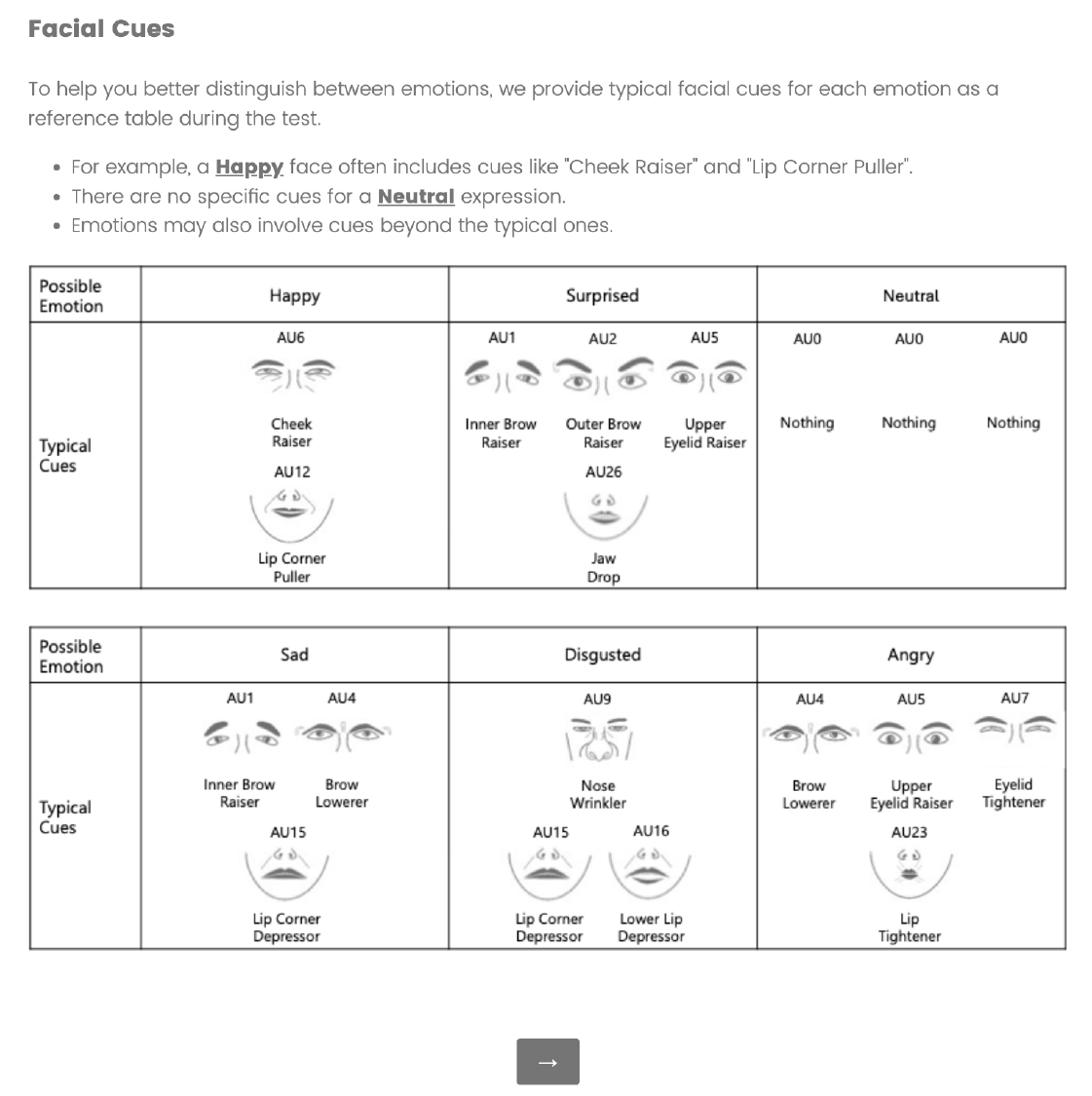}
    \caption{Tutorial on Action Units (AUs) and their correlation with face expressions.}
    \label{appendix-figure:Tutorial_AU}
\end{figure*}

\begin{figure*}[h!]
    \centering    
    \includegraphics[width=12cm]{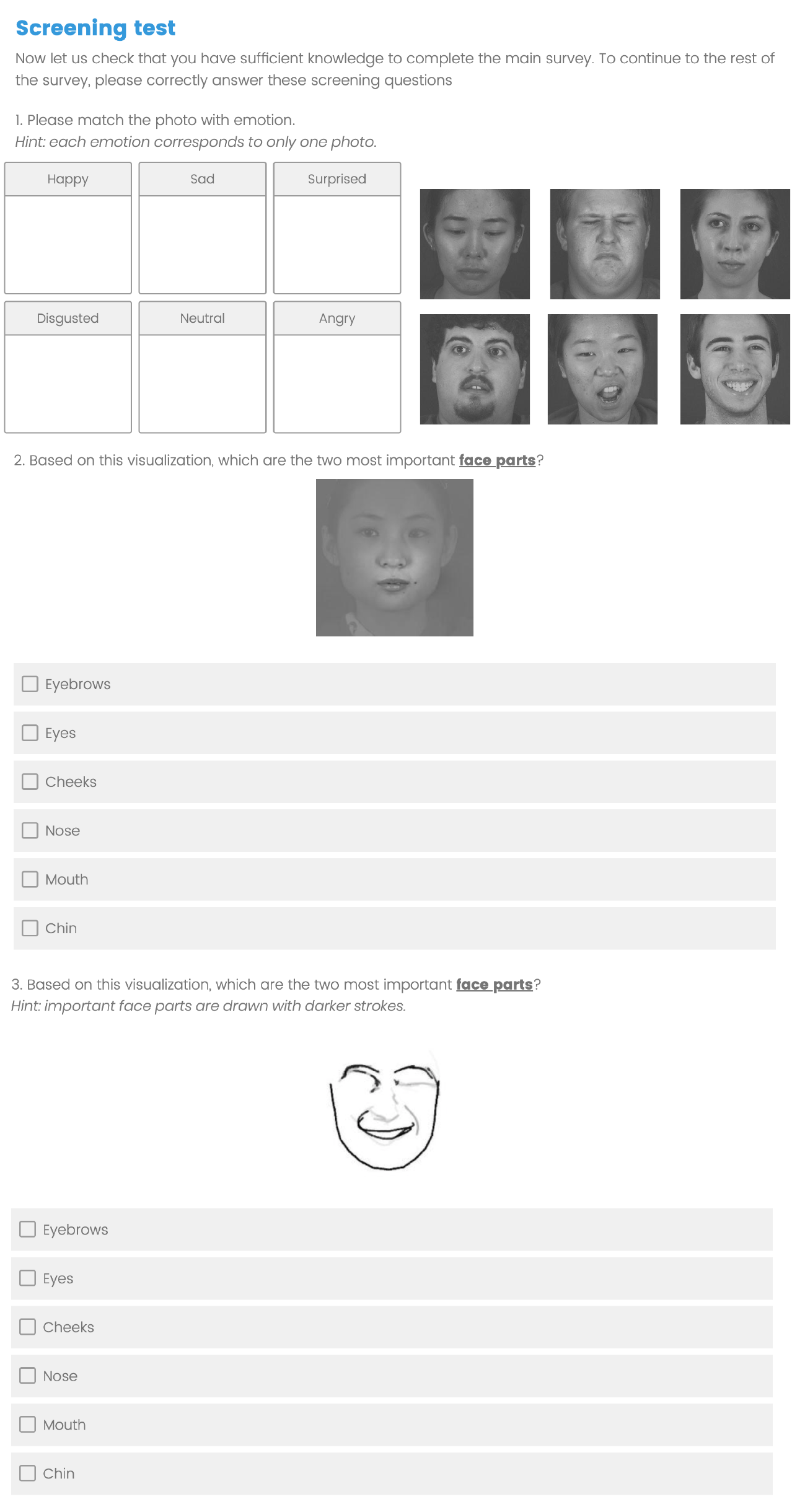}
    \caption{Screening questions on face expression and visualizations.
    }
    \label{appendix-figure:Screening_shared}
\end{figure*}

\begin{figure*}[h!]
    \centering    
    \includegraphics[width=12cm]{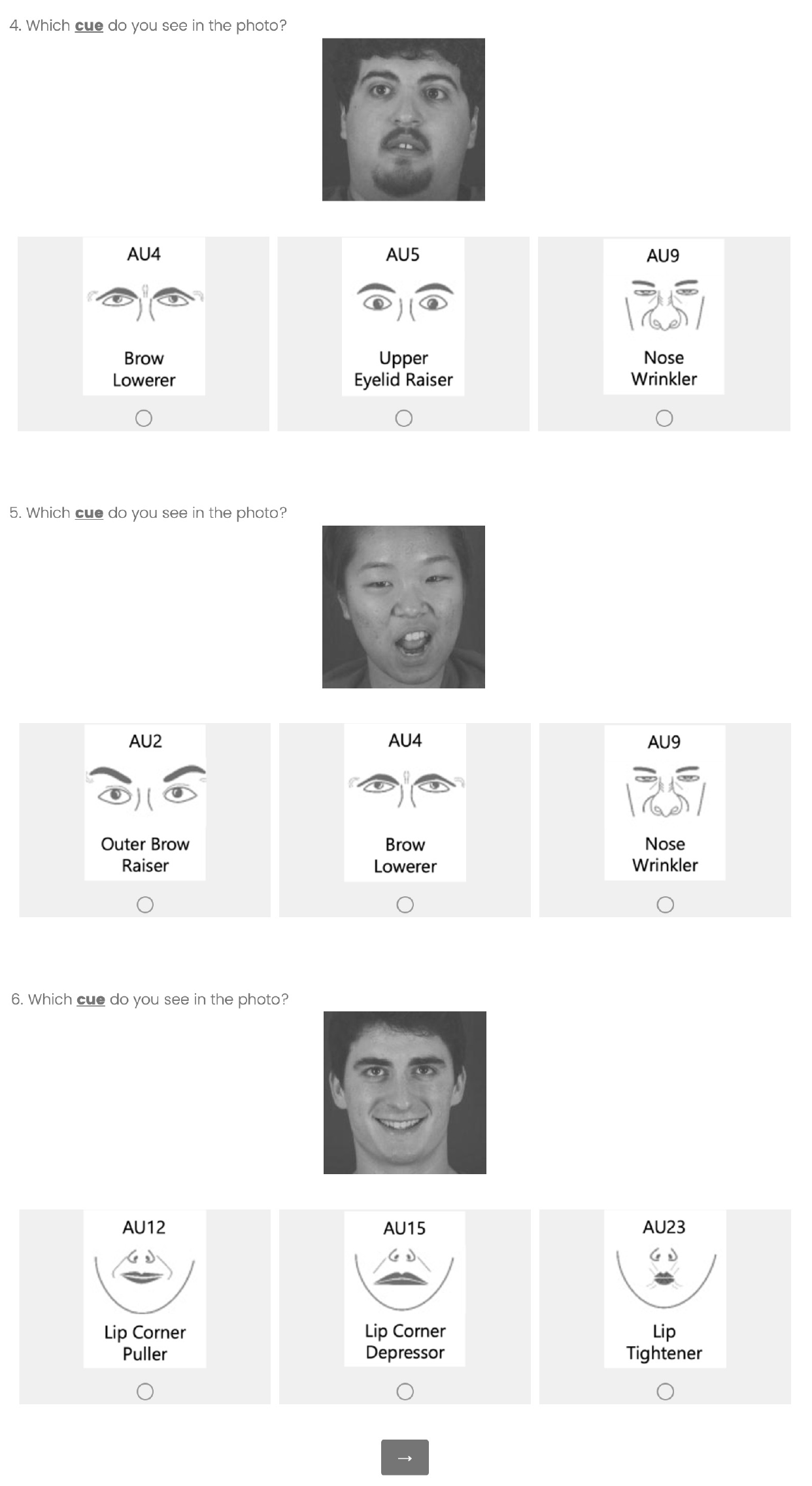}
    \caption{Screening questions to check users' understanding on Action Units.}
    \label{appendix-figure:Screening_AU}
\end{figure*}

\begin{figure*}[h!]
    \centering    
    \includegraphics[width=10cm]{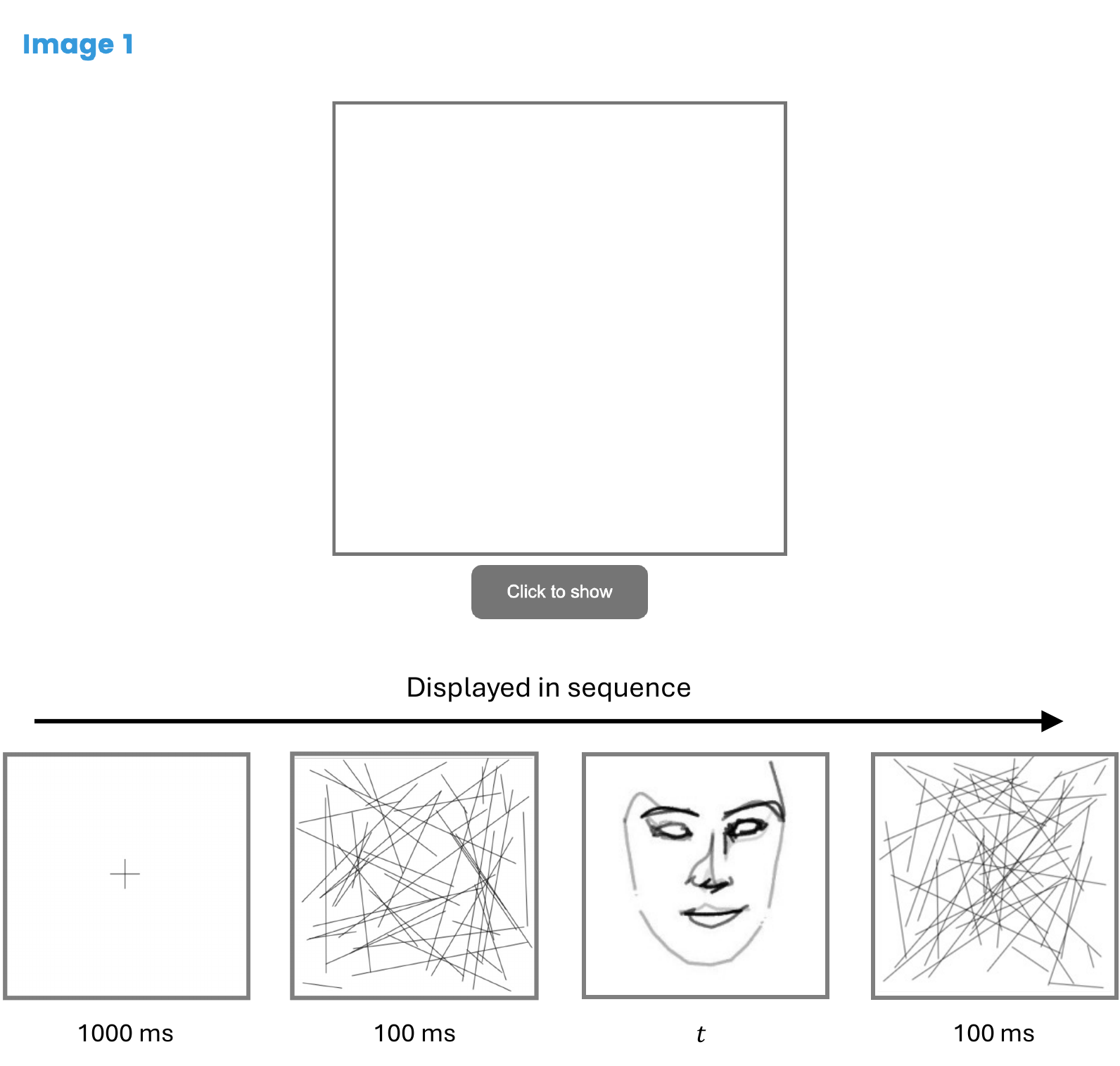}
    \caption{
        Example main study per-image trial with SketchXplain visualization. After ``click to show'', the participant views the first frame with a cross to focus attention, views random lines before and after target visualization to clear memory, views the SketchXplain with $t \in \{33.3, 66.7, 100.0, 133.3, 166.7\}$ ms.}
    \label{appendix-figure:Trial_speed_test}
\end{figure*}

\begin{figure*}[h!]
    \centering    
    \includegraphics[width=12cm]{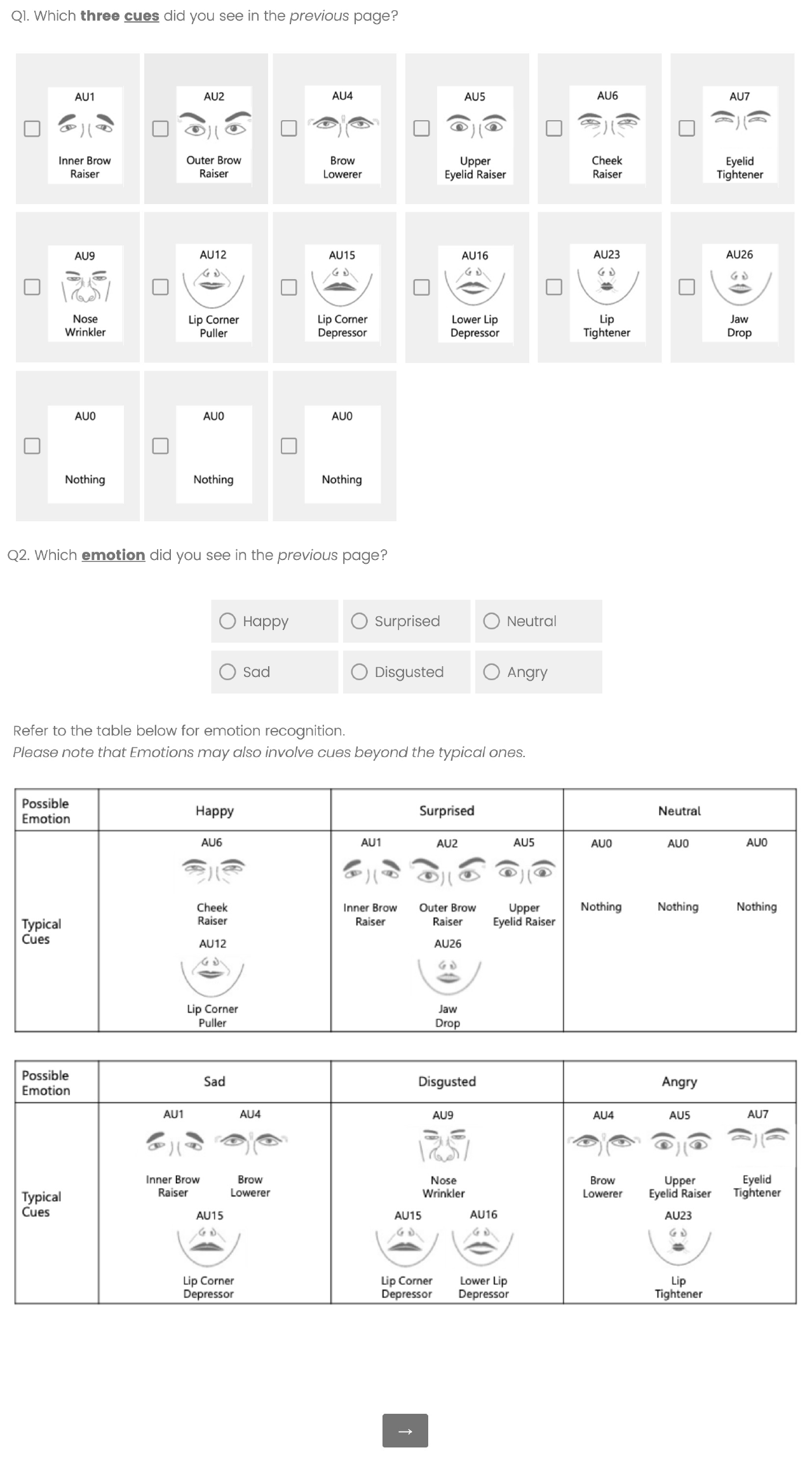}
    \caption{Example main study per-image trial with Action Unit and \fixed{face} expression recognition questions with reference table attached.}
    \label{appendix-figure:Trial_speed_test_q}
\end{figure*}
\clearpage

\begin{multicols}{2}
\subsection{Quantitative User Study on Face Privacy Protection}
\label{sec:privacy_evaluation}

We investigated in a quantitative user study how sketch explanations provide privacy protection by omitting identifiable information in the original \fixed{face} photo.

\subsubsection{Experiment Apparatus and Measures}

The user task was to identify facial attributes (gender and ethnicity) and match identity (from 4 face photos) from various visualizations. Unlike the previous quantitative user study \fixed{(Section~\ref{sec:eval_face_quantitative})}, images were displayed without a time limit. We conducted a within-subjects experiment with Visualization Type and Expression as independent variables.

\subsubsection{Participants}

We recruited another 89 participants from Prolific with the same qualification criteria. 37 passed our screening test; 21 were male \rev{and 16 were female}, with ages 25--66 (Median = 41). They completed the survey in median 23.7 minutes and were compensated UK \pounds 4.00. 

\subsubsection{Experiment Procedure}

\begin{figure*}[h!]
    \centering    
    \includegraphics[width=12cm]{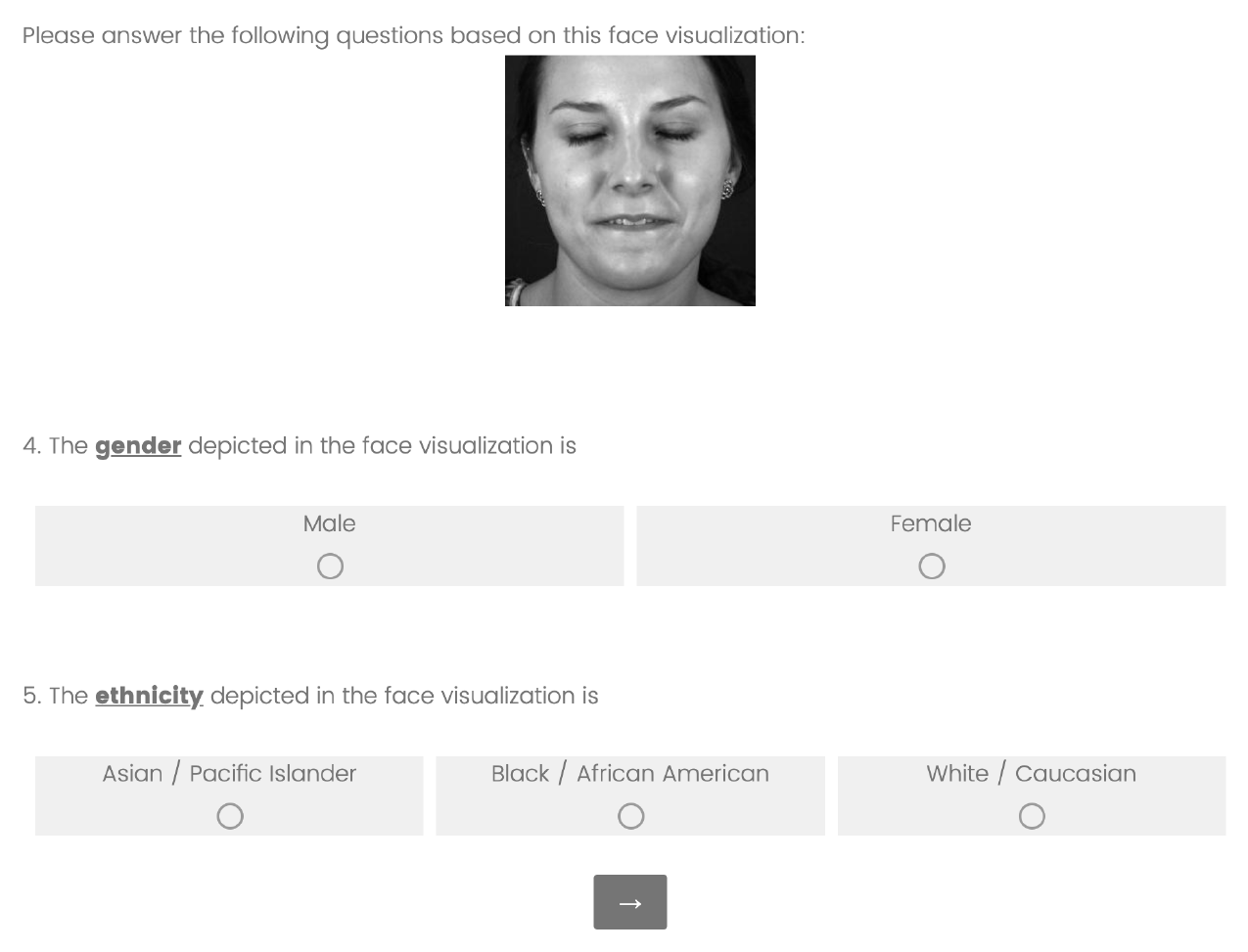}
    \caption{Screening questions to check users' understanding on gender and ethnicity.}
    \label{appendix-figure:Screening_privacy_a}
\end{figure*}

\begin{figure*}[h!]
    \centering    
    \includegraphics[width=12cm]{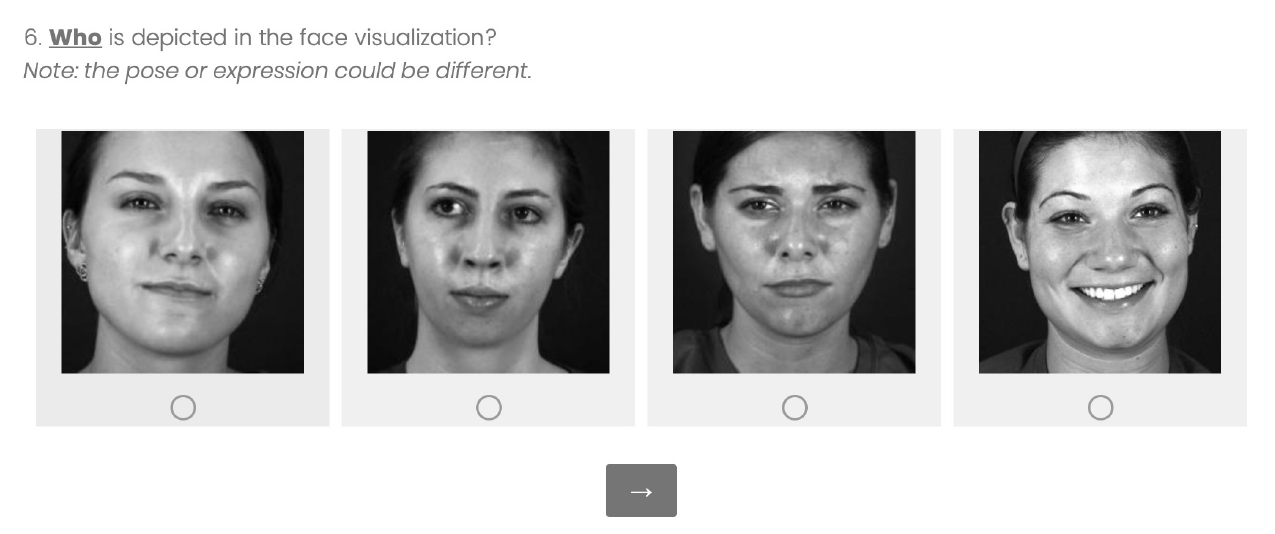}
    \caption{Screening questions to check users' understanding of identity; the question continues on the next page.}
    \label{appendix-figure:Screening_privacy_b}
\end{figure*}

\begin{figure*}[h!]
    \centering    
    \includegraphics[width=11.5cm]{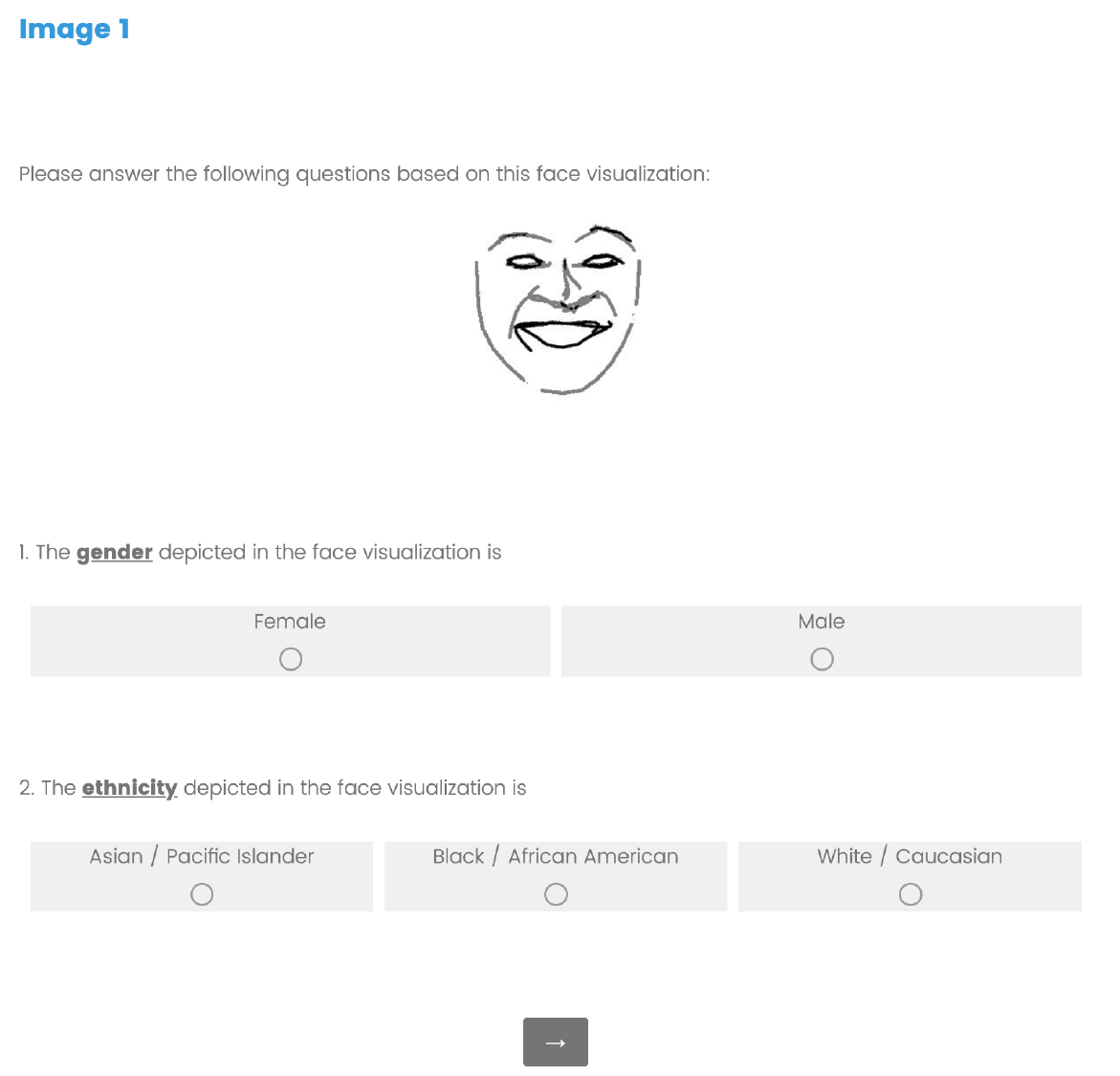}
    \caption{Example main study per-image trial with gender and ethnicity recognition questions.}
    \label{appendix-figure:Trial_privacy_test_q1}
\end{figure*}

\begin{figure*}[h!]
    \centering    
    \includegraphics[width=12cm]{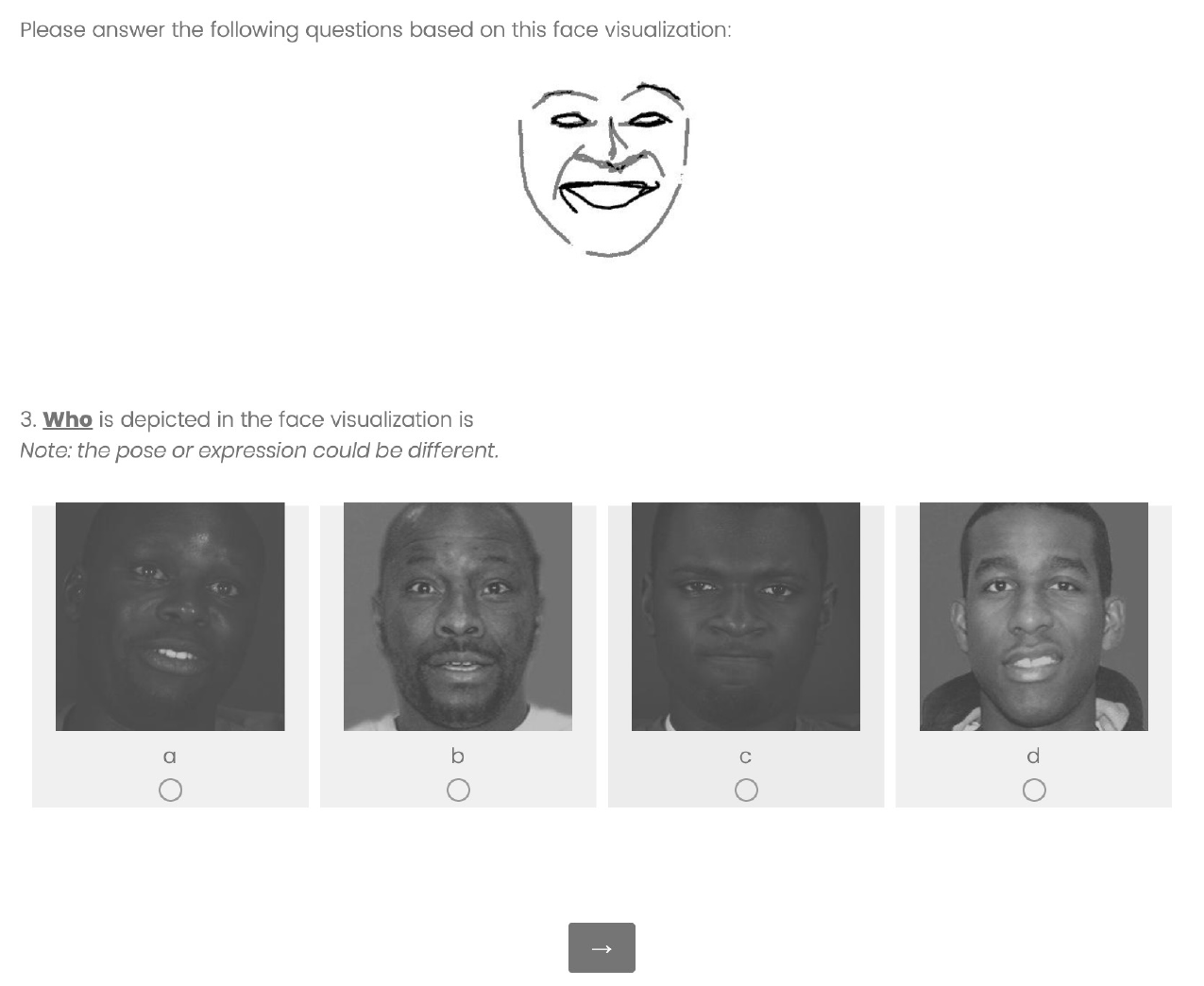}
    \caption{Example main study per-image trial with identification questions.}
    \label{appendix-figure:Trial_privacy_test_q2}
\end{figure*}

Participants followed a procedure similar to the previous quantitative study. 
For screening, they had to 
a) demonstrate ability to interpret \fixed{all} visualizations, 
b) identify facial attributes and identity from photos (Appendix Figs.~\ref{appendix-figure:Screening_privacy_a}--\ref{appendix-figure:Screening_privacy_b}). 
In the main study, for each of 72 trials, participants were asked to identify facial attributes from \fixed{the} visualization on the first page (Appendix Fig.~\ref{appendix-figure:Trial_privacy_test_q1}), identity on the next page (Appendix Fig.~\ref{appendix-figure:Trial_privacy_test_q2}).

\subsubsection{Statistical Analysis and Quantitative Results}

For all dependent variables, we fit a linear mixed-effects model with Expression, Ethnicity, Gender, and Visualization Type as fixed main effects, various fixed interaction effects, and Participant as the random effect. See Appendix Table~\ref{table:statPrivacyStudy} for details. 

\begin{figure*}[!t]
    \centering
    \hspace*{-0.2cm}
    \includegraphics[width=14.5cm]{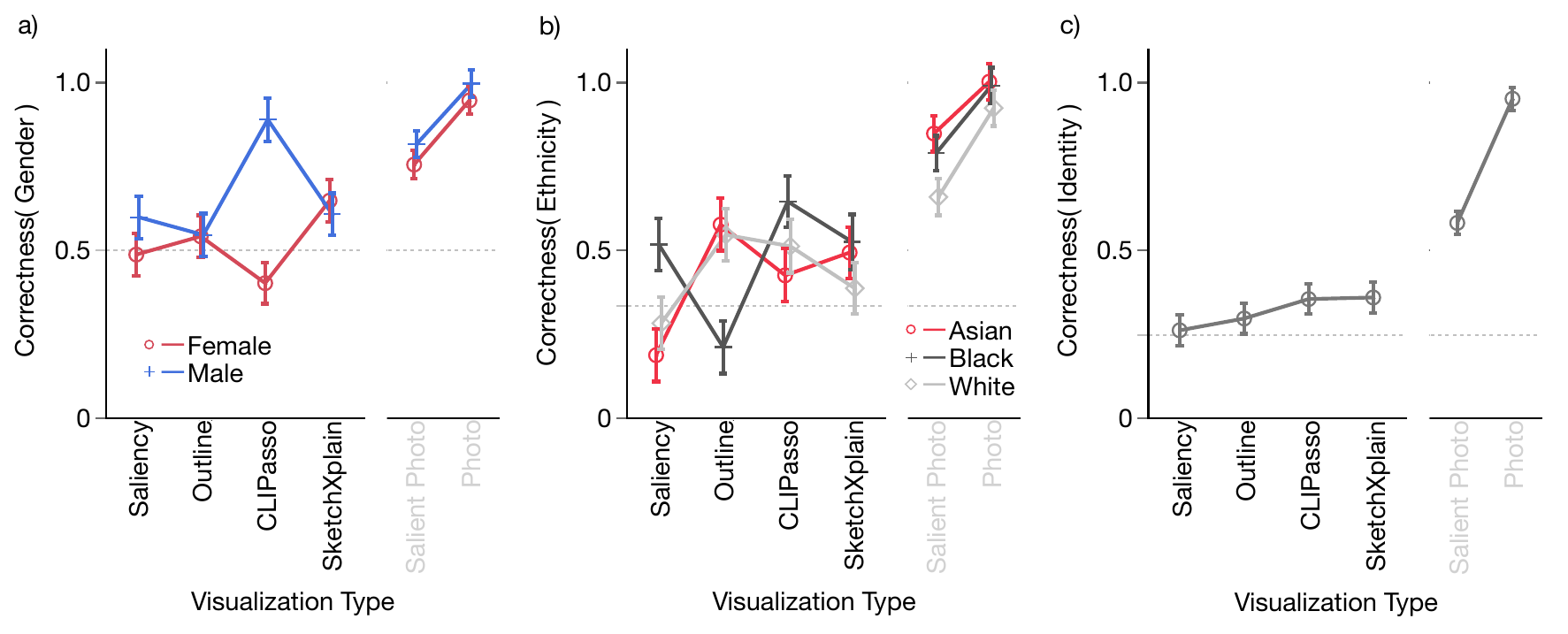}
    \vspace{-0.3cm}
    \caption{
        Results 
        \fixed{from the face privacy protection user study}
        of recognition correctness of sensitive attributes: a) gender, b) ethnicity, and c) identity.
        Grey dotted line represents the random guess correctness (50.0\% for gender, 33.3\% for ethnicity, 25\% for identity).
    }
    \label{fig:privacy_results}
    \vspace{-0.3cm}
\end{figure*}

Appendix Fig.~\ref{fig:privacy_results} shows that
\fixed{\underline{\smash{SketchXplain}}, \underline{\smash{CLIPasso}} and \underline{\smash{Outline}}}
suppressed Gender, Ethnicity and Identity information
almost as well as \underline{\smash{Saliency}}.
\underline{\smash{Salient Photo}} leaked the most information through \fixed{exposed image} pixels.
Interestingly, \underline{\smash{CLIPasso}} leaked more information about males than females.
\end{multicols}
\balance

\begin{table*}[!t]
\footnotesize
\centering
\caption{
    Statistical analysis of responses due to effects one per row as linear mixed effects models for \fixed{face privacy protection user study}. All models had various fixed main and interaction effects (shown as one effect per row) and Participant as a random effect. Rows with grey text indicate non-significant effects. $F$ and $p$ values indicate ANOVA tests and $R^2$ indicates model goodness-of-fit.
}
\begin{tabular}{llrrc}
\hline
\textbf{Response}                       & \textbf{\begin{tabular}[c]{@{}l@{}}Linear \fixed{Mixed} Effects Model\\ (Participant as random effect)\end{tabular}} & \textbf{F} & \textbf{p$>$F} & \textbf{R\textsuperscript{2}} \\ \hline
\multirow{7}{*}{Recall (Gender)}        & {\color{lightgray}Target Expression} +                                                                                       & {\color{lightgray}1.0}        & {\color{lightgray}n.s.}                      & \multirow{7}{*}{.208}        \\
                                        & Target Ethnicity +                                                                                     & 28.6       & $<$.0001          &                               \\
                                        & Target Gender +                                                                                        & 46.8       & $<$.0001          &                               \\
                                        & Visualization Type +                                                                                   & 68.6       & $<$.0001          &                               \\
                                        & Visualization Type $\times$ Target Expression +                                                                  & 1.9        & .0042                     &                               \\
                                        & Visualization Type $\times$ Target Ethnicity +                                                                & 3.6        & $<$.0001          &                               \\
                                        & Visualization Type $\times$ Target Gender                                                                   & 22.4       & $<$.0001          &                               \\ \arrayrulecolor{lightgray} \hline
\multirow{7}{*}{Recall (Ethnicity)}     & Target Expression +                                                                                       & 2.9        & .0126                     & \multirow{7}{*}{.283}         \\
                                        & Target Ethnicity +                                                                                     & 4.9        & .0076                     &                               \\
                                        & Target Gender +                                                                                        & 32.5       & $<$.0001          &                               \\
                                        & Visualization Type +                                                                                   & 140.3      & $<$.0001          &                               \\
                                        & Visualization Type $\times$ Target Expression +                                                                  & 1.8        & .0093                     &                               \\
                                        & Visualization Type $\times$ Target Ethnicity +                                                                & 15.0       & $<$.0001          &                               \\
                                        & Visualization Type $\times$ Target Gender                                                                   & 4.7        & .0003                     &                               \\ \arrayrulecolor{lightgray} \hline
\multirow{7}{*}{Correctness (Identity)} & Target Expression +                                                                                       & 3.8        & .0019                     & \multirow{7}{*}{.273}         \\
                                        & Target Ethnicity +                                                                                     & 20.6       & $<$.0001          &                               \\
                                        & Target Gender +                                                                                        & 4.4        & .0358                     &                               \\
                                        & Visualization Type +                                                                                   & 163.4      & $<$.0001          &                               \\
                                        & {\color{lightgray}Visualization Type $\times$ Target Expression} +                                                                  & {\color{lightgray}1.4}        & {\color{lightgray}n.s.}                      &                               \\
                                        & Visualization Type $\times$ Target Ethnicity +                                                                & 3.7        & $<$.0001          &                               \\
                                        & {\color{lightgray}Visualization Type $\times$ Target Gender}                                                                   & {\color{lightgray}1.2}        & {\color{lightgray}n.s.}                      &                               \\ \arrayrulecolor{black} \hline
\end{tabular}
\label{table:statPrivacyStudy}
\end{table*}

\subsubsection{Survey Screenshots}
\fixed{We present screenshots of the survey questionnaire in the quantitative user study on face privacy protection (Appendix Section~\ref{sec:privacy_evaluation}).}

\clearpage

\onecolumn
\begin{multicols}{2}
\section{Evaluation on Skin Lesion Image Domain}

\fixed{We provide details on applying SketchXplain to explaining skin lesion images (Section~\ref{sec:evaluation_skin_lesion}).}

\subsection{SketchXplain Intermediate Outputs}
\fixed{Appendix Table~\ref{table:demo-lesion-intermediate} shows intermediate outputs from modular steps in SketchXplain for the several skin lesion example instances.}

\end{multicols}
\vspace{-0.3cm}
\begin{table}[t]
    \centering
    \caption{
        Examples of intermediate outputs of SketchXplain for skin lesion instances (rows).
    }
    \vspace{-0.3cm}
    \label{table:demo-lesion-intermediate}
    \includegraphics[width=13.5cm]{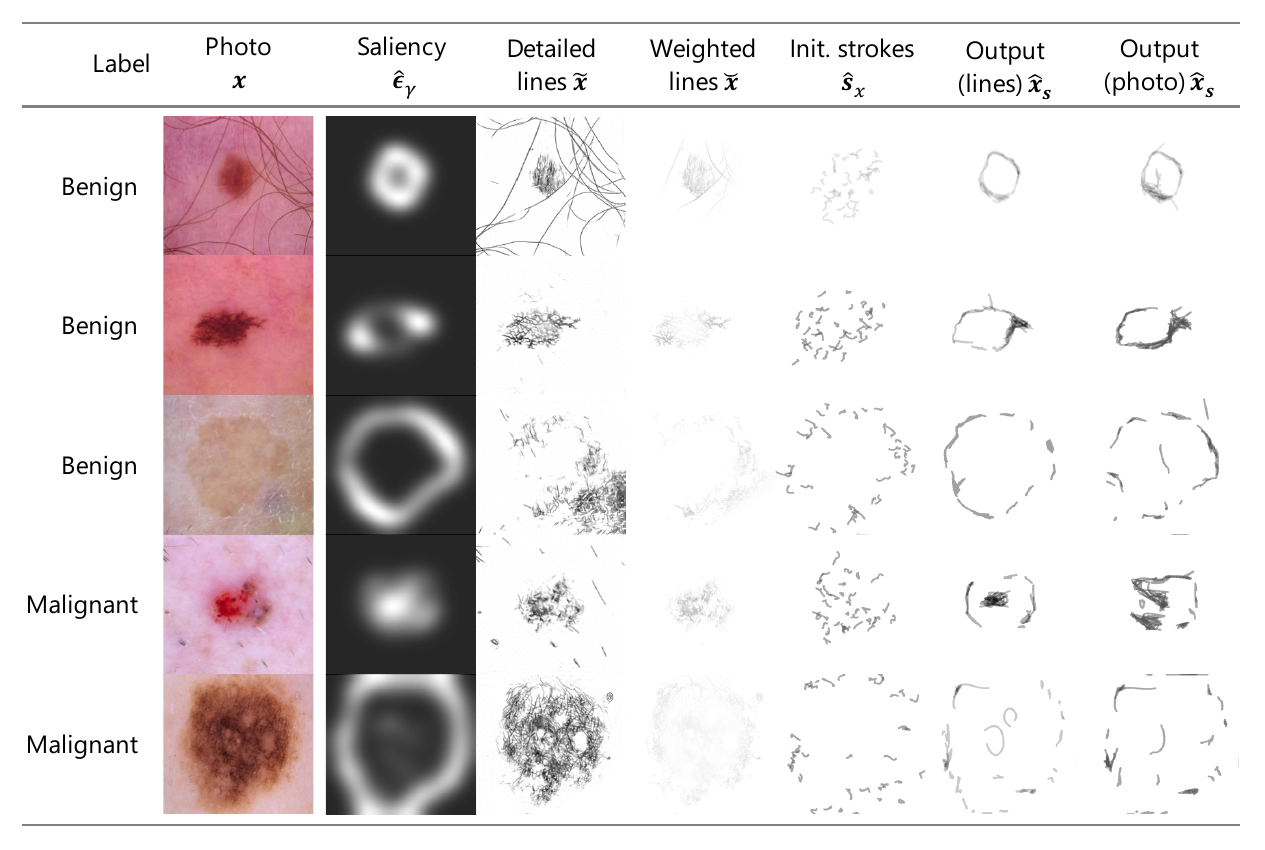}
    \vspace{-0.3cm}
\end{table}
\clearpage


\begin{table*}[t]
\footnotesize
\centering
\vspace{0.1cm}
\caption{
Concepts used in the Label-Free Concept Bottleneck model for different domains. We obtained the concepts for each class using GPT-4o with the prompt: ``Please provide the most important visual attributes to recognize [class].''
}
\renewcommand{\arraystretch}{1.2} 
\begin{tabular}{llp{8.5cm}}
\hline
Domain                       & Class                & Concepts         \\ \hline
\multirow{16}{*}{\begin{tabular}[c]{@{}l@{}}Dog\\Breeds\end{tabular}} 
& Australian Terrier & Hair-covered eyes, Upright ears, Shaggy muzzle, Bushy eyebrows, Topknot on head \\
& Border Terrier     & Folded ears, Wiry coat, Narrow body, Short muzzle, Thick base tail \\
& Samoyed            & Fluffy ears, Curled bushy tail, Thick White coat, Dark almond eyes, Feathered legs \\
& Beagle             & Long floppy ears, Short coat, Large Brown eyes, Square muzzle, White-tipped tail \\
& Shih-Tzu           & Hairy floppy ears, Short flat muzzle, Long flowing coat, Topknot on head, Curled tail over back, Short legs \\
& English Foxhound   & Long drooping ears, Deep chest, Strong straight legs, Long tapered tail \\
& Rhodesian Ridgeback& Floppy ears, Strong legs, Long muzzle, Tapered tail \\
& Dingo              & Pointed upright ears, Bushy tail, Slender long legs, Narrow muzzle, Sandy coat \\
& Golden Retriever   & Fluffy ears, Feathered tail, Long wavy coat, Muscular build \\
& Old English Sheepdog & Hair-covered eyes, Fluffy drooping ears, Shaggy grey-white coat, Docked tail \\ \hline
\multirow{13}{*}{\begin{tabular}[c]{@{}l@{}}Land\\Animals\end{tabular}} 
& Bear               & Rounded ears, Massive paws, Rounded head. \\
& Bison              & Shaggy front, Large hump, Short horns, Massive face \\
& Bull               & Broad body, Short legs, Horns \\
& Buffalo            & Curved horns, Straight back, Low posture, Large muzzle \\
& Cheetah            & Round ears, Long straight tail, Long limbs \\
& Elephant           & Large fan ears, Tusks, Trunk \\
& Giraffe            & Long neck, Spotted coat, Ossicones \\
& Horse              & Elongated face, Flowing tail, Flowing mane \\
& Impala             & Lyre-shaped horns, Sleek body \\
& Lion               & Large mane, Round ears, Tufted tail end \\
& Rabbit             & Long ears, Short round tail, Compact round body \\
& Squirrel           & Large bushy tail, Pointed ears with tufts, Arched back \\
& Zebra              & Striped coat, Mane stands upright, Elongated face \\ \hline
\multirow{8}{*}{\begin{tabular}[c]{@{}l@{}}Land\\Vehicles\end{tabular}} 
& Race Car           & Spoiler, Racing stripes, Air intake, Wheels, Windshield \\
& Police Car         & Siren lights, Badge, Wheels, Windshield \\
& Luxury Car         & Elegant design, Spoiler, Wheels, Side-mirrors, Windshield \\
& Minivan            & Sliding doors, Large windows, Wheels, Side-mirrors, Windshield \\
& Dump Truck         & Dump bed, Hydraulic arms, Heavy tires \\
& Fire Truck         & Water hose, Ladder, Emergency lights, Wheels, Windshield \\
& Truck              & Cargo bed, Wheels, Side-mirrors, Windshield \\
& Scooter            & Small wheels, Handlebar, Seat \\ \hline
\end{tabular}
\label{table:free_label_concepts}
\end{table*}

\begin{multicols}{2}

\section{Evaluation on General Image Domains}
\label{sec:general_image}

To investigate the generalizability of sketch explanations beyond face expressions \fixed{and skin lesions}, we applied SketchXplain to three other domains (dog breeds, land animals, and land vehicles). 
These span visual granularity and explanatory concepts.
We describe data preparation, extensions to SketchXplain, results from both modeling studies and a qualitative study.
We defer the quantitative evaluation of general sketches to future work, due to the heterogeneity of user goals for open domain images.
Nevertheless, we draw qualitative insights from a qualitative study for the potential uses and benefits of sketch explanations in general.

\subsection{Data Preparation}
We used
12,954 images of 10 dog breeds (Australian Terrier, Border Terrier, Samoyed, Beagle, Shih-Tzu, English Foxhound, Rhodesian Ridgeback, Dingo, Golden Retriever, Old English Sheepdog) from the ImageWoof dataset~\cite{howard2020imagenette}, 
21,874 images of 13 land animals (Bear, Bison, Bull, Buffalo, Cheetah, Elephant, Giraffe, Horse, Impala, Lion, Rabbit, Squirrel, Zebra) from the iNaturalist dataset~\cite{van2018inaturalist}, and 
9,199 images of 8 land vehicles (Race Car, Police Car, Luxury Car, Minivan, Dump Truck, Fire Truck, Truck, Scooter) from ImageNet.
For simplicity, these classes were selected such that the class objects were clearly shown in the photos, and their concepts can be simply identified and drawn; further work is needed to extract and depict more difficult concepts.

\begin{figure*}[htbp]
    \centering
    \includegraphics[width=11.5cm]{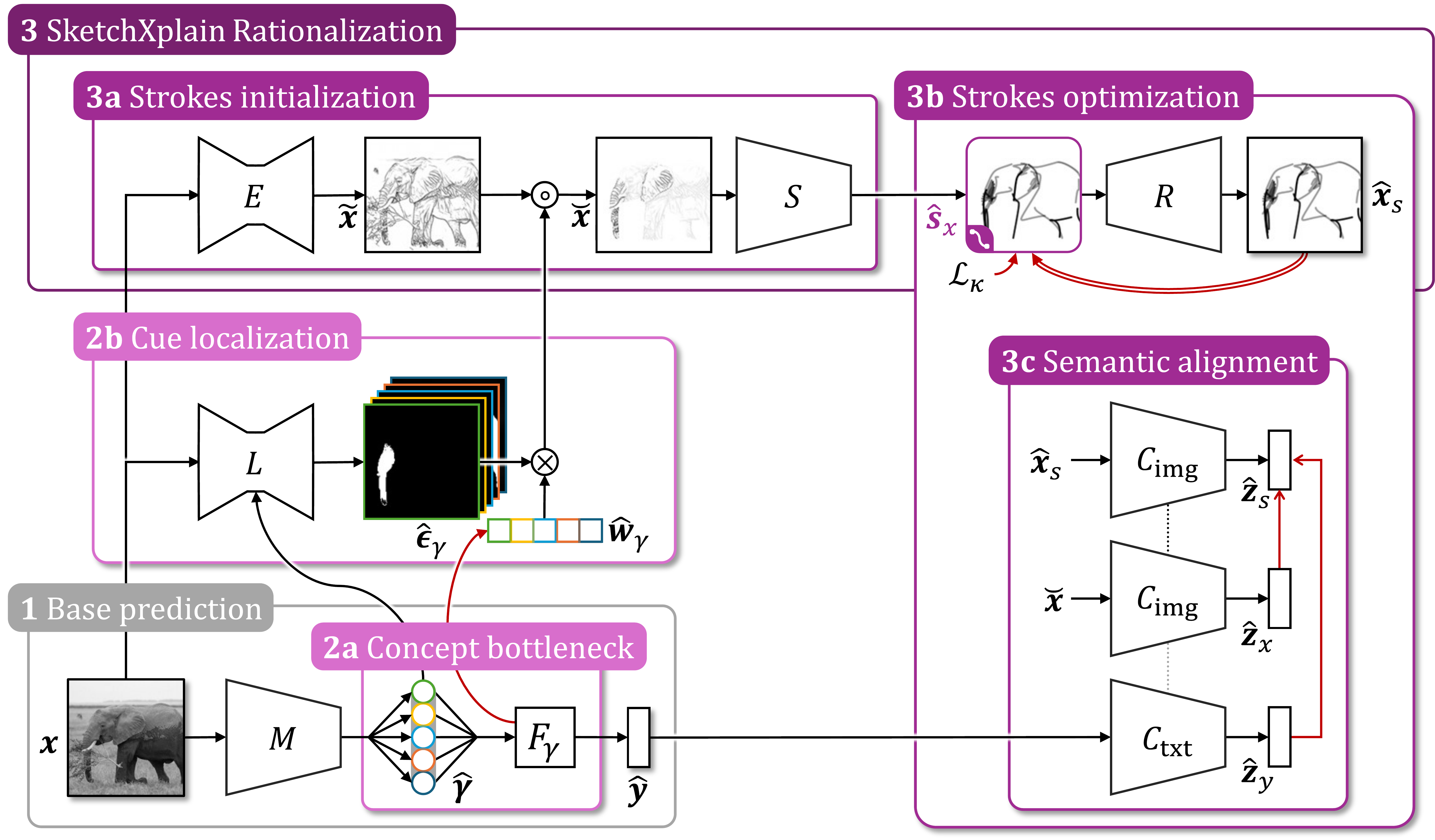}
    \caption{
        SketchXplain architecture adapted to general images. 
        The concept bottleneck model $M$ is substituted by \fixed{a CLIP visual encoder~\cite{radford2021learning}}, and cue localization $L$ by OWL-ViT~\cite{minderer2022simple} and Segment Anything~\cite{kirillov2023segment}.
        Other components are unchanged from Fig.~\ref{fig:architecture}.
    }
    \label{fig:architecture-general}

    \begin{minipage}{\textwidth}
        \centering
        \caption{
            Examples comparing visualizations (cols) for explaining Dog Breeds, Land Animals and Land Vehicles (rows). 
        }
        \includegraphics[width=10.5cm]{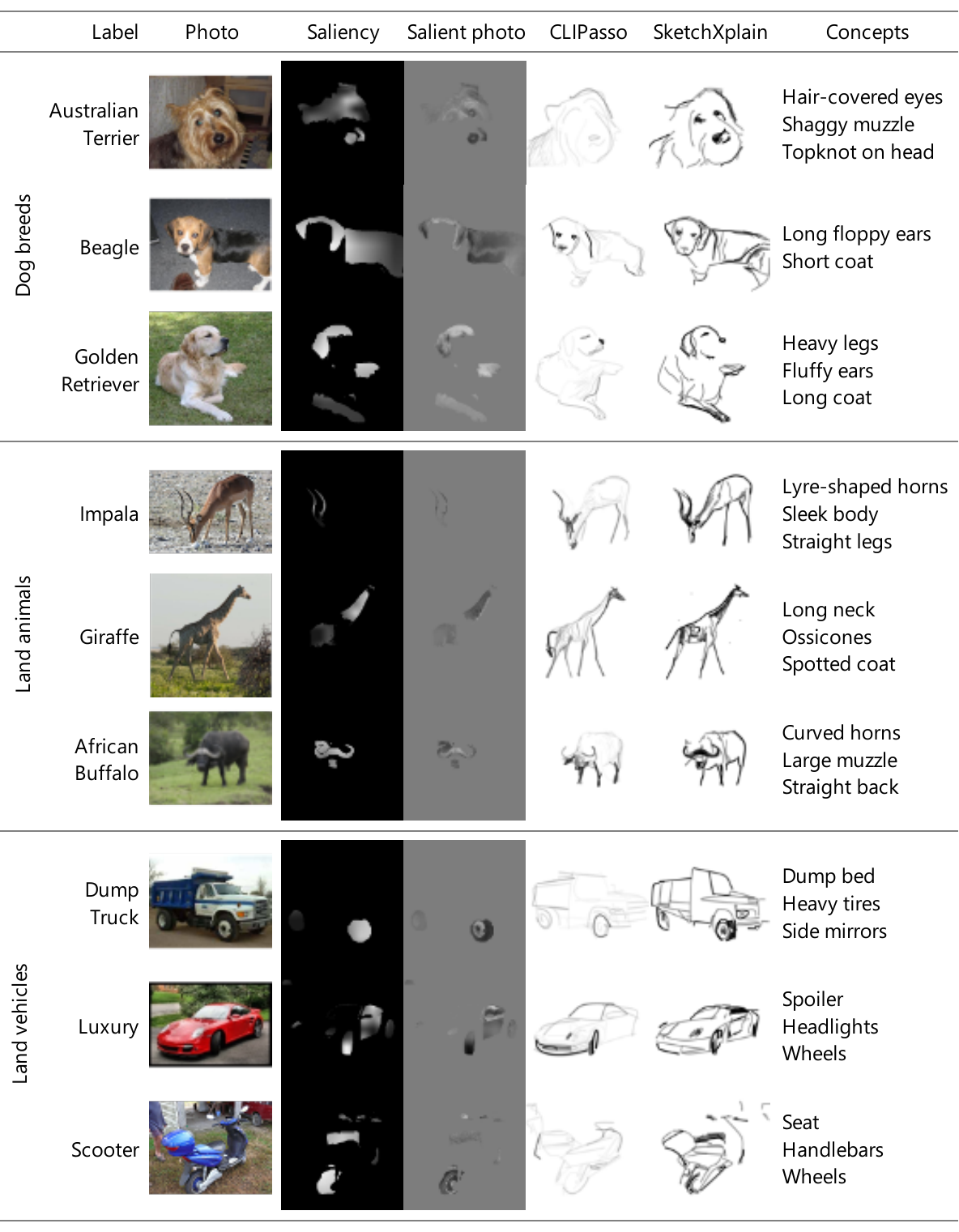}
        \label{table:demo-general}
    \end{minipage}

\end{figure*}

\subsection{Adapting SketchXplain for General Images}
\label{sec:general_image_sketchxplain}
Due to the increased heterogeneity of open-domain, general images, we adapted the concept bottleneck model (Step 2a) and cue localization (Step 2b) in SketchXplain to segment diverse concepts \fixed{(Appendix Fig.~\ref{fig:architecture-general})}.


\subsubsection{Semi-supervised Concept-bottleneck Model}
Unlike the BP4D face expression dataset, the datasets of general images do not have concept labels.
We prompted the large language model GPT-4o~\cite{achiam2023gpt} using the same prompt template as~\cite{oikarinen2023label} to generate a list of relevant concepts for each class \fixed{(Appendix Table~\ref{table:free_label_concepts})}.
Next, we trained a semi-supervised Label-Free Concept Bottleneck Model (LF-CBM)~\cite{oikarinen2023label} that co-trained two tasks for unsupervised learning of the concept labels, and supervised learning on the class labels.
At inference time, this model performs Step 2a to predict the concepts $\hat{\gamma}$, then uses them to predict the class $\hat{y}$.

\subsubsection{Open-domain Cue Localization}
While AU concepts tend to have fixed locations for the forward-looking faces in the BP4D dataset,
concepts are heterogeneously placed in general images.
This makes cue localization more challenging.
Thus, we substituted the prior saliency map approach with two sub-steps:
i) open-domain object detection with Owl-ViT~\cite{minderer2022simple} to extract the bounding boxes for each concept (if present) from the input image, and
ii) open-domain segmentation with Segment Anything Model (SAM)~\cite{kirillov2023segment} to segment a pixel mask of the concept within the bounding box.
We found that without object detection, SAM would spuriously select distant locations for the concept, thus the first sub-step was necessary.
The segments of all concepts are weighted by importance determined by the gradients from the concept bottleneck and aggregated.
The weighted segments were used as a mask to select edges to initialize the sketch strokes (Step 3a).
\balance
\subsection{Modeling Results}

\begin{figure*}[!h]
    \centering
    \includegraphics[width=13.0cm]{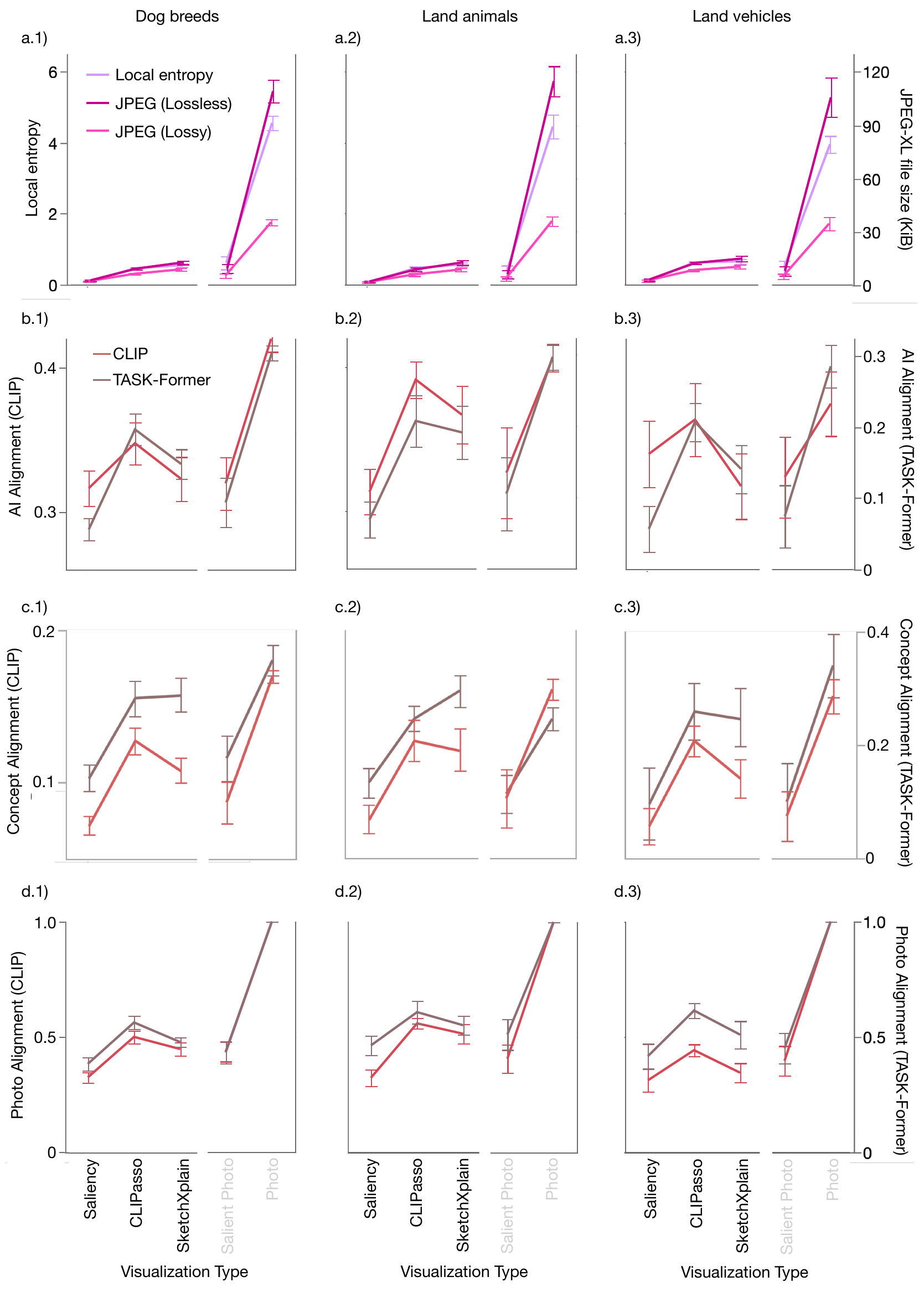}
    \vspace{-0.3cm}
    \caption{
        Results of modeling evaluation across Visualization Types 
        on visual complexity (row a) and coherence (rows: b, c, d) and   
        for general images of Dog Breeds (first column: a.1--d.1), Land Animals (a.2--d.2), and Land Vehicles (a.3--d.3). For each row, all graphs have the same left y-axes and the same right y-axes.
    }
    \label{fig:results_general_all}
    \vspace{-0.3cm}
\end{figure*}

We trained SketchXplain on 80\% of the dataset and tested it on the rest.
Class prediction accuracy was 91.8\% for Dog breeds, 88.1\% for Land animals, and 82.3\% for Land vehicles.


As in Section~\ref{sec:eval_face_modeling}, we conducted a modeling study for each dataset to compare the coherence and simplicity of SketchXplain against other visualizations. 
Appendix Table~\ref{table:demo-general} and Appendix Fig.~\ref{fig:results_general_all} show example explanations and the quantitative results, respectively.

In general, results for general images were similar to those for face expressions. 
\fixed{Regarding \textbf{coherence},} CLIP and TASK-Former \fixed{scores had similar trends}, except for \fixed{Appendix Fig.~\ref{fig:results_general_all}}c.1--c.3 where SketchXplain had higher Concept Alignment \fixed{as measured by} TASK-Former \fixed{(brown color lines)}.
\underline{\smash{SketchXplain}} sketch explanations were equally aligned to the photo, AI prediction and concept as other visualizations, except for \underline{\smash{Saliency}}, which was the lowest due to omitting fine visual details.
\fixed{Furthermore,} \underline{\smash{Salient Photo}} alignment was relatively lower for general images than for face images, 
perhaps due to the \fixed{salient regions being too small, fragmented, and scattered (see Appendix Table~\ref{table:demo-general}) for reliable CLIP representation.}
%
\fixed{Regarding} \textbf{visual complexity}, \underline{\smash{SketchXplain}} sketches had \fixed{low complexity (high simplicity) and} similar to \fixed{\underline{\smash{CLIPasso}} sketches.}
\underline{\smash{Saliency}} and \underline{\smash{Salient Photo}} \fixed{had lower complexity, possibly due to the small segment fragments.}


\subsection{Qualitative User Study}

Having shown that SketchXplain can faithfully generate sketches of other domains,
we next conducted a qualitative user study to explore the potential use cases, usefulness, and challenges in interpreting sketch explanations of general images.

\subsubsection{Method and Procedure}

We recruited 15 participants through university mailing lists and personal contacts. They were 8 females and 7 males, with ages 21--31 years old \fixed{(Median = 24.0)}.
Notably, 5 participants had a background in art or design\fixed{, who have experience with sketches}. 
We showed participants sketch explanations of various Visualization Types for \fixed{12} image instances \fixed{(4 per domain)}.
Participants described what they saw, why it mattered, and how they might use or share it. 
Due to the diversity of image domains, unlike for face sketches, we asked open-ended questions such as: 
\begin{itemize}[noitemsep, topsep=0pt]
    \item ``Why do you think the model focused on this part?'', 
    \item ``What information here feels important to you?'', 
    \item ``How would you use this explanation?'', and
    \item ``How would you share this explanation with others?''. 
\end{itemize}
Sessions were conducted over Zoom.
With consent, we recorded audio utterances and screen interactions.
\fixed{The experiment took 25--35 minutes and each participant was compensated \$7.80 USD in local currency.}

\subsubsection{Findings}
We conducted a thematic analysis of the recorded interviews, focusing on
i) perceived value of sketch explanations (SketchXplain and CLIPasso) and 
ii) possible \fixed{future} applications.


In general, \fixed{unlike} \underline{\smash{Photo}} which captured many details, sketch explanations provided an \textbf{abstraction} of visual content that \textit{``should be easy to understand, focusing on two or three features at a time''}~[P10] for recognizing an object from the image. 
\fixed{Sketches} were also \textit{``much easier to remember''}~[P13], \fixed{and} could facilitate \fixed{AI understanding} by \textit{``showing the thought and idea process''}~[P9]. 
Specifically, \underline{\smash{SketchXplain}} focused on 
\fixed{\textbf{coherent} visual cues},
such as floppy ears or a muzzle to \textit{``break down the dog into the right silhouette''}~[P1],
\fixed{the grille and headlights as \textit{``what makes it a Jeep''}~[P1]}, 
or ossicones and neck to confirm \textit{``what makes it a giraffe''}~[P3]. This \fixed{selective simplicity}  \textit{``[helped] to know which part to focus on''}~[P7],
\fixed{and facilitates \textbf{quick interpretation} \textit{``[made] it easier to identify the animal quickly''}~[P11]}.
In contrast, \underline{\smash{CLIPasso}} tended to \textit{``capture as much detail as possible''}~[P13] that could confuse interpretation.
However, participants noted the limitation that \underline{\smash{SketchXplain}} sometimes tended to \textbf{oversimplify details}, mentioned by P10 \textit{``too abstract, [because] some important lines [were] missing''}. This suggests that SketchXplain could be improved with fine-tuning to domain-specific concepts. 

\fixed{To formatively explore potential uses of sketch explanations, we asked participants about hypothetical uses for other applications.}
Sketches were seen as \textit{``good for anatomy or biology pathways''}~[P11], helpful to \textit{``identify dogs or birds for beginners''}~[P3], and clear enough to \textit{``show why it’s a swallow: tail and beak''}~[P3]. 
\textbf{Creative uses} included extracting \textit{``Jeep grille and headlights that carry brand identity''}~[P1] and \textit{``traits to share on a mood board so everyone sees the same key cues''}~[P1]. \textbf{Everyday communication} examples ranged from \textit{``a simple map for which subway exit to take''}~[P13] to \textit{``website layout sketches to organize rows of text and images''}~[P14].
Participants emphasized the \textbf{modularity} of sketches, which allowed refining or complementing them. 
\textbf{Editing} practices included \textit{``draw over directly''}~[P14], \textit{``use thicker strokes for changes''}~[P12], and \textit{``color-code corrections''}~[P12]. 
\textbf{Annotations} were also suggested: \textit{``combine a sketch with a short explanation''}~[P9], or \textit{``use arrows and brief text labels to make intent clear''}~[P9].
\fixed{Clearly, participants drew from their everyday or training experiences to articulate how sketch explanations should follow visualization communication design principles~\cite{agrawala2011design}.}

\balance
\end{multicols}
\end{document}